\documentclass[prd]{revtex4}
\usepackage{epsf,epsfig}

\def\dalemb#1#2{{\vbox{\hrule height .#2pt
        \hbox{\vrule width.#2pt height#1pt \kern#1pt
                \vrule width.#2pt}
        \hrule height.#2pt}}}

\def\hF{\hat F}
\def\tA{\widetilde A}
\def\tcA{{\widetilde{\cal A}}}
\def\tcF{{\widetilde{\cal F}}}
\def\hA{\hat{\cal A}}
\def\cF{{\cal F}}
\def\cA{{\cal A}}
\def\wdg{{\sst \wedge}}

\def\0{{\sst{(0)}}}
\def\1{{\sst{(1)}}}
\def\2{{\sst{(2)}}}
\def\3{{\sst{(3)}}}
\def\4{{\sst{(4)}}}
\def\5{{\sst{(5)}}}
\def\6{{\sst{(6)}}}
\def\7{{\sst{(7)}}}
\def\8{{\sst{(8)}}}

\def\tA{\widetilde A}

\def\wtd{\widetilde}

\let\w=\omega

\def\nn{\nonumber} \def\bd{\begin{document}} \def\ed{\end{document}}
\def\ds{\documentstyle} \let\fr=\frac \let\bl=\bigl \let\br=\bigr
\let\Br=\Bigr \let\Bl=\Bigl
\let\bm=\bibitem
\let\na=\nabla
\let\pa=\partial \let\ov=\overline
\newcommand{\be}{\begin{equation}}
\newcommand{\ee}{\end{equation}}
\def\ba{\begin{array}}
\def\ea{\end{array}}
\def\ft#1#2{{\textstyle{{\scriptstyle #1}\over {\scriptstyle #2}}}}
\def\fft#1#2{{#1 \over #2}}
\def\del{\partial}
\def\sst#1{{\scriptscriptstyle #1}}
\def\oneone{\rlap 1\mkern4mu{\rm l}}
\def\ie{{\it i.e.\ }}
\def\etc{{\it etc.\ }}
\def\via{{\it via}}
\def\semi{{\ltimes}}
\def\cv{{\cal V}}
\def\str{{\rm str}}
\def\jm{{\rm j}}
\def\im{{\rm i}}

\def\hF{\hat F}
\def\tA{\widetilde A}
\def\tcA{{\widetilde{\cal A}}}
\def\tcF{{\widetilde{\cal F}}}
\def\hA{\hat{\cal A}}
\def\cF{{\cal F}}
\def\cA{{\cal A}}
\def\wdg{{\sst \wedge}}

\newcounter{fixy}
 \newenvironment{fixy}[1]{\setcounter{figure}{#1}}
{\addtocounter{fixy}{1}}
\renewcommand{\thefixy}{\arabic{fixy}}
\renewcommand{\thefigure}{\thefixy\alph{figure}}
\begin{document}
 \begin{abstract}
We consider solutions of the four dimensional Einstein-Yang-Mills system
with a negative cosmological constant $\Lambda=-3g^2$,
where $g$ is the nonabelian gauge coupling constant. This theory corresponds to
a consistent truncation of ${\cal N}=4$ gauged supergravity and may be uplifted
to $d=11$ supergravity.
A systematic study of all known solutions is presented as well as new configurations
corresponding to rotating regular dyons and rotating nonabelian black holes.
The thermodynamics of the static black hole solutions is also discussed.
The generic EYM solutions present a nonvanishing magnetic flux at infinity and should
give us information about the structure of  a CFT in a background SU(2) field.
We argue that the existence of these configurations violating the no hair conjecture
is puzzling from the AdS/CFT point of view.
\end{abstract}

\title{
NONABELIAN SOLUTIONS IN AdS$_4$ \\AND $d=11$ SUPERGRAVITY}

\author{{\bf Robert B. Mann},$^{1,2}$~~ {\bf Eugen Radu}$^{3}$%
 ~~and ~~{\bf D. H. Tchrakian}$^{3,4}$
\\
\vspace{0.5cm}
$^{1} ${\small {Department of Physics, University of Waterloo, Waterloo, Ontario N2L
3G1, Canada}}\\
$^{2}${\small {Perimeter Institute for Theoretical Physics, Ontario N2J
2W9, Canada}}\\
$^{3} ${\small {Department of Mathematical Physics, National University of
Ireland, Maynooth, Ireland}}\\
$^{4} ${\small {School of Theoretical Physics -- DIAS, 10
Burlington Road, Dublin 4, Ireland}}
}

\maketitle

\section{INTRODUCTION}

Recently a tremendous amount of interest has been focused on anti-de Sitter (AdS) spacetime.
This interest is mainly motivated by the proposed correspondence between physical effects
associated with gravitating fields propagating in AdS spacetime and those of a conformal
field theory (CFT) on the boundary of AdS spacetime \cite{Witten:1998qj,Maldacena:1997re}.

A precise formulation of the AdS/CFT correspondence is made
in equating the generating
function of the correlation functions in the CFT
with the string/gravity partition function on the AdS space
\cite{Witten:1998qj, Gubser:1998bc}
\begin{eqnarray}
\label{frel}
Z_{AdS}[h ,\Psi _{0}]=\int_{[h ,\Psi _{0}]}D\left[ g\right] D\left[
\Psi \right] e^{-I\left[ g,\Psi \right] }=
\left\langle \exp \left(
\int_{\partial {\cal M}_{d}}d^{d}x\sqrt{\gamma}
{\cal O}_{[h ,\Psi_{0}]}\right) \right\rangle =Z_{CFT}[h ,\Psi _{0}].
\end{eqnarray}
The induced boundary metric and matter fields are respectively
denoted by $h $ and $\Psi _{0}$ symbolically, with $\mathcal{O}$ a
conformal operator defined on the boundary of $AdS_{d+1}$. The integration
is over configurations $\left[ g,\Psi \right] $ of metric and matter fields
that approach $[h ,\Psi _{0}]$ when one goes from the bulk of
$AdS_{d+1} $ to its boundary. This conjecture has been verified for several
important examples, encouraging the expectation that an understanding of
quantum gravity in an asymptotically AdS  spacetime  (AAdS)
can be obtained by studying its holographic CFT dual, defined on the
boundary of spacetime at infinity.

In this context,  various  AAdS  solutions of the Einstein equations
coupled with matter fields have been studied in the literature.
However, although the gauged supergravity AdS theories generically
contain Yang-Mills (YM) fields, most of the studies in the
literature have been restricted to the case of Abelian matter
content in the bulk. At the same time, a number of results obtained
for an asymptotically flat (AF) spacetime clearly indicate that a
variety of well-known, and rather intuitive, features of
self-gravitating Maxwell fields are not shared by nonabelian gauge
fields. In particular, and in contrast to the Abelian situation,
self-gravitating YM fields can form particle-like configurations
\cite{Bartnik:1988am}. Moreover, the Einstein-Yang-Mills (EYM)
equations also admit black hole solutions that are not uniquely
characterized by their mass, angular momentum, and YM charges
\cite{Volkov:sv}. Therefore the uniqueness theorem for electrovacuum
black hole spacetimes  ceases to exist for EYM systems.

As proven by some authors \cite{Winstanley:1998sn,Bjoraker:2000qd},
even the simple spherically symmetric EYM-SU(2) system with a
negative cosmological constant $\Lambda$ in four spacetime
dimensions presents a number of surprising results. A variety of
well known features of AF self-gravitating nonabelian solutions are
not shared by their AAdS regular and black hole counterparts.

Although the picture one finds is very much $\Lambda$-dependent,
it is still possible to identify some general features.
For example, there is always a continuum of regular and black hole
solutions in terms of the adjustable shooting parameters
that specifies the initial conditions at the origin or at the event horizon,
rather then discrete points.
Depending on the value of $\Lambda$, the spectrum has a finite number of continuous branches.
Furthermore, there exist nontrivial solutions that are stable against spherically symmetric
linear perturbations, corresponding to stable configurations with nonavanishing nonabelian charges.
The solutions are classified by nonabelian electric and magnetic charges
and the ADM mass.
When the parameter $\Lambda$ approaches zero,
an already-existing branch of monopoles and dyon solutions
collapses to a single point in the moduli space \cite{Hosotani:2001iz}.
At the same time new branches of solutions emerge.
A general study of these configurations together with a stability analysis
is presented in \cite{Sarbach:2001mc, Breitenlohner:2003qj}.

These spherically symmetric solutions have been generalized in various
directions.  Axially symmetric solutions are
discussed in \cite{Radu:2001ij}-\cite{Radu:2002rv},
NUT-charged configurations  and topological black holes
with nonabelian fields  were considered in \cite{Radu:2002hf,VanderBij:2001ia}.
Spherically symmetric, five dimensional solutions of the
EYM system with negative cosmological constant were examined in
\cite{Okuyama:2002mh}, this result being
extended recently to $d>5$ \cite{Radu:2005mj}.

Although further research is clearly necessary, at least some of the
EYM-SU(2) solutions, emerging as consistent reduction of $d=11$
supergravity on a seventh dimensional sphere
\cite{Cvetic:1999au,Pope:1985bu}, have relevance in AdS/CFT context.
In this case, the ratio between the four-dimensional cosmological
constant and the gauge coupling constant $g$ is fixed by
$\Lambda/g^2=-3$. Apparently the bulk/boundary correspondence for
AAdS EYM configurations has not received much attention in the
literature. 
Although all such solutions containing
non-abelian fields are classical they may have a role to play in the
full quantum theory. If the AdS/CFT correspondence conjecture is
indeed correct, it should either be able to account for solutions to
the EYM system from the CFT viewpoint or be able to
demonstrate why the conjecture does not apply to them.

The lack of attention given to AAdS EYM solutions is presumably due
to the notorious absence of closed form solutions in the presence of
nonabelian matter fields in the bulk.
(Very few such exact solutions exist 
e.g. \cite{Chamseddine:1997nm}, which features an effective negative
cosmological constant but is not AAdS, and,
\cite{Fuster:2005qt} for $\Lambda=0$.)
However, one can analyse their
properties by using a combination of analytical and numerical
methods, which is enough for most purposes. Euclidean solutions of
the $\Lambda=-3g^2$ EYM model have been discussed in
\cite{Maldacena:2004rf}, in a different context, however. The
authors of Ref. \cite{Maldacena:2004rf} considered Euclidean
wormhole solutions, with an $S^3$ conformal infinity.

The Lorentzian  solutions we consider here are very different.
Their conformal infinity is the product
of time and a two-dimensional sphere, plane, or hyperboloid.
In the black hole case, these are the nonabelian counterparts of the
well known AdS$_4$ Einstein-Maxwell solutions.
For a $R\times S^2$
boundary structure, there are also globally regular, particle-like solutions.
The existence of  configurations violating the no-hair conjecture
is puzzling from the AdS/CFT point of view.
Since several distinct solutions with the same set of boundary at infinity
data  may exist, it is not clear  how the dual CFT distinguishes between them.

In the first part of this paper we discuss the features
of the AAdS nonabelian solutions with $\Lambda=-3g^2$.
Although some of these solutions are  already known in literature,
 their properties have not been discussed for this particular value of the cosmological constant.
New types of EYM configurations with a nonvanishing angular momentum are presented as well.
All these configurations have a higher dimensional interpretation, solving the equations of motion of
$d=11$ supergravity.

The second part of this paper attempts a discussion of these solutions
in an AdS/CFT context.  The boundary stress tensor
and the associated conserved charges are computed in Section 4,
where we also present a discussion of black hole thermodynamics.
We further argue that studying the $\Lambda=-3g^2$ EYM-AdS
system should give us information about the structure of
the CFT in a background SU(2) field.  We conclude with Section 5.
The Appendices contain a discussion of technical aspects of solutions'
construction.

\section{GENERAL FRAMEWORK}
We start by discussing the embedding of the EYM solutions in M-theory.
It is known that the standard ${\cal N}=4$ SO(4) gauged supergravity in $d=4$ can be viewed
as a reduction of $d=11$ supergravity on $S^7$ \cite{Cvetic:1999au}.
The bosonic sector of this theory contains two SU(2) fields $F_{\mu \nu},~{\tilde F}_{\mu \nu}$,
a dilaton $\phi$ and an axion $\chi$ with the Lagrangian density
\begin{eqnarray}
{\cal L} &=&  {\cal R}  - \frac{1}{2} \partial_\mu\phi \partial^\mu \phi - \frac{1}{2}
e^{2\phi}\partial_\mu\chi \partial^\mu \chi - V(\phi,\chi)\,
- \frac{1}{2} e^{-\phi}\, {F_{ \mu \nu}^a}  F^{a \mu \nu} \nn\\
&& -\frac{1}{2} \fft{e^\phi}{1+\chi^2\, e^{2\phi}}\, {  \wtd
F^{a}_{\mu \nu}}  \wtd F^{a\mu \nu}\, -
\frac{1}{2\sqrt{-g}}\chi\,\epsilon_{\mu \nu \rho \sigma} F^{a \mu
\nu}  F ^{a\rho \sigma} +\frac{1}{2\sqrt{-g}} \frac{\chi\,
e^{2\phi}}{1+\chi^2\, e^{2\phi}}\, \epsilon_{\mu \nu \rho \sigma}
\wtd F^{a \mu \nu} \wtd F^{a\rho \sigma},
\end{eqnarray}
 where the
potential $V(\phi,\chi)$ is
\begin{eqnarray}
V(\phi,\chi) = -2g^2\, (4+ 2 \cosh\phi + \chi^2\, e^\phi).
\end{eqnarray}
It can easily be seen that
$\chi=\phi=0$ is a consistent truncation of this theory
for $F_{ \mu \nu}={  \wtd F_{\mu \nu}}$
 and, as a result one finds the action
\begin{equation}
\label{action}
I=\frac{1}{4}\int d^{4}x\sqrt{-g}[ \mathcal{R} - 2 \Lambda
-2Tr(F_{\mu \nu }F^{\mu \nu })]
\end{equation}
 where the field strength tensor $F_{\mu \nu}=\frac{1}{2} \tau^a F_{\mu\nu}^{a}$
 is
\begin{equation}
F_{\mu \nu} =
\partial_\mu A_\nu -\partial_\nu A_\mu + i \left[A_\mu , A_\nu \right]
\ , \label{fmn}
\end{equation}
and the gauge field
$A_{\mu} = \frac{1}{2} \tau^a A_{\mu}^{a},$ with $\tau^a$ an SU(2) basis
written in terms of Pauli matrices (the value of the gauge coupling $g$
has been set to one without loss of generality).
The gauge field transforms as
$
A_{\mu}' = U A_{\mu} U^\dagger + i (\partial_\mu U) U^\dagger
$
under a SU(2) gauge transformations $U$.

The value of the cosmological constant as read from $V(\phi=0,\chi=0)$ is $\Lambda=-3$.

As usual, to ensure well-defined Euler-Lagrange field equations,
one adds to the action principle (\ref{action}) the
Hawking-Gibbons surface term \cite{Gibbons:1976ue},~~ $I_{{\rm
surf}}=-\frac{1}{2}\int_{\partial\mathcal{M}} d^{3}x\sqrt{-h}K$
where $K$ is the trace of the extrinsic curvature for the boundary
$\partial\mathcal{M}$ and $h$ is the induced metric of the
boundary. This term does not affect the equations of motion but it
is relevant in the discussion  of  solutions' mass and boundary
stress tensor.

Variation of the action (\ref{action})
 with respect to the metric $g^{\mu\nu}$ leads to the Einstein equations
\begin{equation}
\label{einstein-eqs}
R_{\mu\nu}-\frac{1}{2}g_{\mu\nu}R +\Lambda g_{\mu\nu}  = 2 T_{\mu\nu},
\end{equation}
where the YM stress-energy tensor is
\begin{eqnarray}
\label{tik}
T_{\mu\nu} = 2{\rm Tr}
    ( F_{\mu\alpha} F_{\nu\beta} g^{\alpha\beta}
   -\frac{1}{4} g_{\mu\nu} F_{\alpha\beta} F^{\alpha\beta}).
\end{eqnarray}
Variation with respect to the gauge field $A_\mu$
leads to the YM equations
\begin{equation}
\label{YM-eqs}
D_{\mu}F^{\mu\nu}=0,
\end{equation}
where $D_{\mu}=\partial_{\mu} + i\left[A_\mu , \cdot \right]$.

In this paper we will consider solutions of the EYM equations possesing at
least two Killing vectors $\xi=\partial_t$, $\eta=\partial_{\varphi}$, corresponding to
a stationary, axially symmetric spacetime.
For the time translation symmetry, we choose a natural gauge
such that the matter fields have no time dependence, $\partial A/\partial t$=0.
However, a rotation around the $z-$axis
can be compensated by a gauge rotation
${\mathcal{L}}_\varphi A=D\Psi$ \cite{Forgacs:1980zs},
with $\Psi$ being a Lie-algebra valued gauge function.
This introduces an integer $n$ in the matter ansatz (which is a constant of motion)
and implies the existence of a potential $W$ with
\begin{eqnarray}
\label{relation}
F_{\mu \varphi} =& D_{\mu}W,
\end{eqnarray}
where $W=A_{\varphi}-\Psi$.
The integer $n$  represents the winding number with
respect to the azimuthal angle $\varphi$.
While $\varphi$ covers the trigonometric circle once,
the fields wind $n$ times around.
The qualitative features of the solutions obtained in the abelian
sector of the theory are insensitive to $n$; however, the winding number
determines to some extent the properties of the nonabelian solutions.

Using the formulas in \cite{Cvetic:1999au}
with the two equal gauge fields, $A^a = {\tilde A}^a$, we can uplift any configuration
that extremizes the action principle (\ref{action}) to $d=11$
supergravity. The eleven dimensional metric ansatz reads
\begin{eqnarray}
  ds^2_{11} = ds_4^2 + 4d\xi^2 +
\cos^2\xi \sum_a (\Theta^a-A^a_\mu dx^\mu)^2 + \sin^2\xi
\sum_a (\tilde{\Theta}^a-A^a_\mu dx^\mu)^2.
\end{eqnarray}
The antisymmetric tensor field $\hat{F}_{(4)}$, which appears in the action principle of the
$d=11$ supergravity, can be read from \cite{Cvetic:1999au, Pope:1985bu}
\begin{eqnarray}
\hat{F}_{(4)} =
-3\epsilon_{(4)}+\sqrt{2}\sin\xi\cos\xi d\xi\wedge
(\Theta-\tilde{\Theta})^a\wedge *F_2^a +\cr +{\sqrt{2}\over 4}\epsilon_{abc}
[\cos^2\xi(\Theta-A)^a\wedge
(\Theta-A)^b + \sin^2\xi(\tilde{\Theta}-A)^a\wedge (\tilde{\Theta}-A)^b]  \wedge *F_2^c
\end{eqnarray}
where $\Theta^a$, $\tilde{\Theta}^a $ are SU(2) right
 invariant one forms on two 3-spheres $S^3 , \tilde{S}^3$.

\section{$\Lambda=-3$ EYM SOLUTIONS}
The field equations (\ref{einstein-eqs}), (\ref{YM-eqs}) contain a
large variety of solutions. First, any solution $(g_{\mu
\nu},~{\cal A}_\mu)$ of the Einstein-Maxwell (EM) theory with
$\Lambda=-3$ can be viewed as a solution of the EYM equations
(\ref{einstein-eqs}), (\ref{YM-eqs}) by taking $A_\mu={\cal A}_\mu
T$ where $T$ belongs to the Lie algebra of the nonabelian gauge
group \cite{Yasskin}. The properties of these AAdS abelian
solutions have been discussed by various authors (for generic
$\Lambda$), as well as their relevance in AdS/CFT context.

Here we restrict ourselves to the pure nonabelian case.
As expected, a number of EYM configurations with $\Lambda=0$
are found to possess AAdS
counterparts, with very different properties, however.
For $\Lambda<0$, there are also rotating regular EYM solutions,
which  do not survive in the AF limit.

In this Section we present a discussion
of the   EYM solutions which arise as a truncation of ${\cal N}=4$ SO(4) gauged supergravity,
both regular and black hole
configurations being considered.
Some of these solutions have been already presented in the literature,
however, without discussing the case of interest $\Lambda=-3$.
Also, to simplify the general picture we will not consider dyon solutions,
except for the rotating case where Einstein equations
require the presence of a YM electric field.

The mass-energy of these solutions is computed by using the formalism presented
in Section 4. Also, we will present here general features of these solutions,
without entering into technical details, which are the subject of
the Appendices A-D.

\subsection{Regular configurations}
\subsubsection{The $\Lambda=-3$ Bjoraker-Hosotani monopole solutions}
We start with the simplest case, corresponding to spherically symmetric solutions.
These are the AdS generalizations of the Bartnik-McKinnon configurations
found in \cite{Bjoraker:2000qd} by Bjoraker and Hosotani, within a metric ansatz
\begin{eqnarray}
\label{sph}
ds^2 = \frac{dr^2}{H(r)}
 + r^2 (d\theta^2 + \sin^2 \theta \, d\varphi^2)-\frac{H(r)}{p^2(r)} dt^2
\end{eqnarray}
where
\begin{eqnarray}
\nonumber
H(r) = 1 - \frac{2   m(r)}{ r}-\frac{\Lambda}{3} r^2 ,
\end{eqnarray}
$m(r)$ corresponding to the local mass-energy density.

The static, spherically symmetric SU(2) YM ansatz is obtained for a winding number
$n=1$ and can be parametrized
by one real function   $\omega(r)$
\begin{eqnarray}
\label{A1}
A=\frac{1}{2} \Big\{  w(r)\tau_1d\theta
+(\cot\theta\tau_3+w(r)\tau_2)\sin\theta d\varphi\Big\},
\end{eqnarray}
which implies the field strength tensor expression
\begin{eqnarray}
F=
\frac{1}{2}\Big\{
w^{\prime}  dr\wedge (\tau_1 d\theta + \tau_2 \sin\theta d\varphi)
-(1-w^2 )\tau_3 d\theta\wedge\sin\theta d\varphi \Big\},
\label{YM-static1}
\end{eqnarray}
where a prime denotes a derivative with respect to $r$.
For this purely magnetic  ansatz, the EYM equations take the form
\begin{eqnarray}
\label{eq-hoso}
m'=\w'^2 H+\frac{(w^2-1)^2}{2 r^2},
~~~~
w''=\frac{w'}{H} \left ( \frac{2\Lambda r}{3}
-\frac{2m}{r^2}+\frac{(w^2-1)^2}{r^3} \right)
+\frac{w (w^2-1)}{r^2 H},
\end{eqnarray}
the equation for $p$ decoupling from the rest,
$p'/p=-2w'^2/r$.

No exact solutions of this system are known yet, so the equations
must be solved numerically. The solutions with a regular origin
have the following behaviour at $r=0$
\begin{eqnarray}
\label{origin}
w(r) = 1-br^2 +O(r^4),~~m(r)=2b^2r^3+O(r^4),~~p(r)=p_0(1-4b^2r^2)+O(r^4),
\end{eqnarray}
where $b,~p_0$ are real constants.
We are here interested in solutions with AdS asymptotics, which implies
the following expansion at large $r$
\begin{eqnarray}
\nonumber
&&m(r)=
M+\left(\frac {\Lambda C_1^{2}}{3}-\frac {1}{2} (
\omega _{\infty }^{2}-1) ^{2}\right)\frac {1}{r} +O\left( \frac {1}{r^{2}} \right),
~~
\omega (r) = \omega _{\infty } +\frac {C_1}{r}
+O\left( \frac {1}{r^{2}} \right),
\\
\label{asympt}
&&p(r)=1+ \frac{C_1^2}{r^4}+O\left( \frac {1}{r^5}\right),
\end{eqnarray}
where $\omega_0$, $M$ and $C_1$   are constants determined by
numerical calculations. $M$ corresponds to the ADM mass of the
solutions, while
 $\omega_0$ determines the value of the magnetic charge, $Q_m=|1-\omega_0^2|$ (see Appendix C).
For $\Lambda=0$,  $\omega_0=\pm 1$ are the only allowed values, $(M,~b)$
being restricted to a discrete family indexed by the node number of the gauge function
$\omega(r)$ \cite{Bartnik:1988am}.

As found in \cite{Bjoraker:2000qd}, the AAdS solutions
are much less restricted.
By varying the  parameter $b$ which enters the expansion at the origin (\ref{origin}),
a continuum of monopole solutions are found.
These configurations are regular in the entire
space. The overall picture depends on the value of the cosmological constant;
for $\Lambda=-3$, solutions with AAdS asymptotics are found for only one branch with
 $-0.557< b < 1.31$.
In Figure 1, we plot the asymptotic quantities $M$ and $\omega_0$
 as well as the value of the metric function $p$ at the origin as a function
 of the parameter $b$. One can see that $p(0)$ diverges as $b \to b_{max}$ ($i.e.$ $g_{tt}(0) \to 0$),
 while $M(b_{max}),~\omega(b_{max})$ remain finite, while
all these parameters appear to diverge as $b \to b_{\min}$.
The critical solutions have been studied in \cite{Breitenlohner:2003qj}
and are not of interest here.

A configuration is uniquely characterized by the asymptotic parameters $M,\omega_0$.
 As a characteristic feature of $\Lambda=-3$ globally regular
 configurations,
we notice the existence of zero- and one-node monopole
solutions only.
The nodeless solutions are of particular interest because, as discussed in \cite{Bjoraker:2000qd},
they are stable against linear perturbations.
 The solution with $b\simeq 0.619$ has $\omega_0=0$ and, similar to the
't Hooft-Polyakov monopole, has magnetic charge $Q_m=1$.
In this case the AdS boundary conditions play a role similar to the Higgs field in the AF case.
As seen in Figure 1,
for values of $b$ in the interval $ 0.62 < b < 1.31$ the gauge function
$\omega$ crosses the $r-$axis,
approaching a negative value at infinity.
Although for $0.635 \leq b \leq 1.309$ there are two configurations with the same
asymptotic value of the gauge function,
these are distinguished by the value of $M$, with $M(b_2)<M(b_1)$ if $b_2<b_1$.

Typical configurations are displayed  in Figure 2 for three diferent values of the
parameter $b$. The solution with $b=1.3$ has $\omega_0=-0.045$,
corresponding to a near critical configuration with $g_{tt}(0)=-0.0032$.

\subsubsection{Axially symmetric monopoles}
It is well known that, for $\Lambda=0$, an  SU(2) YM-theory coupled
to  gravity  also possesses AF static
axially symmetric globally regular solutions
\cite{Kleihaus:1997mn}, labeled by
a winding number $n>1$.
The AAdS generalizations of these configurations were reported in
\cite{Radu:2001ij}, featuring very different properties, as expected.
Here we present a discussion of these solutions for the special case $\Lambda=-3$.

As discussed in Appendix A, the minimal axially symmetric YM ansatz is
parametrized by four functions $H_i(r,\theta)$. These magnetic potentials
satisfy a suitable set of boundary conditions at the origin, at infinity
and on the symmetry axis imposed by finite energy, regularity and symmetry requirements.
The spherically symmetric YM ansatz is recovered for a unit winding number $n=1$,
two vanishing gauge potentials $H_1=H_3=0$ while the other two are equal $H_2=H_4=\omega(r)$.

For large $r$, the configuration becomes spherically symmetric,
with $H_2=H_4=\omega_0$, the asymptotic value of the other two functions being zero.
The magnetic charge of these solutions is $Q_M=n|1-\omega_0^2|$.
The expression of the gauge ansatz and the boundary conditions are presented in
Appendices A, B; see also Appendix D for a discussion of the numerical
procedure we used to find these solutions.

The static axially symmetric EYM solutions are obtained within a metric ansatz
\begin{equation}
\label{axial}
ds^2= \frac{m}{f} \big( \frac{d r^2}
{1-\frac{\Lambda}{3} r^2}+ r^2 d \theta^2 \big)
           +  \frac{l}{f} r^2 \sin ^2 \theta d\varphi^2
       - f(1-\frac{\Lambda}{3} r^2) dt^2,
\end{equation}
where the metric functions
$f$, $m$ and $l$ are only functions of
$r$ and $\theta$. Here $r$ is the radial coordinate,
$t$ is a global time coordinate,
$(\theta,\varphi)$ being the usual coordinates on the sphere,
with $0\leq \theta\leq\pi,0\leq \varphi<2 \pi$.
The expansion of the metric functions as $r\to \infty$ is
\begin{eqnarray}
\label{as1}
f=1+\frac{f_1+f_2 \sin^2 \theta}{r^3}+O(\frac{1}{r^5}),~~
m=1+\frac{m_1+m_2 \sin^2 \theta}{r^3}+O(\frac{1}{r^5}),~~
l=1+\frac{l_1+l_2 \sin^2 \theta}{r^3}+O(\frac{1}{r^5}),
\end{eqnarray}
which leads to AAdS solutions. Here $f_1,~f_2$ are undetermined constants, while
$l_1=m_1=2f_1/3,$ $l_2= 6f_2/17,$ $m_2= 14f_2/17.$
As discussed in Section 4, the mass $M$ of these configurations
is encoded in the parameters $f_1,~f_2$
\begin{eqnarray}
\label{ct-mass}
M= \frac{\Lambda}{3}\left(\frac{2f_1}{3}+\frac{8f_2}{17}\right).
\end{eqnarray}
A spherically symmetric spacetime is recovered for  $l=m$ and
$f_2=m_2=l_2=0$ (i.e. no  angular dependence).
The coordinate transformation between the resulting line
element and the more familiar Schwarzschild-like form (\ref{sph}) is discussed
in Ref. \cite{Radu:2001ij}.

The results of the numerical integration
indicate that every $\Lambda=-3$
spherically symmetric regular solution appears to present axially symmetric
generalizations.
For any given winding number, we find only one branch of solutions
classified
by the mass and the value of the parameter $\omega_0$ (we have studied solutions up to $n=4$).
Similar to the $n=1$ case, the functions $H_2$ and $H_4$ are nodeless or present one node only.
The gauge potentials $H_1,~H_3$ always present a complicated $\theta-$dependence,
while one finds usually a small angular dependence for $H_2,~H_4$.
The metric functions $f$, $m$ and $l$ do not exhibit a strong angular dependence,
while $m$ and $l$ have a rather similar shape.
As expected, the angular dependence of the metric and matter functions
increases with $n$.
Also, the values at the origin of the metric functions decreases with $n$.
The typical profiles of the metric and gauge functions are similar to those exhibited
in \cite{Radu:2001ij} and we will not present them here.

In Figure 3 we plot the mass $M$  as a function of the $\omega_0$
for various winding numbers. We see that the $n>1$ branches generally follow the picture found for $n=1$
(with higher values of mass, however).

The energy density of the matter fields $\epsilon$, shows a
pronounced peak along the $\rho$-axis and decreases monotonically
along the $z$-axis (with $z=r \cos \theta,~\rho=r \sin \theta$).
The contours of equal energy density $\epsilon=-T_t^t$ are
two-torii and squashed two-spheres. The peak of the energy density
along the $\rho$-axis
slightly shifts outward with increasing $n$ and increases in
height. For a fixed value of $\omega_0$, the mass of the
solutions, $M(n)$, increases with $n$. For example, for
$\omega_0=0.05$, one has $M(1)=1.014$, $M(2)=2.467$, $M(3)=4.747$
and so on. As a general feature, the particle-like nonabelian
solitons are less massive than the extremal RNAdS solutions with
the same magnetic charge.

We have found  it difficult to obtain axially symmetric
generalizations of the spherically
symmetric solutions near limits of the $b$-interval,
with large errors for the functions.
A different metric parametrization appears to be necessary.

\subsubsection{Rotating regular solutions}
An interesting physical question is whether these static nonabelian regular
solutions can be
generalized to include an angular momentum.
For the AF case, contrary to results from perturbation theory \cite{Brodbeck:1997ek},
no rotating generalizations
of the BK solutions seem to exist.
In this case, the $A_t$ components of the gauge field act like an isotriplet Higgs field
with negative metric, and by themselves would cause the other components of the gauge field
to oscillate as $r \to \infty$ \cite{VanderBij:2001nm}, which would imply an infinite
mass.
Rotating solutions are found by including in the theory a triplet
Higgs field with a vacuum expectation value
greater than the asymptotic value of $A_t$ \cite{Paturyan:2004ps}.

For $\Lambda<0$, there are no boundary conditions to exclude a nonabelian solution
with nonzero electric potential; dyon EYM solutions already exist
in the spherically symmetric case \cite{Bjoraker:2000qd}.
The existence of dyon solutions without a Higgs field
is a feature for AdS spacetime;
if $\Lambda \ge 0$ the electric part of the gauge fields
is forbidden \cite{Bjoraker:2000qd, Galtsov:1989ip}.
This makes possible the existence of rotating regular configurations, too.
A discussion of these solutions in a fixed AdS background is presented in \cite{Radu:2002rv},
where is argued that they survive
in the presence of gravity.
However, the general picture appears to be very complicated,
crucially depending on the value of the cosmological constant.
Here we analyse the solutions' properties for $\Lambda=-3$.

These rotating EYM solutions are found for a
 metric form  generalizing (\ref{axial}) for an extradiagonal metric
component $g_{\varphi t}$,
which satisfies also the circularity condition  \cite{wald}
\begin{eqnarray}
\label{rot}
ds^2= \frac{m}{f} ( \frac{d r^2}{1-\frac{\Lambda}{3}r^2}+ r^2 d \theta^2 )
+  \frac{l}{f} r^2 \sin ^2 \theta (d\phi+\frac{\Omega}{r}dt)^2 - f(1-\frac{\Lambda}{3}r^2) dt^2,
\end{eqnarray}
where the four metric functions $f$, $m$, $l$ and $\Omega$
depend only on the coordinates $r$ and $\theta$.
At infinity, the asymptotic form of the metric functions $f,~l,~m$ is still
given by (\ref{as1}) while the metric function associated with rotation decays as
\begin{eqnarray}
\label{rot1}
\Omega=\frac{j_1+j_2 \sin^2 \theta}{r^3}+O(\frac{1}{r^4}),
\end{eqnarray}
$j_1,~j_2$ being two real constants.
As discussed in Section 4, the mass-energy $M$ of the rotating solutions
is still given by (\ref{ct-mass}), while their angular momentum is
\begin{eqnarray}
\label{ct-J}
J= \frac{\Lambda}{3}\left(\frac{j_1}{2}+\frac{2j_2}{5}\right).
\end{eqnarray}
The YM ansatz contains in this case six functions: four magnetic potentials
$H_i$ $(i=1,..,4)$ and two electric potentials $H_5,~H_6$.
The boundary conditions satisfied by the magnetic
potentials are similar to static case. For
  the electric potentials we impose at infinity
$H_{5}=V \cos \theta,~H_{6}=V \sin \theta$.
The parameter $V$ corresponds to the asymptotic magnitude of the
electric potential $A_t(\infty)$ and determines the properties
of these solutions.
In the abelian case, by using a suitable
gauge transformation one can set set $V=0$
(or any other value) without any loss of generality. In the nonabelian theory,
however, such a gauge transformation would render the whole configuration
time-dependent.
In this case, $V$ enters
the asymptotic expansion for the nonabelian field strength and has a physical relevance
(although it does not contribute to the electric YM charge).
As discussed in Appendix C, the value
$V=0$ implies a purely magnetic, static configuration
$A_t=\Omega=0$.
Technical details on these solutions, including the boundary conditions are presented in Appendix B, D.

Globally regular rotating solutions are found by
staring with a purely magnetic EYM configuration with a given $\omega_0$
and increasing the value of $V$.
Here we consider only rotating solutions with the lowest winding number $n=1$, although
we obtained a number of configurations with $n=2$ also.

The branch structure of the spherically symmetric solutions
is preserved in the presence of rotation.
The rotating configurations depend on two continuous parameters: the value $\omega_0$
of the magnetic potentials $H_2,~H_4$ at infinity (which fixed the magnetic charge $Q_m=n|1-\omega_0^2|$)
and the magnitude of the
electric potential at infinity $V$.

For a given value of the magnetic charge, a branch of rotating dyon solutions emerges
smoothly from every spherically monopole solution
and extends up
to some maximal value of $V$ beyond which gravity becomes
too strong for regular dyons to persist.
We notice that the value at the origin of the metric function $f$ decreases
with increasing $V$ and tends to zero as
$V$ approaches the critical value $V_{max}$ (which is $\omega_0$ dependent),
corresponding to the formation
of a horizon.
With increasing $V$, the dyon becomes more and more deformed.
The mass, angular momentum and electric charge increase with $V$
and we find again a maximal value for the magnitude of the electric potential at infinity $V_{max}$.
Alternatively, we may keep fixed the magnitude of the electric potential at infinity
and vary the parameter $\omega_0$.
Again, it has proven difficult  to obtain rotating
generalizations of the spherically
symmetric solutions near the limits of the $b$-interval.
  In Figure 4 we present the properties of typical branches of solutions for
a fixed value of $\omega_0$ (Figure 4a) and for a fixed $V$ (Figure 4b).

All solutions we have found present nonvanishing nonabelian electric and magnetic charges,
representing rotating dyons.
A vanishing $Q_e$ implies a nonrotating, purely magnetic configuration.
However, we find  dyon solutions with vanishing total angular momentum
(e.g. $J=0$ for $\omega_0=0.895$, $V=2$) that are
not static (locally $T_{\varphi}^t \neq 0$).

Similar to the static case, the functions $H_2$ and $H_4$ are nodeless
or present one node only,
although they have a small $\theta$ dependence, while $H_1$, $H_3$ and the electric
potentials depend on $\theta$ in a complicated way.
The metric functions $f,l,m$ present a rather small angular dependence.

For all configurations, the energy density of the solutions
has a strong peak along the $\rho$ axis,
and it decreases monotonically along the symmetry axis, without being possible to
clearly distinguish any individual component.

\subsection{Black hole configurations}
The spherically symmetric  black holes
found by Winstanley in \cite{Winstanley:1998sn} were the first nonabelian solutions
with  AdS asymptotics presented in the literature.
More details on these configurations have been presented in
\cite{Bjoraker:2000qd}, including dyonic black holes.
These solutions obviously violate the no-hair conjecture and
present static axially symmetric generalizations
that are absent in the abelian sector.

Also, it is well known that for $\Lambda<0$, the
EM theory has
black hole solutions for which the topology of the horizon
is an arbitrary genus Riemann surface.
The EYM-SU(2) counterparts of these solutions
with a nonspherical event horizon topology have been
discussed in \cite{VanderBij:2001ia}.

The properties of the AAdS nonabelian solutions are strikingly different from
those valid in the AF case.
For example,  black holes with $\Lambda<0$ exist
for continuous intervals of the parameter
space (the value of the gauge field on the event horizon), rather than discrete points.
Also, there are configurations for which the gauge field has no zeros.
Moreover, some of these configurations are stable within a perturbation theory
approach.

Since the features of these configurations depend on the value
of the cosmological constant, we present in this Section
an analysis of their properties for $\Lambda=-3$.
Apart from some classes of configurations    already known in the
literature,
we present here numerical arguments for the existence of rotating
black holes with nonabelian hair, generalizing for an SU(2) field
the Kerr-Newman-AdS solution.
\subsubsection{$n=1$ static solutions}
We start by discussing the better known spherically symmetric solutions
and their topological black hole counterparts.

These solutions are obtained within a metric ansatz generalizing (\ref{sph})
for a nonspherically symmetric topology
of the event horizon
\begin{equation}
\label{metric}
ds^{2}=\frac{dr^{2}}{H(r)}+r^2d \Omega_k^2-\frac{H(r)}{p^2(r)} dt^{2}
\end{equation}
where
\begin{equation}
H(r)=k-\frac{2m(r)}{r}-\frac{\Lambda r^2}{3}.
\end{equation}
Here $d \Omega_k^2=d\theta^{2}+f_k^{2}(\theta) d\varphi^{2}$
is the metric on a two-dimensional surface $\Sigma$ of constant curvature $2k$.
$r$ is the radial coordinate for which
$r\to \infty$ defines the asymptotic region.
The discrete parameter $k$ takes the values $1, 0$ and $-1$
and implies the form of the function $f_k(\theta)$
\begin{equation}
\label{fk}
f_k(\theta)=\left \{
\begin{array}{ll}
\sin\theta, & {\rm for}\ \ k=1 \\
\theta , & {\rm for}\ \ k=0 \\
\sinh \theta, & {\rm for}\ \ k=-1.
\end{array} \right.
\end{equation}
The topology of a constant
$(t,r)$ slice is $H^2_g$.
When $k=+1$, the universe takes on the familiar
spherically symmetric form, and the $(\theta, \varphi)$ sector has
constant positive curvature. When $k=0$, the $\Sigma$ is a flat
surface, while for $k=-1$, the $(\theta, \varphi)$ sector is a
space with constant negative curvature, also known as a hyperbolic
plane. When $\Sigma$ is closed, we denote its area by $V_k$. Also,
we set $V_k/4 \pi=1$ in all numerical data we present in this
paper.

The construction of the SU(2) connection
is presented in  \cite{VanderBij:2001ia}.
Taking into account the symmetries of the line element (\ref{metric}) we find
\begin{equation}
\label{Atop}
A=\frac{1}{2} \Big\{
\omega(r) \tau_1  d \theta
+\big(\frac{d \ln f_k}{d \theta} \tau_3
+ \omega(r) \tau_2 \big) f_k d \varphi \Big\},
\end{equation}
which reduces to (\ref{A1}) for $k=1$.

As a result, we obtain a simplified YM curvature
\begin{equation}
F=\frac{1}{2}\Big\{
\omega' \tau_1 dr\wedge d\theta +
f_k \omega' \tau_2 dr\wedge d\varphi +
(w^2-k)f_k \tau_3 d\theta \wedge d\varphi \Big\}.
\end{equation}
The EYM equations reduce to
\begin{eqnarray}
\label{eq-bh}
m'=\omega'^2 H+\frac{(\omega^2-k)^2}{2 r^2},
~~~~
(\frac{H\omega'}{p})'=\frac{\omega(\omega^2-k)}{p r^2},
~~~~
\frac{p'}{p}=-\frac{2}{r}\omega'^2,
\end{eqnarray}
We find the following expansion near the event horizon which is located at $r=r_h>0$,
\begin{eqnarray}
\label{expansion}
m(r)=\frac{r_h}{2}\left(k-\frac{\Lambda r_h^2}{3}\right)+m'(r_h)(r-r_h),~~
\omega(r)=\omega_h+\omega'(r_h)(r-r_h),
\end{eqnarray}
where
\begin{eqnarray}
m'(r_h)=\frac{(\omega_h^2-k)^2}{2 r_h^2},
~~~
\omega'(r_h)&=&\frac{r_h\omega_h(\omega_h^2-k)}
{(k-\Lambda r_h^2)r_h^2-(\omega_h^2-k)^2},
\end{eqnarray}
(since the equations (\ref{eq-bh})  are invariant under the
transformation $\omega \rightarrow - \omega $, it is enough to consider only values of
$\omega _{h}>0$).
For $k=1$, $w(r)=\pm 1$ corresponds to vacuum Schwarzschild-AdS
solution, while $w(r)=0$ is the abelian Reissner-Nordstr\"om-AdS (RNAdS) solution,
with unit magnetic charge.
Note that, apart from embedded abelian configurations,
no extremal solutions with reasonable asymptotics
exist in this case.

The condition for a regular event horizon is
$
H'(r_h) >0
$
and places a bound on $\omega_h$
\begin{eqnarray}\label{mhbound}
2m'(r_h ) =\frac {\left( \omega_h^{2}-k \right) ^{2}}{r_h ^{2}}
<k-\Lambda r_h ^{2},
\end{eqnarray}
and implies positiveness of the quantity $\omega'(r_h)$.
For $\Lambda=-3$, this relation implies $\omega_h<\omega_{h(max)}$, with
\begin{eqnarray}\label{whbound}
\omega^2_{h(max)}= k+r_h\sqrt{3r_h^2+k},
\end{eqnarray}
(with a minimal event horizon radius $r_h=1/\sqrt{3}$ for $k=-1$).
The asymptotics as $r \to \infty$ are still given by (\ref{asympt}) for any value of $k$.
As argued in Section 4, the configurations' mass is given by $MV_k/4 \pi$,
with $M$ the asymptotic value of the metric function $m(r)$.

The Hawking temperature of the solutions is evaluated from the surface gravity $\kappa$ as given by
 $T_H=\kappa /(2 \pi)$, with
\begin{eqnarray}
\label{kappa}
\kappa^2 =
-(1/4)g^{tt}g^{ij}(\partial_i g_{tt})(\partial_j g_{tt})\Big|_{r=r_h}.
\end{eqnarray}
This implies
\begin{eqnarray}
\label{TH-n1}
T_H=\frac{k-2m'(r_h)-\Lambda r_h^2}{4 \pi r_h p(r_h)}.
\end{eqnarray}
For every considered value of $r_h$, we find regular black hole solutions
with $w_h$ taking values in
only one interval $0<\omega_h< \omega_h^c$,
where $\omega_h^c$ is always smaller than $\omega_{h(max)}$.
There are also solutions for which $\omega_{0}>1$
although $\omega_h<1$.
The behavior of the metric functions $m$ and $p$ are
qualitatively similar for any value of $k$. However
the gauge field behaviour depends on the topology of
the event horizon.
For $k=0,-1$, in contrast to the spherically symmetric case,
we find only nodeless solutions.
This can be analytically proven by
integrating the equation for $\omega$,
$(H\omega'/p)'=\omega(\omega^2-k)/p r^2$ between $r_h$ and $r$;
thus we obtain $\omega'>0$ for every $r>r_h$.
For $k=1$, solutions where $\omega$ crosses the axis can exist for small enough values
of $r_h$.  For large values of
$r_h$ (e.g. $r_h=1$), only nodeless solutions are found in this case too.
In this case, for sufficiently small $\omega_h$, all field variables remain
close to their values for the abelian configuration
 with the same $r_h$.
Significant differences occur for large enough values of $\omega_h$
and the effect of the nonabelian field on the geometry
becomes more and more pronounced.

In contrast to the picture found  in \cite{VanderBij:2001ia},
the $k=-1,~\Lambda=-3$ configurations always  have $m>0$. The black holes
therefore only occur with positive values of the mass.
Typical solutions in this case are presented in Figure 5.
In Figure 6, $M$, $\omega_0$, the value of the metric function
$p(r)$ on the event horizon and the Hawking temperature
are plotted as a function of $\omega_h$ for $k=0,~\pm 1$ black holes
and several values of $r_h$. Note that as the temperature approaches
zero ($i.e.$ as the solutions approach extremality), these physical quantities
all diverge.

In Figures 7-8 we plot the mass $M$ and the value of the
gauge potential at the event horizon $w_h$ for several
value of  the asymptotic value of the magnetic gauge potential
as a function of the event horizon radius.
For $k=1$, the corresponding solution with a regular origin is
approached as $r_h \to 0$.
For topological black holes,
we noticed the existence of a minimal event horizon radius $r_c$,
for any given $w_0$.
The Hawking temperature vanishes as $r_h \to r_c$ and a naked singularity develops,
while the mass stays finite.

The discussion  in \cite{Winstanley:1998sn}, \cite{VanderBij:2001ia}
on the stability of these black hole solutions within the perturbation
theory can easily be applied to $\Lambda=-3$.
It follows that  all $k=0$ solutions are stable; the $k=-1$ solutions  with $\omega_{0}>1$
are also stable as well as the nodeless spherically symmetric solutions.

\subsubsection{Static, axially symmetric  black holes}
Similar to the regular case, the $k=1,~n=1$ solutions discussed above
 admit static axially symmetric generalizations.
 (Static $k=0,-1$ topological black holes with a winding number $n>1$ are also
 likely to exist but the corresponding EYM
 ansatz has not yet been considered in the literature.)

The situation for a nonabelian field is very different from
the EM theory, where the static black hole solution is spherically symmetric (or, for $\Lambda<0$ belongs to
one of the three cases (\ref{fk}), with the same amount of symmetry) .

The properties of the AAdS axially symmetric EYM black
holes were addressed in \cite{Radu:2004gu},
however without consideration of the case $\Lambda=-3$.
The metric ansatz in this case is given again by (\ref{axial}),
with a gauge potential $A_i$ written in terms of four functions $H_i$.
We require  the horizon of
the black hole to reside at a surface of constant radial coordinate $r=r_h$,
where $g_{tt}(r_h)=0$.
Similar to the procedure in the regular case,
axially symmetric
solutions are obtained by extending the $n=1$
 configurations to higher values of the winding number.
These solutions are AAdS and have a regular event horizon but for $n>1$
 they are not spherically symmetric and
the event horizon gets deformed away from spherical symmetry.

The boundary conditions at infinity and on the symmetry axis
 are similar to those used in the regular case;
in particular, the asymptotic expansion (\ref{as1}) is still valid
(see also Appendix B).

The surface gravity $\kappa$ turns out to be constant at the
horizon, as required by the zeroth law of black hole
thermodynamics. To derive its expression we use the asymptotic
expansion near the event horizon in $\delta = (r-r_{h})/r_{ h}$
\begin{eqnarray}
\label{expan-h}
\nonumber
f(r,\theta)=f_2(\theta) \delta^2 + O(\delta)^3,
~~~
m(r,\theta)=m_2(\theta) \delta^2 + O(\delta)^3,
~~~
l(r,\theta)=l_2(\theta) \delta^2 + O(\delta)^3.
\end{eqnarray}
From $T_H=\kappa /(2 \pi)$,
we finds for the Hawking temperature
\begin{equation}
\label{temp}
T_H=\frac{f_2(\theta) (1- \Lambda r_h^2/3)}{2 \pi r_h \sqrt{m_2(\theta)} }
\ ,
\label{tempe}
\end{equation}
which is constant as a consequence of the $(r,~\theta)$ Einstein
equation  implying $f_2  m_{2, \theta}=2m_2 f_{2,\theta}.$

For the line element (\ref{metric}),
the  area $A$
of the event horizon  is given by
\begin{equation}
\label{area}
A = 2 \pi \int_0^\pi  d\theta \sin \theta
\frac{\sqrt{l_2(\theta) m_2(\theta)}}{f_2(\theta)} r_h^2,
\end{equation}
which allows a computation of the black hole entropy $S=A/4$.

The axially symmetric black hole solutions depend on two  continuous
parameters $(r_h,~\omega_0)$
as well as the winding number $n>1$.

The behavior of the solutions is
in many ways similar to that of the axially symmetric solitons.
Again, starting from a spherically symmetric black hole configuration
with given $\omega_0$ we obtain higher winding number generalizations.
Axially symmetric generalizations seem to exist for every
spherically symmetric black hole solution.
For a fixed winding number, the solutions form a branch,
which can be indexed
by the mass and the nonabelian magnetic charge $Q_m=n|1-\omega_0^2|$.
This branch  follows the picture found for $n=1$
(with higher values of mass, however).
This is in sharp contrast to the $\Lambda=0$ case,
where only a discrete set of solutions is found \cite{Kleihaus:1997ws}.
Also, the Kretschman scalar $K=R_{ijkl}R^{ijkl}$
remains finite for every $(r \geq r_h,~\theta)$.
One finds that the deviation from spherical symmetry increases with growing $n$.

Once we have a solution, the horizon variables such as $T_H,A$ are calculated in a
straightforward way from (\ref{temp}),~(\ref{area}).
The mass of the solution is computed by using the relation (\ref{ct-mass}), extracting the
values of the coefficients $f_1,~f_2$ from the asymptotics of the metric functions.
In Figure 9 we plot the
  mass $M$, the Hawking temperature and the entropy
 as a function of $\omega_0$
for  black hole  monopole solutions with $r_h=1$ and $n=1,2,3$.

The gauge functions $H_2,~H_3,~H_4$  start always at (angle dependent)
nonzero values on the event horizon.
For $r_h=1$, we find only solutions where $H_2,~H_4$ do not cross the $r$ axis.
The gauge function $H_2$ is always almost spherically symmetric,
while the gauge functions  $H_1$ and $H_3$ are much smaller
than the functions $H_2$ and $H_4$.

For the considered solutions,
the metric functions $m, ~f, ~l$
do not exhibit a strong angular dependence.
These functions start with a zero value on the event horizon and
approach rapidly the asymptotic values.
The functions  $m$ and $l$ have a rather similar shape, while the ratio $m/l$ indicating
the deviation from spherical symmetry is typically close to one,
except in a region near the horizon.
The typical profiles of the metric and gauge functions we find for $\Lambda=-3$
are similar to those presented
in \cite{Radu:2004gu} for other values of the cosmological constant.

The horizon has $S^2$ topology, but geometrically is not a sphere,
since its circumference along the equator $L_e$ turns out to be
different from that along a meridian $L_p$
\begin{eqnarray}
\label{circ}
L_e = \int_0^{2 \pi} { d \varphi
\left.
\sqrt{ \frac{l}{f}} r \sin\theta \right|_{r=r_{h}, \theta=\pi/2} }
 = 2 \pi r_{h} \left. \sqrt{\frac{l_2(\theta)}{f_2(\theta)}}
  \right|_{\theta=\pi/2}
\ ,
\\
\label{lp}
L_p = 2 \int_0^{ \pi} { d \theta \left.
 \sqrt{ \frac{m  }{f}} r
 \right|_{r=r_{h}, \varphi=const.} }
 = 2 r_{h} \int_0^\pi { d \theta
 \sqrt{\frac{m_2(\theta)}{f_2(\theta)}} }
\ .
\end{eqnarray}
However,
for these static solutions,
one finds a small deviation from spherical symmetry
(as measured by the ratio  $L_e/L_p$),
at the level of few percent.

\subsubsection{Rotating black holes}
AF rotating hairy black holes in EYM theory were obtained in
\cite{Kleihaus:2002ee}, within the standard Lewis-Papapetrou
parametrization of the metric and a gauge field ansatz consistent
with the circularity and Froebenius conditions. These solutions
possess three global charges: mass, angular momentum and
nonabelian electric charge. Although they possess nontrivial
magnetic gauge fields outside the event horizon, they do not carry
a nonabelian magnetic charge.

Obviously,  the static AAdS solutions should also possess rotating counterparts,
representing nonabelian generalizations of the
Kerr-Newman-AdS solution.
However, the construction of such hairy rotating solutions represents a very difficult
task since
it involves the solution of a large number of coupled nonlinear partial differential
equations for the metric and gauge field functions and a much richer set of
possible boundary conditions as compared to the AF case.

Here we present the first set of rotating black hole configurations with $\Lambda=-3$.
The ansatz we used in this case is similar to that employed
for the regular rotating solitons,
with the same asymptotic expansion as $r\to \infty$.
In particular, the expressions (\ref{as1}), (\ref{rot1})
are valid in this case too,
as well as the expressions (\ref{ct-mass}), (\ref{ct-J})
of the mass energy and angular momentum.
The boundary conditions and details on the numerical
integration are presented in Appendices B, D.

For a given winding number,
the rotating nonabelian black hole solutions depend on four continuous
parameters: two geometric parameters - $(r_h,~\Omega_h)$
representing the event horizon radius and the value of the metric function $\Omega$
at the horizon respectively,
and two parameters associated with the gauge field: $\omega_0$ which gives the
magnetic charge and $V$ which is the asymptotic value of the electric nonabelian
potential  $A_t(\infty)$.
Not surprisingly, different from the AF case, rotating black holes are found also for $V \neq 0$.

The complete classification of the solutions in the
space of these four physical parameters
is a considerable task, whose scope is beyond the aim of this paper.
We have studied mainly rotating
configurations with  $r_h=1$ and several values of $w_0$, although a number of solutions have
been found for other values of $r_h$.
Although rotating black hole solutions should exist for any
value of $n$, we restrict here to
a unit value of the winding number.

The properties of the horizon can be computed similar to the static case.
The surface gravity is obtained from
\begin{equation}
\kappa^2 = -1/4 (D_\mu \chi_\nu)(D^\mu \chi^\nu)
\end{equation}
where the Killing vector
$\chi = \xi -(\Omega_h/r_h) \eta$ ($\xi=\partial_t$,
$\eta=\partial_\varphi$)
is orthogonal to and null on the horizon.
It can be proven that the expansion near the event horizon (\ref{expan-h})
remains valid in the rotating case, which implies the expression (\ref{temp}) for
the Hawking temperature. We further consider the area $A$ of the
black hole horizon computed according to (\ref{area}),
and the deformation of the horizon, quantified by the ratio $L_e/L_p$
of the circumferences along the equator and the poles.

To construct a rotating solution, we start from the corresponding
static black hole configuration with $\Omega_h=0, ~V=0$. As we
increase ($\Omega_h,~V$) from zero via the boundary conditions,
while keeping $r_h$ fixed, a first branch of solutions forms. For
a given $V$, this branch ends at a critical value $\Omega_h$,
which  depends on the value of ($\omega_0,~r_h$), and the
numerical errors increase dramatically for
$\Omega_h>\Omega_{h(cr)}$ rendering the solutions increasingly
less reliable. As $\Omega_h \to \Omega_{h(cr)}$, the geometry
remains regular with no event horizon appearing for $r>r_h$, and,
the mass and angular momentum approach finite values. As found in
\cite{Kleihaus:2002ee} for $\Lambda=0$, a second branch of
solutions bends backward toward $\Omega_h=0$; there the mass and
angular momentum diverge with $\Omega_h^{-1}$ in the limit
$\Omega_h \to 0$. Therefore we expect a similar picture in  the
AAdS case. However the numerical construction of such
configurations presents a considerable numerical challenge beyond
the scope of the present work. Also, the existence of other
branches of AAdS  rotating solutions, not necessarily connected to
the static configurations, might be possible.

The picture gets simpler if we study the dependence of the solutions as a function of $V$
(the magnitude of the electric YM potential at infinity) for fixed ($\omega_0,~r_h,\Omega_h$).
In this case the solutions share a number of common properties
with the rotating regular counterparts.
In Figure 10 we plot the  mass, angular momentum,  electric charge and the contribution of
the electric field to the total mass  as a function of $V$ for
a fixed value of $\omega_0$.
As seen in this picture, the mass, angular momentum and electric charges increase with $V$
and we find again a maximal value for the magnitude of the electric potential at infinity.

The functions $H_2$ and $H_4$ are always nodeless,
although they have a small $\theta$ dependence, while $H_1$, $H_3$ and the electric
potentials depend on $\theta$-angle in a complicated way.
The metric functions $f,l,m$ present a
rather small angular dependence, the metric function $\Omega$
presenting a strong dependence of $\theta$ for small values of $V$.

For all configurations, the energy density of the solutions
has a strong peak along the $\rho$ axis,
and it decreases monotonically along the symmetry axis.
In contrast to the rotating regular case, we found no locally rotating solutions
with vanishing total angular momentum, although such configurations are likely to exist.

Further details on these rotating black hole solutions, as well as a discussion
of the  dependence of the solutions on the value of the
cosmological constant, will be presented elsewhere.
\section{A COMPUTATION OF PHYSICAL QUANTITIES}
\subsection{The counterterm formalism}

The mass, angular momentum and action of the solutions discussed in Section 3 is found by using the
counterterm formalism proposed by Balasubramanian and Kraus \cite{Balasubramanian:1999re}
to compute conserved quantities
for a spacetime with a negative cosmological constant.
This technique was inspired by AdS/CFT correspondence and consists in
adding suitable counterterms
$I_{ct}$
to the action. These counterterms are built up with
curvature invariants of a boundary $\partial \cal{M}$ (which is sent to
infinity after the integration)
and thus obviously they do not alter the bulk equations of motion.

The following counterterms are sufficient to cancel divergences in
four dimensions \cite{MannMisner} , for vacuum solutions with a
negative cosmological constant (to agree with the standard
conventions in literature, we set the usual factors $1/16 \pi G$
in the action principle (\ref{action}) and $1/8 \pi G$  for the
Gibbons-Hawking boundary term)
\begin{eqnarray}
\label{ct}
I_{\rm ct}=-\frac{1}{8 \pi G} \int_{\partial {\cal M}}d^{3}x\sqrt{-h}\Biggl[
\frac{2}{ \ell}+\frac{ \ell}{2}\rm{R}
\Bigg]\ .
\end{eqnarray}
Here $\rm{ R}$ is the Ricci scalar for the boundary metric $h$,
while $ \ell^2=-3/\Lambda=1$.

Using these counterterms one can
construct a divergence-free stress tensor from the total action
$I{=}I_{\rm bulk}{+}I_{\rm surf}{+}I_{\rm ct}$ by defining
\begin{eqnarray}
\label{s1}
{\rm T}_{\mu \nu}&=& \frac{2}{\sqrt{-h}} \frac{\delta I}{ \delta h^{\mu \nu}}
=\frac{1}{8\pi G}(K_{\mu \nu}-Kh_{\mu \nu}-\frac{2}{\ell}h_{\mu \nu}+\ell E_{\mu \nu}),
\end{eqnarray}
where $E_{ab}$ is the Einstein tensor of the intrinsic metric $h_{ab}$.
The efficiency of this approach has been demonstrated in a broad range of examples,
the counterterm subtraction method being developed for its own interest and applications.
If there are matter fields on $\cal{M}$ additional counterterms
may be needed to regulate the action (see $e.g.$ \cite{Radu:2004xp}
for such an example in EYM-dilaton theory).
However, we find that for a pure SU(2) nonabelian matter content
in four dimensions, the prescription  (\ref{ct}) removes all divergences
(a different situation is found for the five dimensional AAdS
nonabelian solutions where the counterterm method fails and
logarithmic divergences are presented in the
total action and the expression of mass \cite{Okuyama:2002mh}).

Having obtained the boundary energy-momentum tensor, one can
determine the conserved charges corresponding to the Killing
vectors as explained in \cite{Balasubramanian:1999re,MannMisner}.
The usual prescription is to first pick a spacelike surface
$\Sigma$ on the boundary with metric $\sigma_{ab}$. The boundary
metric is written in the following form
\cite{Balasubramanian:1999re}
\begin{eqnarray}
\label{b-AdS}
h_{\mu \nu}dx^{\mu} dx^{\nu}=-N_{\Sigma}^2dt^2
+\sigma_{ab}(dx^a+N_{\sigma}^a dt) (dx^b+N_{\sigma}^b dt).
\end{eqnarray}
The conserved charge associated to a symmetry generated by the Killing vector
$\chi^{\mu}$ is
\begin{eqnarray}
\label{charge-b}
Q_{\chi}=\int_{\Sigma}d^2 x \sqrt{\sigma}{\rm T}_{\mu \nu}n^\mu \chi^\nu.
\end{eqnarray}
where $n^{\mu}$ is a timelike unit normal to $\Sigma$.

The conserved charge associated with
time translation is the mass, the angular momentum
being  the charge associated with the Killing vector $\partial_{\varphi}$.

\subsection{$n=1$ static configurations}
We consider first the case of $n=1$ spherically symmetric
and topological black hole static configurations.
The results we find by using the asymptotic expressions (\ref{asympt})
for the boundary stress tensor at large $r$ are
\begin{eqnarray}
\label{BD4}
\nonumber
{\rm T}_{\theta \theta} &=&\frac{1}{8 \pi G}\frac{\ell M}{r}
+\frac{1}{32 \pi G}\frac{\ell (\ell^2-4(w_0^2-k)^2)}{r^2}
+O\left(\frac{1}{r^3}\right),
\\
{\rm T}_{\varphi \varphi}&=&\frac{1}{8 \pi G}\frac{\ell M}{r}
f_k^2(\theta)
+\frac{1}{32 \pi G}\frac{\ell (\ell^2-4(w_0^2-k)^2)}{r^2}f_k^2(\theta)
+O\left(\frac{1}{r^3}\right),
\\
\nonumber
{\rm T}_{tt}&=&\frac{1}{8 \pi G}\frac{2M}{\ell r}
+\frac{1}{32 \pi G}\frac{\ell^4 -4\ell^2(w_0^2-k)^2-8C_1^2}{r^2}
+O\left(\frac{1}{r^3}\right).
\end{eqnarray}
Thus the leading order terms in this expression boundary stress
tensor are similar to the (topological-) Schwarzschild-AdS$_4$ black holes.
The presence of nonabelian matter is manifest  in the second order
of this expansion only.

It can easily be verified that the mass of these solutions
computed from (\ref{charge-b}) is given by
$MV_k/4 \pi G$, where $V_k$ is the area of the surface $\Sigma$.
Obviously, these solutions have a vanishing angular momentum.

\subsection{Axially symmetric configurations}
We consider now the general case of a axially symmetric, rotating
spacetime described by the line element (\ref{rot}).
All relevant expressions can easily be derived by using the
the asymptotic form of the metric functions (\ref{as1}), (\ref{rot1}).
The boundary metric in this case is the Einstein universe,
$ds^2=\ell^2(d \theta^2+\sin^2 \theta d \varphi^2)-dt ^2$.

One finds in this way the large-$r$ expansion
\begin{eqnarray}
\label{BD-axial}
\nonumber
\nonumber
{\rm T}_{\theta \theta} &=&-\frac{1}{8 \pi G \ell~ r}
\left(\frac{2f_1}{3}+\frac{6f_2}{17}\sin^2 \theta \right)+O\left(\frac{1}{r^2}\right),
\\
{\rm T}_{\varphi \varphi}&=&-\frac{\sin^2 \theta}{8 \pi G \ell ~r}
\left(\frac{2f_1}{3}+\frac{18f_2}{17}\sin^2 \theta \right)
+O\left(\frac{1}{r^2}\right),
\\
\nonumber
{\rm T}_{tt}&=&-\frac{1}{8 \pi G \ell^3 ~r}
\left(\frac{4f_1}{3}+\frac{24f_2}{17}\sin^2 \theta \right)
+O\left(\frac{1}{r^2}\right)
\\
\nonumber
{\rm T}_{t \varphi}&=&-\frac{3}{16 \pi G \ell  r}
\left((j_1+j_2 \sin^2 \theta )\sin^2 \theta\right)
+O\left(\frac{1}{r^2}\right)
\end{eqnarray}
Direct computation shows that this stress tensor is traceless.
This result is expected from the AdS/CFT correspondence, since even dimensional
bulk theories with $\Lambda<0$ are dual to odd dimensional CFTs that have a
vanishing trace anomaly.
The corresponding expressions for the static case are found by taking
$j_1=j_2=0$ in the above relations.
For spherically symmetric configuration found within the metric ansatz (\ref{axial}),
$f_2=0$ and the angular dependence vanishes, as expected.

By using this relation, we  find that the mass and angular momenta of the  solutions are given by
the relations (\ref{ct-mass}) and (\ref{ct-J}) respectively.
For static solutions,  the parameter $f_2$ describes the deviation of the solutions
from the spherically symmetry.

\subsection{Euclidean action and entropy}
One can use the counterterm expression (\ref{ct}) to compute the
regularized gravitational action
and the prove that the entropy of the AAdS hairy black holes
is one quarter of the event horizon area.

Here we start by constructing the path integral \cite{Gibbons:1976ue}
\begin{eqnarray}
\label{Z1}
Z=\int D[g]D[\Psi]e ^{-iI[g,\Psi]},
\end{eqnarray}
integrating over all metrics and matter fields between  some given initial and final
hypersurfaces, $\Psi$ corresponding to the SU(2) potentials.
By analytically continuing the time coordinate $t \to i\tau$,
the path integral formally converges, and in the leading order one obtains
\begin{eqnarray}
\label{Z2}
Z \simeq e^{-I_{cl}}
\end{eqnarray}
where $I_{cl}$ is the classical action evaluated on the equations of motion
of the gravity/matter system.
Since the Euclidean approach becomes  problematic for nonabelian solutions
with an electric potential
\cite{Brihaye:2006nk}, we restrict here to compute the action
of static, purely magnetic solutions.
We should also remark that
the variation of the action (\ref{action})  gives the correct equations of motion only if the
 gauge potential $A_\mu$ is held fixed on the boundary $\partial M$.
This imposes the boundary condition $\delta A_{\mu}=0$ on $\partial M$,
which for purely magnetic
solutions fixes  the value of magnetic charge.

The globally regular solutions have an arbitrary periodicity $\beta$
of the Euclidean time coordinate.
In the black hole case,
the value of $\beta$ is found  by demanding regularity
of the Euclideanized manifold as $r \to r_h$.
It can easily be verified that the Hawking temperature
expression $T_H=1/\beta$ found in this way coincides with that given
by the surface gravity computation.

The physical interpretation of this formalism is that
the class of regular stationary metrics forms an ensemble of
thermodynamic systems at equilibrium temperature
$T_H$ \cite{Mann:2002fg}.
$Z$ has  the interpretation of partition function and we can
define the free energy of the system
$F=-\beta^{-1} \log Z$.

Therefore
\begin{eqnarray}
\label{i1}
\log Z=-\beta F=S-\beta M,
\end{eqnarray}
or
\begin{eqnarray}
\label{i2}
S=\beta M-I_{cl},
\end{eqnarray}
straightforwardly follows, with $S$ the entropy of the system.

To compute $I_{cl}$, we make use of the Einstein equations,
replacing the $R-2\Lambda$ volume term with
$2R_t^t-16\pi G T_t^t$.
For a purely magnetic ansatz ($A_t=0$), the term $T_t^t$  exactly cancels the matter field
Lagrangian in the bulk action
 $L_m=-1/2Tr(F_{\mu \nu }F^{\mu \nu })$.
 The Ricci component $R_t^t$ is computed
by integrating the Killing identity
$\nabla^a\nabla_b K_a=R_{bc}K^c,$
for the Killing vector $K^a=\delta^a_t$.
The divergent contribution given by the surface integral
term at infinity in $R_t^t$ is also canceled by
$I_{\rm{surface}}+I_{ct}$, yielding a finite expression of the action.
For the metric ansatz (\ref{metric}) describing spherically symmetric
and topological black holes static configurations
one finds
\begin{eqnarray}
\label{itot1} I_{cl}=\beta\frac{MV_k}{4 \pi}-\frac{r_h^2V_k}{16\pi
G}
\end{eqnarray}
while the corresponding expression for static axially symmetric configurations described by
the metric ansatz (\ref{axial}) is
\begin{eqnarray}
\label{itot2}
I_{cl}=\beta\left(\frac{\Lambda}{3G}\bigg(\frac{2f_1}{3}+\frac{8f_2}{17}\bigg)
-\frac{r_h}{4G}\bigg(1-\frac{\Lambda r_h^2}{3}\bigg)\int_0^{\pi}
d \theta \sin \theta \sqrt{l_2(\theta)}
\right).
\end{eqnarray}
The correponding expressions in the globally regular case are found by taking $r_h \to 0$ in the
above relations.

Replacing now in (\ref{i2}) (where $M$ is the mass-energy computed in Section 3.2, 3.3), we find
$S=0$ in the absence of an event horizon, while
the entropy of the black hole solutions
is one quarter of the event horizon area, as expected.

\subsection{The thermodynamics of $n=1$ static black holes}
Based on the numerical results presented
in Section III, we attempt here a discussion of the
thermodynamic properties
the $\Lambda=-3$ EYM
black solutions.   To our knowledge these have
not  been  previously considered.

For simplicity, we will restrict our considerations to $n=1$ static configurations
($i.e.$ spherically symmetric or topological black holes).
Thus we shall analyze black hole thermodynamics  in a canonical ensemble,
holding the temperature $T$ and the magnetic potential at the boundary at
infinity ($i.e.$ the magnetic charge) fixed.
The associated potential is the Helmholz free energy $F$.

The response function whose sign determines the thermodynamic
stability is the heat capacity
\begin{eqnarray}
\label{C}
C=T\left(\frac{\partial S}{\partial T}\right)_{Q_m}.
\end{eqnarray}
Stability follows from $C\geq 0$ given the fact that black
holes radiate at higher temperatures when they are smaller.

The behaviour of the specific heat $C$ can be easily
understood from the state equation $S=S(T)$
at fixed magnetic charge $Q_m$.
In Figure 11 we plot these curves for several values of $Q_m$ for $k=1, 0$ and $-1$
 black holes with unit winding number.

For $k=1$, the usual Schwarzschild-AdS behaviour is reproduced: the curves
first decrease
toward a minimum, corresponding to the  branch of small unstable
black holes, then increase along the branch of large stable black holes.
This is in strong contrasts with the behaviour of the
Abelian RNAdS solutions, which present
only one branch and approach the extremal limit as $T \to 0$ (see Figure 11a).
Note that the  $k=1$  solutions with small values around zero of the gauge
potential at infinity appear to present a complicated
thermodynamic structure.
However these solutions are unstable to
small perturbations and we will not consider them here.

As seen in Figure 11b,c,
the heat capacity (\ref{C}) is always positive
 for hairy black holes
with zero or negative curvature horizon,
(this appears to be valid for any $w_0$).
As a result, the $k=0,1$
topological black hole solutions
are always thermodynamically locally stable.

It is instructive to plot also the  free energy $F=I/\beta$ as a function
of temperature and various values of the magnetic charge (Figure 12).
One can see again that for $k=1$ there are
always two black hole radii associated with each temperature, for any value of $Q_m$.
Correspondingly, the smaller branch is unstable having negative
specific heat.
However, the action of the $k=1$ solutions
becomes positive for some critical value of the event horizon radius,
for any value of the nonabelian magnetic charge.

In the vacuum case, this indicates the existence of a phase transition.
When the free energy is negative, the Schwarzschild -AdS black hole
phase is dominant over the thermal AdS background phase.
When the free energy changes its sign, the Hawking-Page
phase transition between the AdS black hole and the thermal AdS
background takes place \cite{hawking1}.

As seen in Figure 8, for $k=1$ monopole configurations the  globally
regular solution has minimal energy in its asymptotic class of
solutions (with a given $Q_m$) and so is the thermal background.
Thus a phase transition
should exist between the large black hole solutions and the corresponding globally
regular configurations.

In the absence of matter fields,
the action of $k=0, -1$ black holes is always negative.
Therefore these configurations can also be globally stable
and
there is no phase transition.
The inclusion of YM fields changes this behaviour and
the low temperature solutions have a positive action.

\subsection{On the boundary CFT}

Restricting to static solutions, we find from
$\gamma_{\mu \nu}=\lim_{r \rightarrow \infty} \frac{\ell^2}{r^2}h_{\mu \nu}$
the following background metric upon which the dual field theory resides
\begin{equation}
\label{l1}
\gamma_{\mu \nu}dx^\mu dx^\nu=\ell^2(d \theta^2+f_k^2(\theta)d \varphi^2)-dt ^2,
\end{equation}
For $k=1$ this is describes a $2+1$ dimensional Einstein universe;
for $k=0$ it is $2+1$ flat space, while $k=-1$ describes the three
dimensional open static universe.

It would be desirable to compute some quantities in this
background and to compare the results with the bulk
predictions.
The main problem is that, even in the vacuum case,
the AdS$_4$/CFT$_3$ correspondence is much less understood that the
AdS$_5$/CFT$_4$ case.
For example, in  \cite{Klebanov:2002ja},
Klebanov and Polyakov proposed a duality between a theory
of massless higher spin gauge fields in AdS$_4$ spacetime
on the one hand and the $O(N$) vector model at large $N$
in three dimensions on the other.
However, although the details of the boundary CFT will depend on the
details of the bulk supergravity theory, the generic properties
are expected to be independent of the precise features of the
theory.

Here we should
remark that as found in Section 3, the nonabelian matter field in the bulk does not
approach asymptotically a pure gauge configuration.
For static solutions, the boundary form of the unit winding number nonabelian potential is
\begin{equation}
\label{A0}
A^{(0)}=\frac{1}{2} \{
\omega_0\tau_1  d \theta
+\left(\frac{d \ln f_k(\theta)}{d \theta} \tau_3
+ \omega_0\tau_2 \right) f_k(\theta) d \varphi \},
\end{equation}
with $w_0$ a real constant.
The corresponding  boundary gauge field expression
for $k=1$ and a winding number $n>1$ is
\begin{equation}
\label{A0-ax1}
A^{(0)}=\frac{1}{2} (1-w_0)\{
\tau_\varphi^n d\theta
 -n\sin\theta \tau_\theta^n  d\varphi
\},
\end{equation}
where $\tau_\varphi^n(\theta,\varphi)$, $\tau_\theta^n(\theta,\varphi)$
are suitable combinations of the Pauli matrices, whose form in given in Appendix A.
One can see that the winding number enters the
boundary gauge field  expression.

From the AdS/CFT correspondence, we expect the nonabelian hairy black holes to be described by
some thermal states
in a dual theory formulated in a background given by (\ref{l1}).
The spherically and axially symmetric solitons will correspond to zero-temperature states
in the same theory. This CFT will interact with
a background  SU(2) field given by (\ref{A0}), (\ref{A0-ax1}).

As conjectured in Ref. \cite{Maldacena:2004rf}, the dual
field theory is the field theory of a stack of $N$ coincident
$M2$ branes with an background
external field coupled to the R-symmetry current.
The bosonic sector of the Lagrangian for a single $M2$ brane is \cite{Maldacena:2004rf}
\begin{equation}
\label{cft}
I=\int d^3x \sqrt{-\gamma}( D_{a}\Phi D^a \Phi +\frac{1}{8}R \Phi^2),
\end{equation}
where $D=\partial-iA^{(0)}$ is the gauge covariant derivative,
and $R$ is the Ricci scalar of the boundary CFT metric $\gamma_{ab}$.

Computing quantum effects for a generic
SU(2) background field  is a difficult task.
Simpler results are found only for the case $w_0=0$.
For spherical symmetry, the expressions
(\ref{A0}), (\ref{A0-ax1}) describe
the field of a Dirac monopole with $n$ units of magnetic charge
\begin{equation}
\label{A0-dirac}
A^{(0)}=\frac{ \tau_3}{2} n\cos \theta  d\varphi.
\end{equation}
(For $n>1$ this is proven
by transforming the gauge connection (\ref{ansatz-YM})
to a special gauge such that $A_\varphi$ has only a $\tau_3$ component.)

A computation of the effective action $I_{eff}$
for a charged singlet
scalar field propagating in a
zero-temperature Euclideanized $d=3$ Einstein-universe background
and interacting with the U(1) field given by (\ref{A0-dirac})
is presented in Appendix E, based on a zeta-function
approach.
From the basic relation (\ref{frel}), we expect $I_{eff}$ to
present a qualitative agreement with the
corresponding bulk computation.
First we note that a straightforward evaluation of (\ref{defIeff}),
(\ref{zetap}) give  positive values for $I_{eff}(n)$, as expected
from the bulk results (with the bulk action
of globally regular solutions $I_B=\beta M$).
It is also interesting to compute the ratio $I_{eff}(n)/I_{eff}(1)$
and to compare with the bulk results.
One finds $e.g.$ $I_{eff}(2)/I_{eff}(1)\simeq 3.89$,
$I_{eff}(3)/I_{eff}(1)\simeq 8.52$ while
 $I_{B}(2)/I_{B}(1)\simeq 2.46$,
$I_{B}(3)/I_{B}(1)\simeq 4.72$.
The discrepancy beteween the bulk and boundary results increases with $n$.

However, on the AdS side, there are two distinct bulk configurations
with  zero temperature and magnetic charge $n$.
First, there is the nonabelian soliton solution with  charge $n$ discussed in the previous
Section.
The second solution is an extremal  RNAdS black hole
with zero temperature and the same
value of magnetic charge.
It is not clear how the CFT knows to distiguish between these
different bulk solutions.
In any case, this is not possible within the model (\ref{cft}).

There is also another problem with the action principle
(\ref{cft}).
Since the $d=4$ Einstein-Maxwell system has
electric-magnetic duality \cite{Hawking:1995ap}, one expects
this duality to also be manifest on the CFT side.
However, its realization in $d=3$ is a rather subtle question
  (see e.g. the discussion in \cite{Leigh:2003gk}).
For example, the U(1) dual of (\ref{A0-dirac})
 is a $A^{(0)}=\tau_3~c~dt/2$ with $c^2=n \sqrt{(-1+\sqrt{1+12n^2})/6}$
 for an extremal RNAdS solution.
We can also perform a computation
similar to that in Appendix E, for an electric $A^{(0)}$.
One can see that, in the zero temperature limit,
the parameter $c$ will not enter the final results,
which are similar to the vacuum case.
Thus we conclude that the model (\ref{cft})
is too simple to  mimic the expected features of the
 boundary CFT.

 However, we can use the AdS/CFT ``dictionary'' to predict
qualitative features of a quantum field theory in the  background (\ref{l1}).
For example,  the expectation value of the dual CFT stress-tensor
can be calculated using the  relation \cite{Myers:1999qn}
\begin{eqnarray}
\label{r1}
\sqrt{-\gamma}\gamma^{ab}<\tau_{bc}>=
\lim_{r \rightarrow \infty} \sqrt{-h} h^{ab}{\rm T}_{bc}.
\end{eqnarray}
Applying this prescription to the
$n=1$ static solutions,
we find the standard form for the stress tensor of a (2+1) dimensional CFT
\begin{eqnarray}
\label{st}
<\tau^{ab}>=\frac{M}{8 \pi \ell^2}[3 u^a u^b+\gamma^{ab}],
\end{eqnarray}
where $u^a=\delta_t^a$.

A similar computation can be done for an axially symmetric configurations in the bulk.
Considering the more general rotating case,
we find the field theory stress tensor
\begin{eqnarray}
\label{st-axial}
<\tau^{a}_b>=<\tau^{a}_b>^{(st)}+<\tau^{a}_b>^{(rot)}
\end{eqnarray}
where $<\tau^{a}_b>^{(st)}$ is a contribution which survives in the static limit
\begin{equation}
\label{st1}
<\tau^{a}_b>^{(st)}=
A\pmatrix{1&0&0\cr 0&1&0\cr 0&0&-2\cr} +
B\pmatrix{1&0&0\cr 0&3&0\cr 0&0&-4\cr},
\end{equation}
where
\begin{eqnarray}
 A=\frac{1}{8 \pi \ell^2}(M+\frac{8f_2}{17}\frac{1}{G \ell^2})
,~~~B=-\frac{1}{8 \pi G \ell^4}\frac{6f_2}{17}\sin^2 \theta,
\end{eqnarray}
and  $x^1=\theta,~x^2=\varphi,~x^3=t$. Here $M,f_2$ are continuous variables
which encode the bulk parameters.
$<\tau^{a}_b>^{(rot)}$ is the part of the CFT induced by the rotation in the bulk and
has the expression
\begin{equation}
 <\tau_{a b}>^{(rot)}=
C\pmatrix{0&0&0\cr 0&0&1\cr 0&1&0\cr},
~~{\rm with~}C=\frac{1}{8 \pi G  }\frac{3 \ell^2}{2}(j_1+j_2 \sin^2 \theta)\sin^2 \theta.
\end{equation}
The CFT stress tensor is covariantly conserved and manifestly traceless.
Even for static configurations in the bulk,
a winding number $n>1$ of the bulk configurations implies  $f_2 \neq 0$ and thus
a $\theta-$dependence
of the dual theory stress tensor (although the boundary metric is spherically symmetric).
This is a unique property of AAdS gravitating nonabelian configurations,
since the boundary stress tensor of an Abelian
solution with the same global charges has $f_2=0$.
This also suggests the dual  theory should also be sensitive to the integer $n$,
and is  much more complex than the simple model (\ref{cft}).

The form (\ref{st-axial}) of the
dual CFT stress-tensor is puzzling from yet another point of view,
since there is no global charge associated with the parameter $f_2$.
However, the expression (\ref{st-axial})  means that the dual CFT
is able to discern between $n>1$ nonabelian
and embedded Abelian bulk solutions with the same set
of boundary data.

\section{CONCLUSIONS}
In this paper we have discussed the basic features of the
nonabelian solutions of a
EYM-SU(2) theory with a negative cosmological constant  $\Lambda=-3g^2$
(where $g$ is the gauge coupling constant of the YM theory).
This theory corresponds to
a consistent truncation of $d=4~{\cal N}=4$ SO(4) gauged supergravity.
Except for the NUT-charged solutions, we have considered the
 $\Lambda=-3g^2$ version of all monopole configurations previously discussed
 in the literature for a generic value
 of the cosmological constant.
This includes both particle-like globally regular and black hole solutions.
These configurations
are asymptotically AdS, possesing a regular origin or a regular event horizon.
Apart from spherically symmetric solutions (or topological
black holes) we presented arguments
for the existence of $\Lambda=-3g^2$ static, axially
symmetric solutions.
These configurations have no counterparts in the abelian theory.
They generalize to higher winding number
the known spherically symmetric solutions,
presenting a nontrivial angle-dependence of matter fields and metric functions.

The main feature of the EYM  AAdS configurations
is the existence of a nonvanishing nonabelian magnetic
flux on the sphere at infinity.
The thermodynamics of the static black hole solutions has been also discussed
to some extent.
Apart from static solutions, we discussed  AAdS configurations with a nonvanishing angular momentum
which have not been presented before in the literature.

All known EYM asymptotically flat configurations
are likely to present $\Lambda=-3g^2$ generalizations.
Thus we expect the existence AAdS counterparts of the EYM configurations
discussed in \cite{Ibadov:2004rt},
satisfying a complicated angle-dependent set of
boundary conditions at infinity.
It would be interesting to construct AAdS nonabelian solutions
which possess only discrete symmetries \cite{Kleihaus:2003tn}
and to find the corresponding boundary stress tensor.

One should remark that all solutions discussed here may be uplifted
to $d=11$ supergravity.
However, the solutions we discussed here are generically not supersymmetric.
Supersymmetric solutions are likely to exist,
but we expect them to present naked
singularities (this is the case of the abelian counterparts with the same amount of
symmetry).
Note that the planar BPS solution of the $\Lambda=-3g^2$ EYM model found in closed form in
\cite{Radu:2004gu} has a naked singularity.

Apart from a discussion of the physical properties
of various bulk EYM configurations, we attempted a preliminary discussion of
these solutions in an AdS/CFT context.
The $\Lambda=-3g^2$ EYM configurations should give us information about the structure of
a dual CFT in a background SU(2) field.
However, a naive computation of the effective action of a charged
scalar field in a U(1) magnetic background field gave some inconclusive results.
Further progress in this direction would require a better knowledge of the
structure of the dual theory defined on the 2+1 dimensional boundary metric.
The results we discussed in the Section IV appear
to indicate that the dual CFT is able to
distinguish between various bulk solutions
with the same set of boundary data at infinity.

We think that these issues deserve further study,
which should lead to a deeper understanding of the AdS/CFT correspondence.

\section*{Acknowledgements}
The work of R.B.M. was supported by the Natural Sciences \& Engineering Research Council of Canada.
The work of E.R. and D.H.T. was carried out in the framework of Enterprise--Ireland
Basic Science Research Project SC/2003/390.



\appendix

\setcounter{equation}{0}
\section{The axially symmetric gauge field ansatz}

The construction of an axially symmetric YM ansatz has been discussed by many authors
starting with Manton \cite{Manton:1977ht}
and Rebbi and Rossi \cite{Rebbi:1980yi}.
The most general axially symmetric
YM-SU(2) ansatz contains  nine magnetic and three electric
 potentials and can be easily obtained in cylindrical coordinates
$x^{\mu}=(\rho,\varphi,z$)
\begin{eqnarray}
\label{A-gen-cil}
A_{\mu}=\frac{1}{2}A_{\mu}^{(\rho)}(\rho,z)\tau_{\rho}^n
        +\frac{1}{2}A_{\mu}^{(\varphi)}(\rho,z)\tau_{\varphi}^n
        +\frac{1}{2}A_{\mu}^{(z)}(\rho,z)\tau_{z}^n,
\end{eqnarray}
where the only $\varphi$-dependent terms are the SU(2) matrices
(composed of the standard $(\tau_1,~\tau_2,~\tau_3)$ Pauli matrices)
\begin{eqnarray}
\label{u-cil}
\tau_{\rho}^n=~~\cos n\varphi~\tau_1+\sin n\varphi~\tau_2,~~
\tau_{\varphi}^n=-\sin n\varphi~\tau_1+\cos n\varphi~\tau_2,~~
\tau_{z}^n=\tau_3.
\end{eqnarray}
This ansatz contains an integer $n$, representing the winding number with
respect to the
azimuthal angle $\varphi$.

Transforming to spherical coordinates $r,~\theta,~\varphi$,
with ($\rho=r \sin \theta$, $z=r\cos \theta$),
it proves convenient to introduce,
without any loss of generality, a new  SU(2)  basis
$(\tau_{r}^n,\tau_{\theta}^n,\tau_{\varphi}^n)$,
with
\begin{eqnarray}
\label{u-sph}
\tau_{r}^n=\sin \theta~\tau_{\rho}^n+\cos \theta~\tau_z^n,
~~
\tau_{\theta}^n=\cos \theta~\tau_{\rho}^n-\sin \theta~\tau_z^n.
\end{eqnarray}
The general expression (\ref{A-gen-cil}) takes the following form in spherical coordinates
\begin{eqnarray}
\label{A-gen-sph}
A_{\mu}=\frac{1}{2}A_{\mu}^{(r)}(r,\theta)\tau_{r}^n
        +\frac{1}{2}A_{\mu}^{(\theta)}(r,\theta)\tau_{\theta}^n
        +\frac{1}{2}A_{\mu}^{(\varphi)}(r,\theta)\tau_{\varphi}^n,
\end{eqnarray}
where
$A_\mu^{a} dx^\mu=A_r^{a} dr + A_\theta^{a} d\theta + A_\varphi^{a} d\varphi+A_t^{a} dt$.
This ansatz is axially symmetric in the sense that a rotation around the $z-$axis
can be compensated by a gauge rotation
${\mathcal{L}}_\varphi A=D\Psi$ \cite{Forgacs:1980zs},
with $\Psi$ being a Lie-algebra valued gauge function.
For the ansatz (\ref{A-gen-sph}),
$\Psi=n \cos \theta  \tau^n_{r}/2 - n \sin \theta  \tau^n_{\theta}/2$.
Therefore we find
$ F_{\mu \varphi} = D_{\mu}W, $
where $W=A_{\varphi}-\Psi$.

We use in this paper a reduced YM ansatz, employed also
in all previous studies on EYM solutions,
with five of the gauge potentials taken identically zero
\begin{eqnarray}
\nonumber
A_{r}^{(r)}=A_{r}^{(\theta)}~=~A_{\theta}^{(r)}~=~A_{\theta}^{(\theta)}
~=~A_{\varphi}^{(\varphi)}~=~A_{t}^{(\varphi)}=0.
\end{eqnarray}
The consistency of this reduction can easily be proven at the
level of the YM field equations.
A suitable parametrization of the six nonzero
components of $A_\mu^{a}$ which factorizes the trivial $\theta$-dependence
 is \cite{Kleihaus:2002ee}
\begin{eqnarray}
\label{ansatz-YM}
 \nonumber
 A_\mu dx^\mu
=
\left[\frac{H_1}{r}dr
 +(1-H_2)d\theta \right]\frac{\tau_\varphi^n}{2}
-n\sin\theta\left[H_3 \frac{\tau_r^n}{2}
            +(1-H_4) \frac{\tau_\theta^n}{2}\right]   (d\varphi-\frac{\Omega}{r} dt)
\\
+
\left[H_5 \frac{\tau_r^n}{2} + H_6 \frac{\tau_\theta^n}{2}\right]
dt
\end{eqnarray}
One may consistently take $\Omega=0$ in this ansatz;
however the inclusion of this metric function simplifies the set of boundary condition
for the rotating configurations.
For $\Omega=H_5 =H_6=0$, the static axially symmetric ansatz  used in previous studied on
AAdS static EYM solutions is recovered; further,  by taking $n=1$,
$H_1=H_3=0$, $H_2=H_4=w(r)$ one finds
a spherically symmetric static ansatz which reduces to (\ref{A1})
after a suitable gauge transformation.

The ansatz (\ref{ansatz-YM})
satisfies also some additional discrete symmetries \cite{Rebbi:1980yi}, \cite{ymh}
(in particular the parity reflection symmetry) and
it is also invariant under Abelian gauge transformations $U$
\begin{equation}
\label{gauge-res}
 U= \exp \left({\frac{i}{2} \tau^n_\varphi \Gamma(r,\theta)} \right).
\end{equation}
 To fix this residual gauge degree of freedom  we choose the usual
gauge condition \cite{Kleihaus:1997ws},\cite{Kleihaus:2002ee}
\begin{eqnarray}
\label{gauge}
\nonumber
r \partial_r H_1 - \partial_\theta H_2 = 0.
\end{eqnarray}
The non-vanishing components of the  field strength tensor are given by
\begin{eqnarray}
F_{r\theta}^{\varphi} & = &
-\frac{1}{r} \left[H_{1,\theta}+ r H_{2,r}\right]
\ ,
\nonumber \\
F_{r\varphi}^{r} & = &
-\frac{\sin\theta}{r}\left[rH_{3,r}-H_1 H_4\right]
\ ,
\nonumber \\
F_{r\varphi}^{ \theta } & = &
 \frac{\sin\theta}{r}\left[rH_{4,r}+H_1(H_3 +{\rm cot}\theta)\right]
\ ,
\nonumber \\
F_{\theta\varphi}^{r} & = &
-\sin\theta\left[ H_{3,\theta}+H_3{\rm cot}\theta +H_2 H_4 -1 \right]
\ , \nonumber \\
F_{\theta\varphi}^{ \theta } & = &
 \sin\theta\left[ H_{4,\theta}+{\rm cot}\theta (H_4-H_2)-H_2 H_3 \right]
,
\\
F_{tr}^{r} & = &
-\frac{1}{r}\left[rH_{5,r}
+ H_1 H_6-\frac{\Omega}{r}\sin\theta(H_1(1-H_4)-H_3+rH_{3,r})
    -\Omega_{,r}\sin\theta H_3\right]
\ ,
\nonumber \\
F_{rt}^{ \theta } & = &
\frac{1}{r}\left[rH_{6,r} - H_1 H_5
        +\frac{\Omega}{r}\sin\theta(H_1 H_3 +(1-H_4)+rH_{4,r})
    -\Omega_{,r}\sin\theta (1-H_4)\right]
\ ,
\nonumber \\
F_{ \theta t}^{r} & = &
\left[H_{5,\theta} - H_2 H_6
     +\frac{\Omega}{r}\sin\theta (H_2(1-H_4)-{\rm cot}\theta H_3-H_{3,\theta})
     -\frac{\Omega_{,\theta}}{r}\sin\theta H_3\right]
\ ,
\nonumber \\
F_{\theta t}^{ \theta } & = &
\left[H_{6,\theta} + H_2 H_5
     -\frac{\Omega}{r}\sin\theta (H_2 H_3+{\rm cot}\theta (1-H_4)
                                     -H_{4,\theta})
     -\frac{\Omega_{,\theta}}{r}\sin\theta  (1-H_4)\right]
\ ,
\nonumber
\\
\nonumber
F_{\varphi t}^{\varphi} & = &
\sin\theta\left[H_5 H_4 +H_6(H_3+{\rm cot}\theta)
-\frac{\Omega}{r}\sin\theta ({\rm cot}\theta(1-H_4)+H_3)\right]
.
\end{eqnarray}
One can easily verify the matter ansatz is compatible
with the metric form (\ref{rot}), since the energy momentum tensor (\ref{tik})
 satisfies $T_{tr}=T_{t\theta}$$=T_{\varphi r}=$$T_{\varphi \theta}=0$.

\section{Boundary conditions}

\subsection{Static axially symmetric solutions}
These solutions are obtained for a truncation $A_t=0$ of the ansatz (\ref{ansatz-YM})
and a metric given by (\ref{axial}).

To obtain AAdS axially symmetric configurations
with a regular origin or event horizon and with the proper symmetries,
we must impose the appropriate boundary conditions.
The boundaries of the system are the origin/event horizon and spacelike infinity,
the $z$-axis and, because of parity reflection symmetry satisfied by the matter fields,
the $\rho$-axis.
The boundary conditions at infinity and along the  $z-$ and the $\rho$-axis
($i.e.$ for $\theta=0,\pi/2$)
are similar for both globally regular and black hole solutions.

We start by setting the boundary conditions at
infinity compatible with the AAdS asumption.
For the metric functions one imposes
\begin{equation}
\label{b2}
f|_{r=\infty}= m|_{r=\infty}= l|_{r=\infty}=1,
\end{equation}
while the boundary conditions for the matter part are
\begin{equation}
\label{b4}
H_2|_{r=\infty}=H_4|_{r=\infty}= \omega_0, \ \ \
H_1|_{r=\infty}=H_3|_{r=\infty}=0,
\end{equation}
where   there are no obvious conditions on the value of $\omega_0$.
For a solution with parity reflection symmetry (the only
type we consider in this paper),
the boundary conditions along the axes are
\begin{eqnarray}
H_1|_{\theta=0,\pi/2}=H_3|_{\theta=0,\pi/2}=0,~~
\partial_\theta H_2|_{\theta=0,\pi/2}
=\partial_\theta H_4|_{\theta=0,\pi/2}
=0,
\\
\nonumber
\partial_\theta f|_{\theta=0,\pi/2}
=\partial_\theta m|_{\theta=0,\pi/2} =
\partial_\theta l|_{\theta=0,\pi/2} =0.
\end{eqnarray}
Therefore we need to consider the solutions only in the region $0\leq \theta \leq \pi/2$.
Regularity on the $z-$axis requires also
\begin{equation}
H_2|_{\theta=0}=H_4|_{\theta=0},~~m|_{\theta=0}=l|_{\theta=0}.
\end{equation}
For globally regular solutions, the boundary conditions imposed at the origin are
\begin{equation}
\label{b1}
\partial_r f|_{r=0}= \partial_r m|_{r=0}= \partial_r l|_{r=0}= 0,
\\
\nonumber
H_{2}|_{r=0}=H_{4}|_{r=0}= 1, \ \ \ H_{1}|_{r=0}=H_{3}|_{r=0}=0.
\end{equation}
The boundary conditions satisfied by the black hole solutions at the event horizon are
\begin{equation}
\label{b11}
f|_{r=r_h}=  m|_{r=r_h}=  l|_{r=r_h}= 0,
~~~
H_{1}|_{r=r_h}=0,~~\partial_r H_{2}|_{r=r_h}=
\partial_r H_{3}|_{r=r_h}=\partial_r H_{4}|_{r=r_h}=0,
\end{equation}

\subsection{Rotating solutions}
These solution are found for  metric form (\ref{rot})
and the YM ansatz (\ref{ansatz-YM}).
Similar to the static case, we consider solutions with
parity reflection symmetry $i.e.$  $0\leq \theta \leq \pi/2$.

A systematic analysis reveals that the boundary conditions for the magnetic potentials
$H_1,..,H_4$ and the metric functions $f,~l,~m$ presented in the static case
remain valid in the presence of rotation.
For the supplementary functions $H_5,~H_6,~\Omega$ we imposed
\begin{eqnarray}
\label{new-inf1}
\Omega|_{r=\infty}=0,~~H_5|_{r=\infty}=V \cos \theta,~~H_6|_{r=\infty}= V \sin \theta
\end{eqnarray}
at infinity, and
\begin{eqnarray}
\label{new-inf2}
\partial_\theta \Omega|_{\theta=0,\pi/2}=0,
~~\partial_\theta H_5|_{\theta=0}=H_5|_{\theta=\pi/2}=0,~~
H_6|_{\theta=0}=\partial_\theta H_6|_{\theta=\pi/2}=0.
\end{eqnarray}
on the axis.
For globally regular solutions, the following
set of boundary conditions is imposed at the origin
\begin{eqnarray}
\label{new-inf3}
H_5|_{r=0} \sin\theta + H_6|_{r=0} \cos\theta =0,~~
\partial_r H_5|_{r=0} \cos\theta - \partial_rH_6|_{r=0} \sin\theta = 0,
~~\Omega|_{r={0}}=0.
\end{eqnarray}
The black hole solutions are found imposing at the event horizon
\begin{eqnarray}
\label{new-inf4}
r_{\rm h} H_5 |_{r=r_{h}}  +\cos\theta \Omega_{\rm h} =0,~~~
r_{\rm h} H_6 |_{r=r_{h}}  -\sin\theta \Omega_{\rm h} =0,~~\Omega|_{r=r_{h}}=\Omega_h.
\end{eqnarray}

\section{General relations}
Solutions of the field equations are also classified by the
nonabelian electric and magnetic charges $Q_e$ and $Q_m$.
The definition of conserved currents and charges in a non-Abelian Yang-Mills theory
is a problem approached  by different authors in the last decades
(see $e.g.$ \cite{Chrusciel:1987jr}-\cite{Creighton:1995au}), with various solutions.
A gauge invariant definition for the nonabelian charges was proposed in
\cite{Corichi:1999nw}  (see also \cite{Kleihaus:2002ee})
\begin{eqnarray}
\label{charges}
Q_e={1\over 4\pi} \oint d \theta d \varphi |\tilde F_{\theta \varphi}|,~~~
Q_m={1\over 4\pi} \oint d \theta d \varphi | F_{\theta \varphi}|,
\end{eqnarray}
where the vertical bars denote the Lie-algebra norm and the integrals
are evaluated as $r \to \infty$.
The expression for the magnetic charge implied
by this definition is
$Q_m=|k-\omega_0^2|V_k/4 \pi$ for spherically symmetric and topological
black hole solutions and
 $Q_m=n|1-\omega_0^2|$ for axially symmetric configurations.

The magnetic charge defined in this way is equal (up to a sign)
to the expression found by using the usual (gauge dependent) definition
\begin{eqnarray}
 Q_m=\frac{1}{4 \pi}\int dS_k \sqrt{-g} ~Tr\{\tilde{F}^{kt}~T\},
\end{eqnarray}
(with $T=\tau_3$ for
the gauge ansatz (\ref{Atop}) and $T=\tau_r$ for
the axially symmetric generalization (\ref{ansatz-YM})).

In evaluating the electric charge expression one uses the asymptotic
expansion of the electric potential
\begin{equation}
\label{el-as}
H_5\sim\cos \theta \big(V+(c_1 \sin^2 \theta +c_2)/r\big),~
H_6\sim\sin \theta \big(V+(c_3 \sin^2 \theta +c_4)/r\big),
\end{equation}
(with $c_i$ real constants).

The energy density of the solutions is given by the
$tt$-component of the energy momentum tensor $T_{\mu}^{\nu}$.
Of interest here is the electric part of this component, $Tr\{ F_{\mu t} F^{\mu t} \}$
and its integral
\begin{eqnarray}
\label{electric-mass1}
-M_e =\int Tr\{ F_{\mu t} F^{\mu t} \}\sqrt{-g}d^3x,
\end{eqnarray}
which measures the contribution of the nonabelian electric field
to the mass/energy of the system.
Similar to the purely abelian part, by using the YM equations (\ref{YM-eqs}) this integral
can be expressed as a total divergence
\begin{eqnarray}
\label{electric-mass2}
-M_e =\int Tr\{ F_{\mu t} F^{\mu t} \}\sqrt{-g}d^3x=
 \oint_{\infty} Tr \{A_t F^{\mu t} \} dS_{\mu}-\oint_{eh} Tr \{A_t F^{\mu t} \} dS_{\mu}.
\end{eqnarray}
Thus, for globally  regular configurations,
a vanishing magnitude of the electric potentials at infinity
implies a purely magnetic solution.
In contrast, one finds rotating black hole solutions with $V=0$ which are supported by the
event horizon contribution.

Here we remark that the angular momentum of any solution
admits also an expression in terms of surface integrals of the matter fields
\cite{VanderBij:2001nm}.
This can easily be proven by using the existence of a potential $W$ (\ref{relation})
and the YM equation
\begin{eqnarray}
\label{J-matter}
J=\int T_{\varphi}^{t}\sqrt{-g} d^{3}x
= \int 2Tr\{F_{r \varphi} F^{r t}
+F_{\theta \varphi} F^{\theta t}\} \sqrt{-g} d^{3}x=
\oint_{\infty}2Tr\{WF^{\mu t} \} dS_{\mu}-\oint_{eh}2Tr\{WF^{\mu t} \} dS_{\mu}.
\end{eqnarray}
The above relation takes a particularly simple form for globally regular configurations.
By using the gauge field asymptotics, the
total angular momentum in terms of matter field coefficients is
\begin{eqnarray}
\label{J-matter2}
J=\frac{4 \pi n}{3} \left(-\frac{2}{5}c_1-c_2+\omega_0 (\frac{8}{5}c_3+2c_4) \right).
\end{eqnarray}
In the black hole case, (\ref{J-matter2}) should be supplemeted  with the event
horizon contribution.
This  relation can also  be used to test the accuracy of the numerical results.

\section{Numerical algorithm}
The spherically symmetric and topological black hole solutions can
easily be found by using a standard differential equations solver.
Starting with  a suitable set of boundary conditions at the origin/event horizon
the equations are integrated towards $r\to\infty$ using an automatic
step procedure and accuracy $10^{-12}$.
The integration stops when the AdS spacetime asymptotics
are reached with a prescribed accuracy.

The situation is much more complicated for axially symmetric configurations,
which requires the solving of seven (in the static case) or ten
nonliniar partial differential equations.
All axially symmetric solutions presented in this paper
have been found by using a similar approach to that
employed by Kleihaus and Kunz in their studies of
AF nonabelian configurations.
We use in the numerical procedure a
suitable combination of the EYM equations
such that the diferential equations for the metric and gauge functions are diagonal
in the second derivatives with respect to $r$.
These  equations are then discretized on a ($r,~\theta$) grid with $N_r\times N_{\theta}$ points.
The angular coordinate $\theta$ runs
from $0$ to $\pi/2$ and the radial coordinate goes from $r_h$ to some large enough value
$r_{max}$ (typically $r_{max}\simeq 2\times 10^3 \div 5\times 10^3$).
For any type of solution,
we tested  that the relevant quantities are insensitive to the
cut off value $r_{max}$.
The grid spacing in the $r-$direction is non-uniform, while the values of the grid points
in the angular direction are given by $\theta_k=(k-1)\pi/(2(N_{\theta}-1))$.
Typical grids have sizes $150 \times 30$ points.
We monitored also the remaining Einstein equations which are not directly solved, assuring that
they are satisfied with a reasonable accuracy.

In this scheme, a new radial variable is introduced
which maps the semi infinite region $[r_c,\infty)$ to the closed region $[0,1]$
(with $r_c=0$ or $r_h$).
 For the globally regular solutions,
our choice for this transformation was $ x=r/(r+1)$.
For the derivatives this leads to the substitutions
\begin{eqnarray}
r F_{,r}   \longrightarrow  x (1-x) F_{,x} \ \ ,
~~
r^2 F_{,r,r}   \longrightarrow
x^2 \left( (1-x)^2 F_{,x,x}
  - 2 (1-x) F_{,x} \right)
\end{eqnarray}
for any function $F$ in the differential equations.
For the black hole solutions, we employed a new coordinte $x$ defined as
$
x = 1-r_h/r,
$
which leads to the following substitutions in the differential equations
\begin{eqnarray}
r F_{,r}   \longrightarrow    (1-x) F_{,x}
~~~
r^2 F_{,r,r}   \longrightarrow
(1-x)^2  F_{,x,x}
  - 2 (1-x) F_{,x}
\end{eqnarray}
for any function $F$.

The resulting system is solved iteratively until convergence is achieved.
All numerical calculations for axially symmetric configurations
are performed by using the program FIDISOL (written in Fortran),
based on the iterative Newton-Raphson method.
 A detailed presentation of the FIDISOL code is presented in \cite{FIDISOL}.
This code requests the system of nonlinear partial differential equations to be written in the
form
$
P(r,\theta,u,u_{r},u_{\theta}
,u_{r \theta},u_{rr},u_{\theta \theta})=0,
$
(where $u$ denotes the unknown functions) subject to a set of boundary
conditions on a rectangular domain.
The user must deliver to FIDISOL the equations, the boundary conditions,
the Jacobian matrices
for the equations
and the boundary conditions, and some initial guess functions.
The numerical procedure works as follows:
for an approximate solution $u^{(1)}$,
$P(u^{(1)})$ does not vanish.
Next step is to consider an improved  solution
$u^{(2)}=u^{(1)}+\Delta u$, supposing that $P(u^{(1)}+\Delta u)=0$.
The expansion in the small parameter
$\Delta u$ gives in the first order
$
0=P(u^{(1)}+\Delta u) \approx
P(u^{(1)})+\frac{\partial P}{\partial u }(u^{(1)}) \Delta u \ .
$
This equation can be used to determine the correction
 $\Delta u^{(1)}= \Delta u$.
 Repeating the calculations
iteratively ($u^{(3)}=u^{(3)}+\Delta u^{(2)}$ etc), the approximate solutions will converge,
provided the initial guess solution is close enough to the exact solution.
The iteration stops after $i$ steps if the Newton residual $P(u^{(i)})$
is smaller than a prescribed tolerance.
Therefore it is essential to have a good
first guess, to start the iteration procedure.
Our strategy therefore is to use a known solution as guess
and then vary some parameter to produce the next solution.

To obtain  axially symmetric solutions,
we start always with the $n=1$ solution
 as initial guess and increase the value of the relevant parameters slowly.
For static solutions,
the  parameter we vary is the winding number $n$.
The physical values of $n$ are integers.
Rotating configurations are found $e.g.$ by increasing
the magnitude of the electric potential at infinity (for black holes, we vary also $\Omega_h$).
The iteration is done in small steps and eventually converges
with a good enough accuracy. Repeating the procedure one obtains
in this way solutions for requested values of the relevant parameters.
For some of the configurations, we interpolate the resulting
 configurations and use them as a starting guess on a finer grid.

FIDISOL automatically provides also an error  estimate for each  function,
which is the maximum of the discretization error divided by the
maximum of the function  \cite{FIDISOL}.
For the solutions discussed in this paper, the typical  numerical error
for the functions is estimated to be on the order of $10^{-3}$.
The output of the code was analysed and visualised mainly with MATHEMATICA.

\section{Charged scalar field in a Einstein Universe
with a U(1) Dirac monopole background}
We consider the following action principle
for a nonminimally coupled scalar field
interacting with a background U(1)-field $A^{(0)}$
\begin{eqnarray}  \label{s-action}
I[\phi]=-  \int \left(D_\mu\phi\,D^{\mu}\phi^*+M^2\phi \phi^*
+\xi R\phi \phi^*\right)\sqrt{g(x)}\,d^3x,
\end{eqnarray}
where $D=\partial-iA^{(0)}$, $M$ is the scalar field mass and $\xi$ determines the coupling with
the scalar curvature $R=2k/\ell^2$.

The
zeta function approach implies the computation of the the eigenfunctions $%
\phi_N$ and the eigenvalues $\lambda_N$ of the differential second-order
selfadjoint operator $A=-D ^\mu D _\mu+M^2+\xi R$.
Thus we consider the series with $s\in C$ (the prime on the sum means that any
possible null eigenvalues are omitted)
\begin{eqnarray}  \label{zeta-def}
\zeta (s|A)={\sum_{N}}^{\prime }\lambda _{N}^{-s}.
\end{eqnarray}%
As is well-known, this series converges provided Re~$s>D/2$. It is possible
to continue the above sum into a meromorphic function of $s$ that is regular
at $s=0$ \cite{eli2}. In a path integral approach, the effective action for
a scalar field can be formally expressed as the functional determinant of
the operator $A$ as
\begin{eqnarray}
\label{defIeff}
I_{eff}=-{\frac{1}{2}}\ln \det (A/\mu ^{2}),
\end{eqnarray}%
where $\mu $ is an arbitrary renormalization mass scale coming from the
path-integral measure. This determinant however is a formally divergent
quantity and needs to be regularized.

In a zeta function renormalization framework, the regularized determinant
reads
\begin{eqnarray}
\ln \det (A/\mu ^{2})=-\zeta ^{\prime }(0|A)-\zeta (0|A)\ln \mu ^{2}.
\end{eqnarray}%
We note that since $\zeta (0|A)=0$ in odd dimensions (which is our case),
the dependence on the renormalization scale drops out.

The eigenvalue
equation $A \phi_N=\lambda_N \phi_N$ can be solved
 by using the ansatz
\begin{eqnarray}
\phi_N=e^{i(m \varphi-\omega t)}F(\theta),
\end{eqnarray}
where  $F(\theta)$ is a solution
of the equation
\begin{eqnarray}
\label{gen-F}
\frac{1}{\ell^2 f_k(\theta)}\frac{d}{d\theta}(f_k(\theta)
\frac{dF}{d\theta})-\frac{m^2}{\ell^2 f_k^2(\theta)}(m+Q_mf_k'(\theta))^2F
+(\omega^2+M^2+\xi R)F=\lambda_N F,
\end{eqnarray}
satisfying certain boundary conditions at the limits of the $\theta$ interval.
Here we will restrict to the $k=1$ case of a zero-temperature
 $2+1$-dimensional Einstein universe
background and a U(1) field $A^{(0)}=Q_m \cos \theta d \varphi$.
Note that  $m$ should be an integer for $k=1$.

The substitution $F=g/\sqrt{\sin\theta}$, $\theta=2x$
 transforms (\ref{gen-F})
into the quantum mechanical problem of the
Schr\"odinger equation with a P\"oschl-Teller potential
 \begin{eqnarray}
\label{g-eq}
-\frac{d^2 g}{dx^2}+((m-Q_m)^2-\frac{1}{4})\frac{g}{\sin^2 x}
+((m+Q_m)^2-\frac{1}{4})\frac{g}{\cos^2 x}
\\
\nonumber
=
(4Q_m^2+1-4\ell^2(\omega^2+M^2 +\xi R-\lambda_N))g=0,
\end{eqnarray}
whose solutions are well known, see e.g. \cite{Cooper:1994eh}.
 This leads to an eigenvalue expression
\begin{eqnarray}
\lambda _{N}=\omega ^{2}+ \frac{1}{4}\Big(
1+|m+Q_m|+|m-Q_m|+2n
\Big)^2+M^{2}+\xi R-Q_m^2-1/4,
\end{eqnarray}%
where $(N=m,n,\omega )$.
The corresponding eigenfunctions are
\begin{eqnarray}
\phi_N(x)=(\cos \frac{\theta}{2})^{|m+Q_m|+1/2}
(\sin \frac{\theta}{2})^{|m-Q_m|}
P^{(|m-Q_m|,|m+Q_m|)}(\cos \theta)e^{i(m\varphi-\omega t)},
\end{eqnarray}%
with $P^{(a,b)}(x)$ the Jacobi polynomials.

After integrating over $\omega$, we arrive at a
sum  on the form
\begin{eqnarray}
\zeta(s)=
\sqrt{\pi}\frac{\Gamma(s-\frac{1}{2})}{\Gamma(s)}
\sum_{n=0}^{\infty}\sum_{m=-\infty}^{\infty}
(\frac{1}{2}(1+|m+Q_m|+|m-Q_m|)+n)^2+M^2+\xi R-Q_m^2-\frac{1}{4})^{-s+1/2}
\end{eqnarray}
Restricting to the case of a massless, conformally coupled field, the resulting
zeta function can be written as
\begin{eqnarray}
\zeta(s)=
\sqrt{\pi}\frac{\ell^{2s-1}}{\Gamma(s)}
\sum_{k=0}^{\infty}
\frac{\Gamma(s+k-\frac{1}{2})}{k!}
Q_m^{2k}
\sum_{n=0}^{\infty}\sum_{m=-\infty}^{\infty}(n+\frac{1}{2}
(1+|m+Q_m|+|m-Q_m|))^{-2s-2k+1}
\end{eqnarray}
Neglecting an additional contribution arising as a result of
interchanging the order of summation, one
performs first the sum over $(m,n)$, finding the simple approximate expression
\begin{eqnarray}
\label{z-f}
&\zeta(s)\sim
\sqrt{\pi}\frac{\ell^{2s-1}}{\Gamma(s)}
\sum_{k=0}^{\infty}
\frac{\Gamma(s+k-\frac{1}{2})}{k!}
Q_m^{2k}
\bigg
(2\zeta_H(2s+2k-2,Q_m+\frac{3}{2})
\\
\nonumber
&+(2Q_m+1)\left(\zeta_H(2s+2k-1,Q_m+\frac{1}{2})
-\zeta_H(2s+2k-1,Q_m+\frac{3}{2})\right)
\bigg)
\end{eqnarray}
with $\zeta _{H}(s,a)$ the Hurwitz zeta functions, which are
meromorphic functions with a unique simple pole at $s=1$.
In deriving this relation we used also \cite{Elizalde:1988fg}
\begin{eqnarray}
\sum_{m,n=0}^{\infty }[m+n+a]^{-s}=\zeta _{H}(s-1,a)-(a-1)\zeta _{H}(s,a).
\end{eqnarray}%
The zeta function (\ref{z-f}) is analytic throughout the complex $s-$plane except
for $s=3/2-n$ (with $n=0,1,\dots )$ where simple poles appear,
while $\zeta(0)=0$.
The derivative of this function evaluated at $s=0$ is
\begin{eqnarray}
\label{zetap}
&\zeta ^{\prime }(0)\sim
\frac{\sqrt{\pi}}{\ell}\bigg
\{
-2\sqrt{\pi}\bigg[\zeta_H(-2,Q_m+\frac{3}{2})
+(2Q_m+1)\left(\zeta_H(-1,Q_m+\frac{1}{2})
-\zeta_H(-1,Q_m+\frac{3}{2})\right)
\bigg]
\\
\nonumber
&+\sqrt{\pi}Q_m^2\bigg[2\zeta_H(0,Q_m+\frac{3}{2})
+(2Q_m+1)\left(\Psi(Q_m+\frac{3}{2})
-\Psi(Q_m+\frac{1}{2})\right)
\bigg]
\\
\nonumber
&+\sum_{k=0}^{\infty}
\frac{\Gamma(k-\frac{3}{2})}{(k+2)!}
Q_m^{2k+4}
\bigg[
2\zeta_H(2k+2,Q_m+\frac{3}{2})
+(2Q_m+1) \left(\zeta_H(2k+3,Q_m+\frac{1}{2})
-\zeta_H(2k+3,Q_m+\frac{3}{2}) \right)
\bigg]
\bigg\},
\end{eqnarray}%
where $\Psi(x)= d\log \Gamma(x)/{dx}$.

\newpage
\setlength{\unitlength}{1cm}

\begin{picture}(16,16)
\centering
\put(-1,0){\epsfig{file=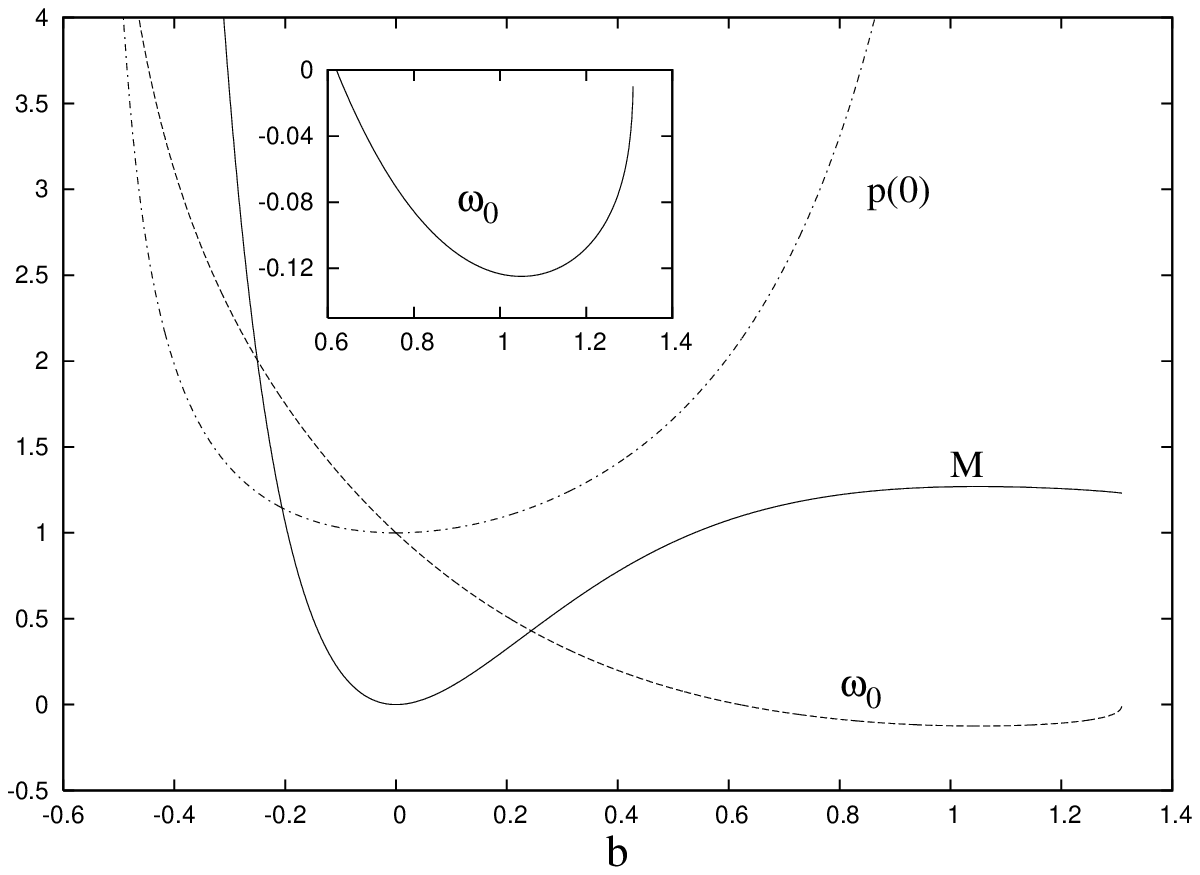,width=16cm}}
\end{picture}
\\
\\
\\
{\small {\bf Figure 1.} \\
The mass-parameter $M$, the asymptotic value of
the gauge function $\omega_0$ and
the value of the metric function
$p(r)$ at the origin
 are represented as a function of the parameter $b$ for spherically symmetric,
 globally regular monopole solutions.}
\newpage
\setlength{\unitlength}{1cm}

\begin{picture}(16,16)
\centering
\put(-1,0){\epsfig{file=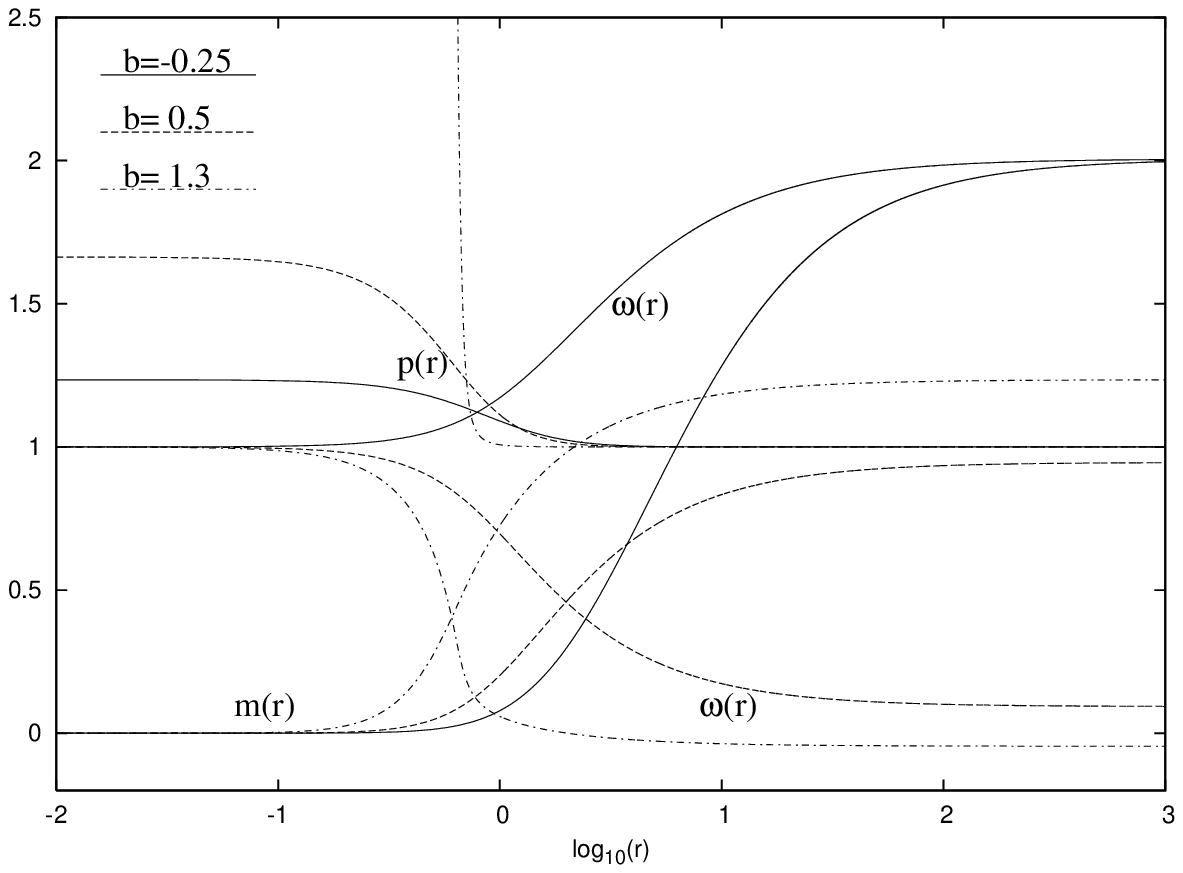,width=16cm}}
\end{picture}
\\
\\
\\
{\small {\bf Figure 2.}
\\
The profiles of typical spherically symmetric globally regular
monopole solutions
are plotted for several values of the parameter $b$
which enters the expansion at the origin of the gauge potential.}

\newpage
\setlength{\unitlength}{1cm}

\begin{picture}(16,16)
\centering
\put(-1,0){\epsfig{file=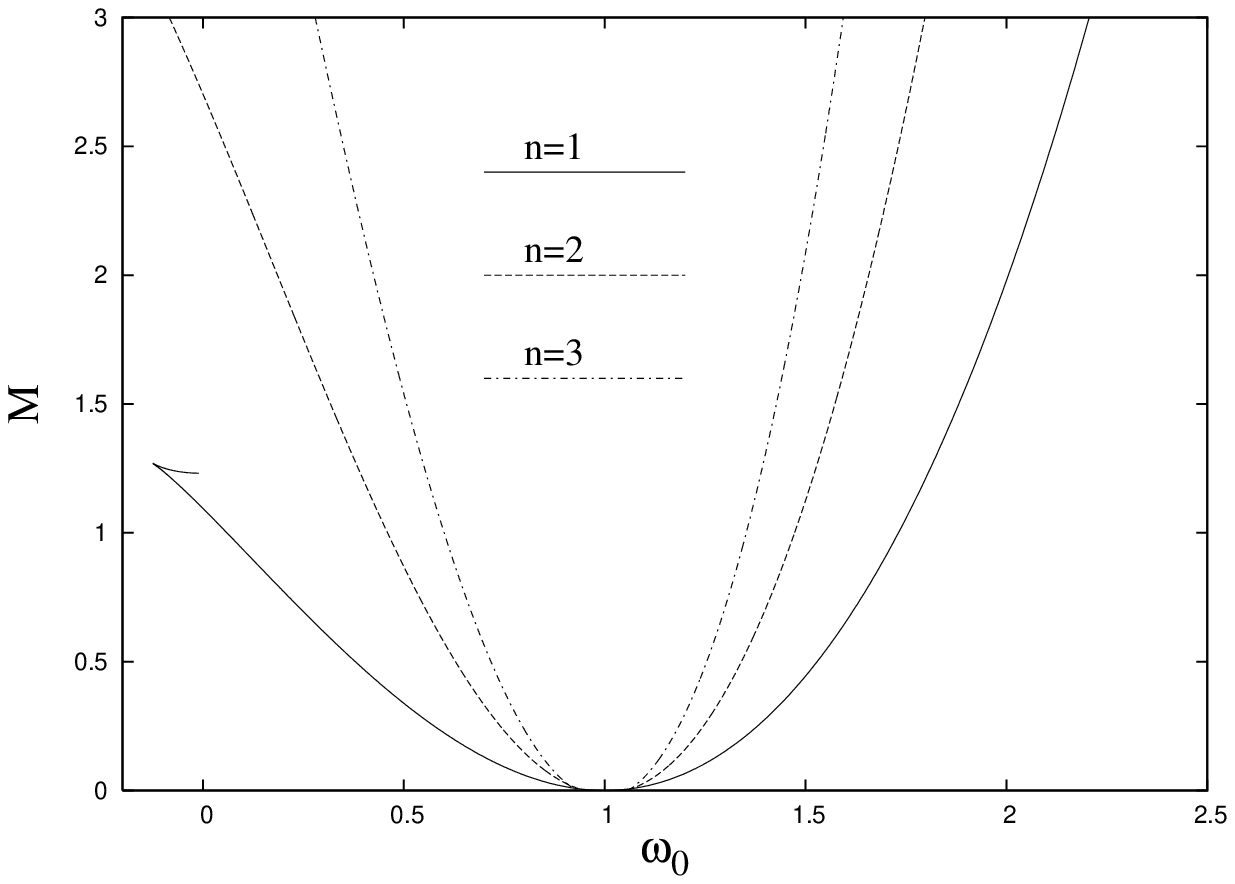,width=16cm}}
\end{picture}
\\
\\
\\
{\small {\bf Figure 3.}
\\
Mass $M$ is plotted as a function of $\omega_0$
for globally regular gravitating monopole static solutions.
The winding number $n$ is also  marked.
The configurations with $n=2,3$ are axially symmetric.}

\newpage
\setlength{\unitlength}{1cm}

\begin{picture}(16,16)
\centering
\put(-1,0){\epsfig{file=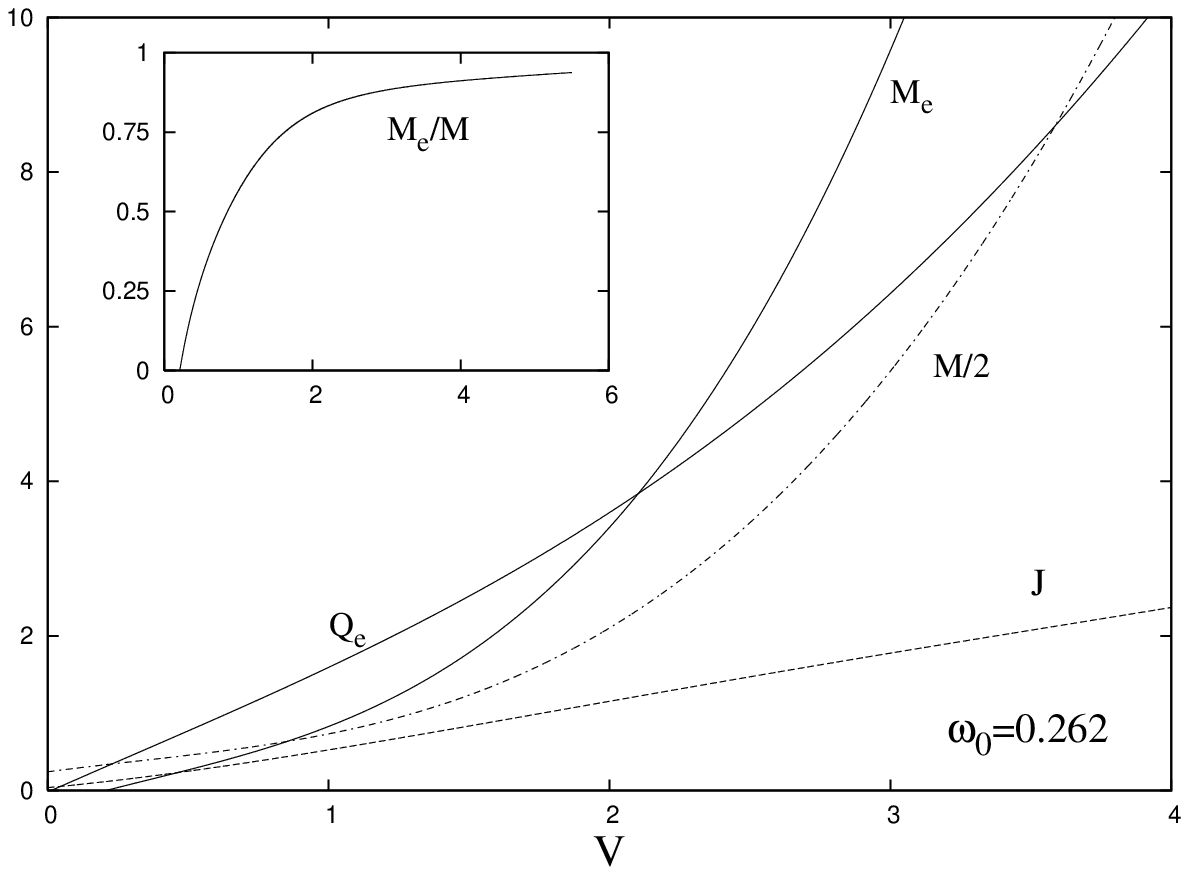,width=16cm}}
\end{picture}
\\
\\
\\
{\small {\bf Figure 4a.} }

\newpage
\setlength{\unitlength}{1cm}

\begin{picture}(16,16)
\centering
\put(-1,0){\epsfig{file=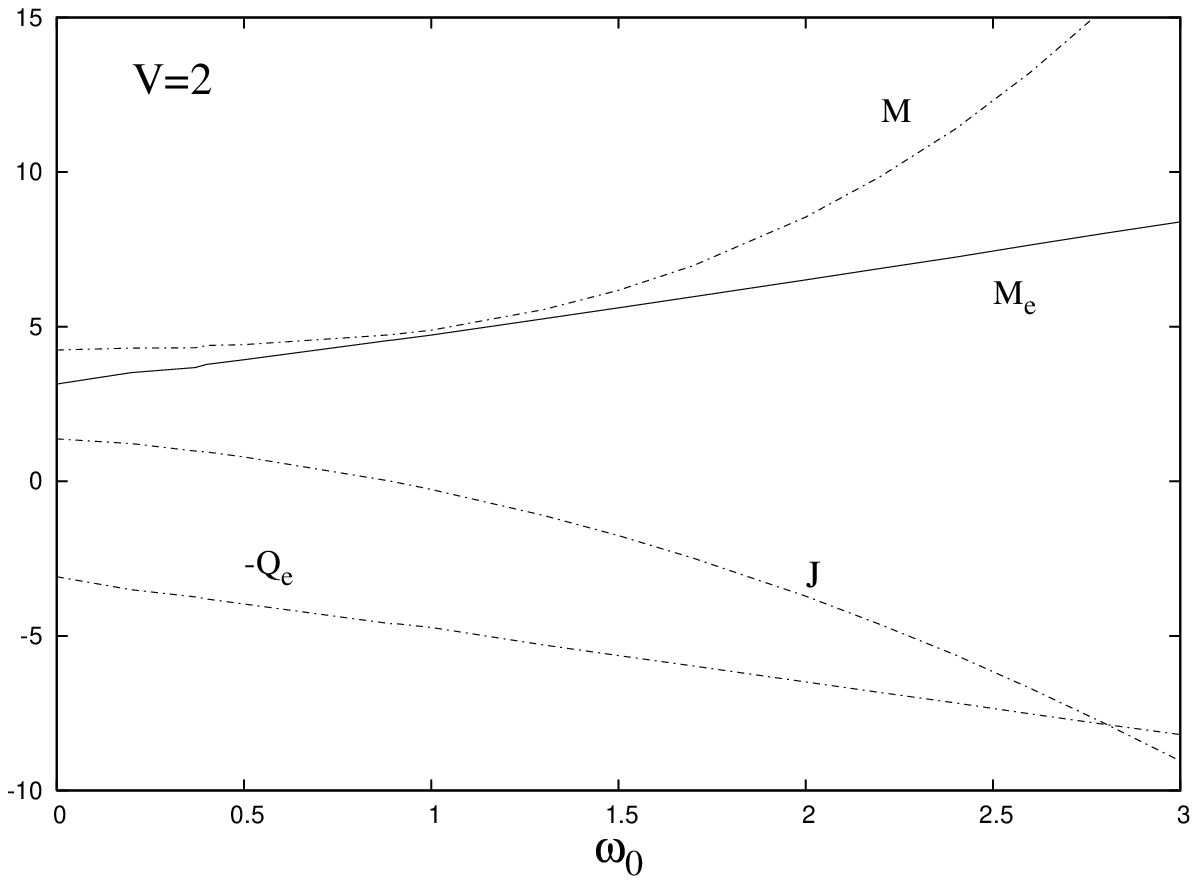,width=16cm}}
\end{picture}
\\
\\
\\
{\small {\bf Figure 4b.}
\\The mass $M$ and the angular momentum $J$
of non-Abelian globally regular rotating solutions
are shown as a function on the parameter $V$ (Figure 4a, for $\omega_0=0.262$)
and the parameter $\omega_0$ (Figure 4b, $V=2$).
Also shown are  the electric charge $Q_e$ and the contribution $M_e$
of the nonabelian electric field  to the total energy
of the system.

\setlength{\unitlength}{1cm}

\begin{picture}(16,16)
\centering
\put(-2,0){\epsfig{file=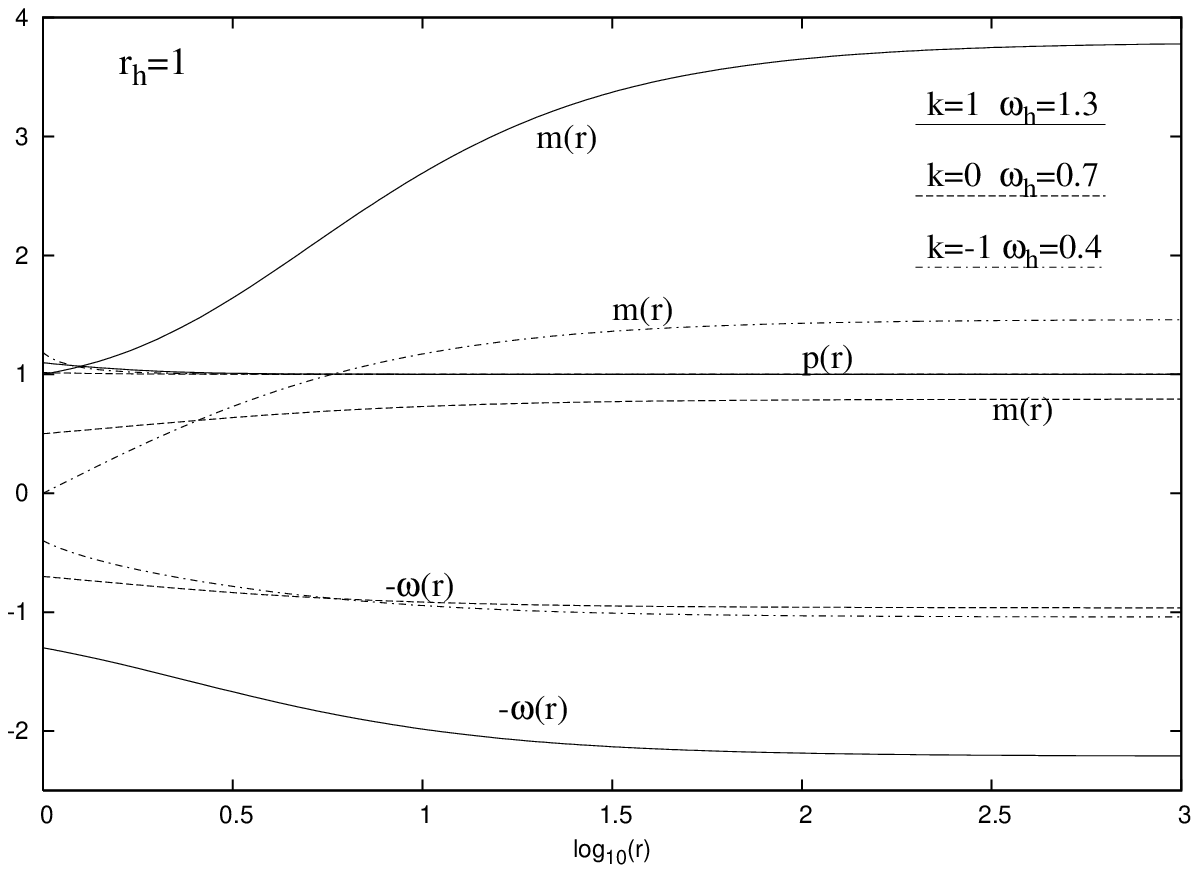,width=16cm}}
\end{picture}
\\
\\
\\
{\small {\bf Figure 5.}
\\
Typical profiels of $n=1$ black hole
monopole solutions. The solutions with $k=0,-1$ correspond to topological
black holes with nonabelian hair.


\setlength{\unitlength}{1cm}

\begin{picture}(16,16)
\centering
\put(-2,0){\epsfig{file=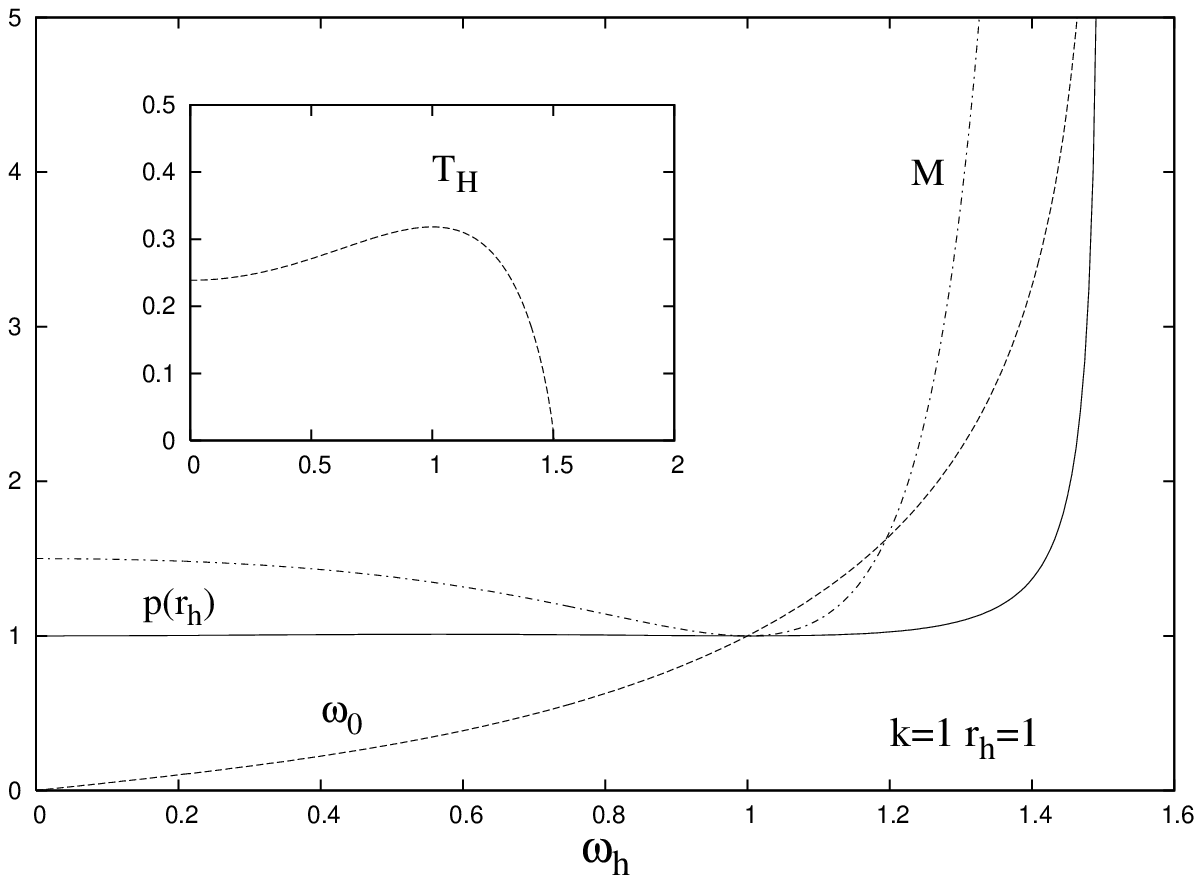,width=16cm}}
\end{picture}
\\
\\
\\
{\small {\bf Figure 6a.}

\setlength{\unitlength}{1cm}

\begin{picture}(16,16)
\centering
\put(-2,0){\epsfig{file=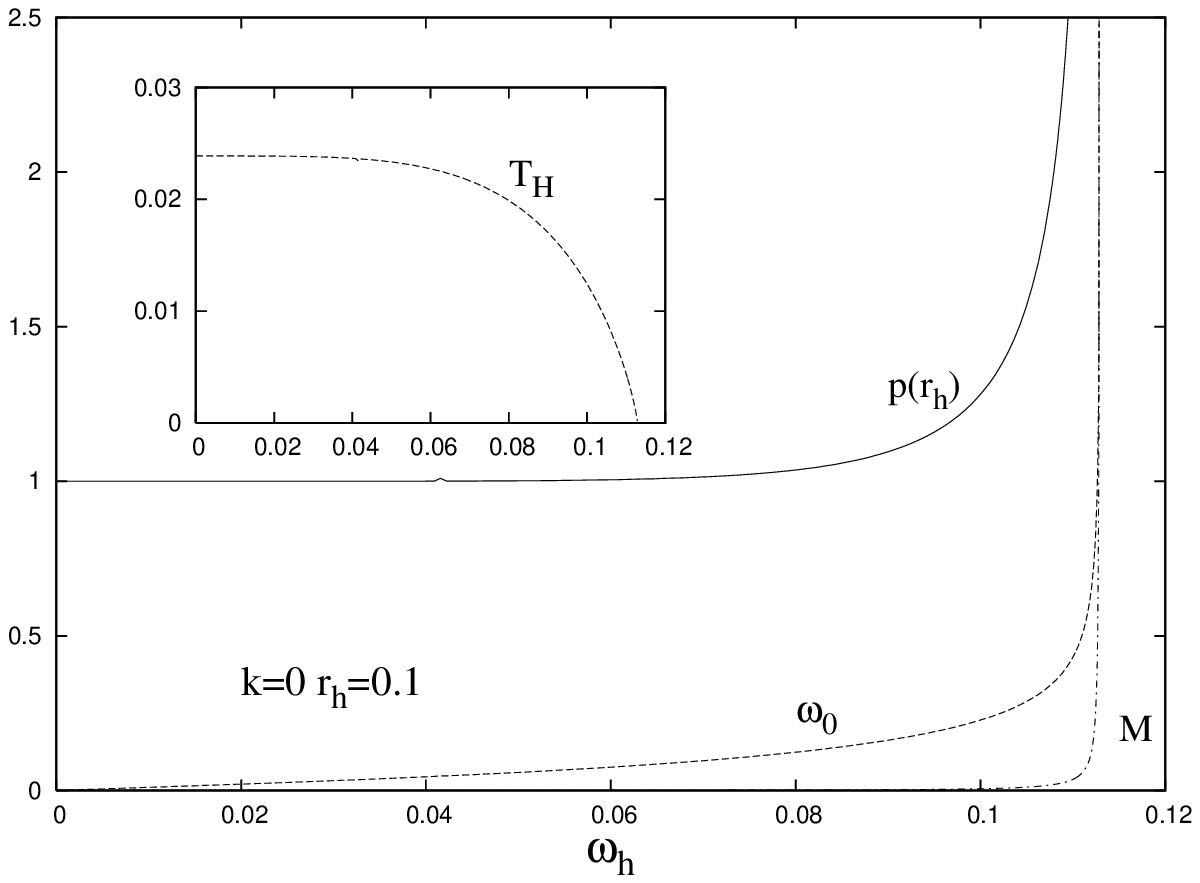,width=16cm}}
\end{picture}
\\
\\
\\
{\small {\bf Figure 6b.}
\setlength{\unitlength}{1cm}

\begin{picture}(16,16)
\centering
\put(-2,0){\epsfig{file=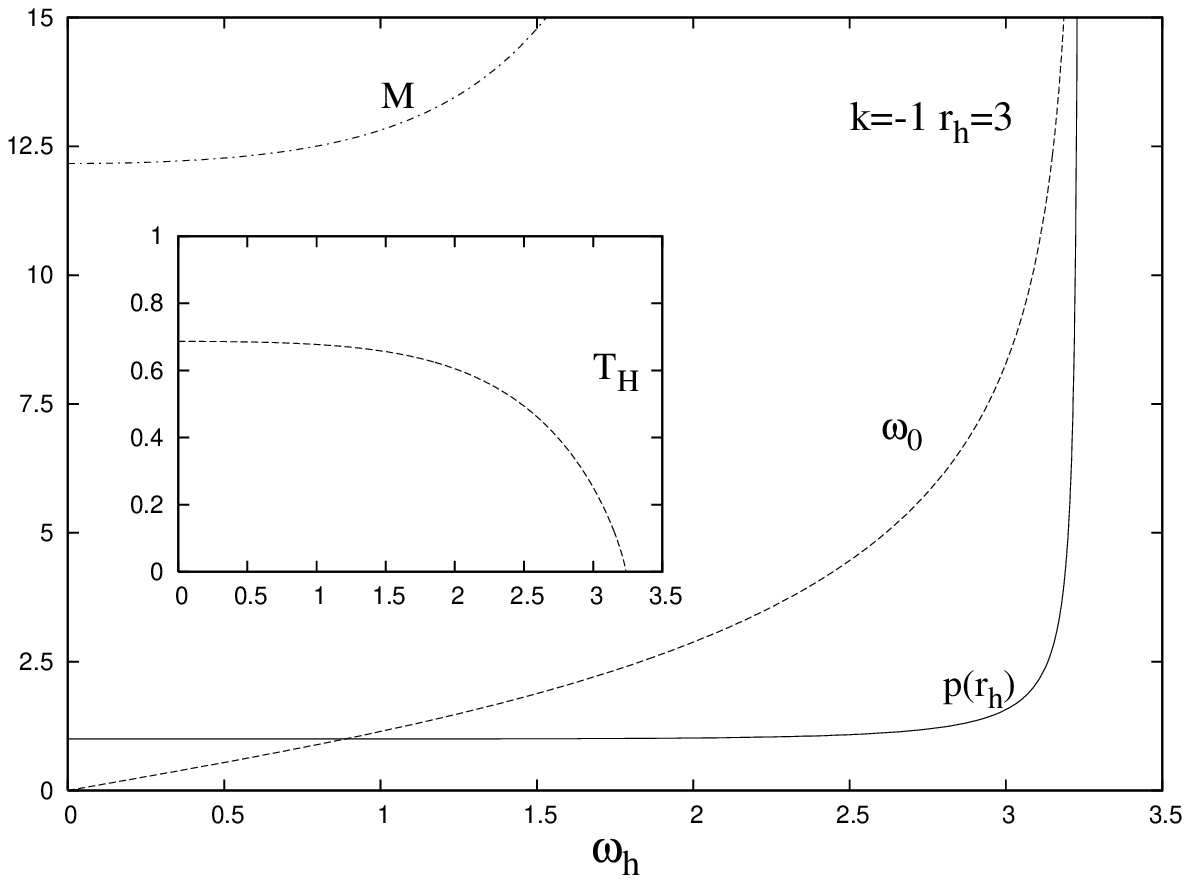,width=16cm}}
\end{picture}
\\
\\
\\
{\small {\bf Figure 6c.}

The mass-parameter $M$, the asymptotic value of
the gauge function $\omega_0$,
the value of the metric function
$p(r)$ at the event horizon and the Hawking temperature
 are represented as a function of the value of the gauge function $\omega$ at the event horizon
for typical monopole black hole solutions.

\newpage
\setlength{\unitlength}{1cm}

\begin{picture}(16,16)
\centering
\put(-1,0){\epsfig{file=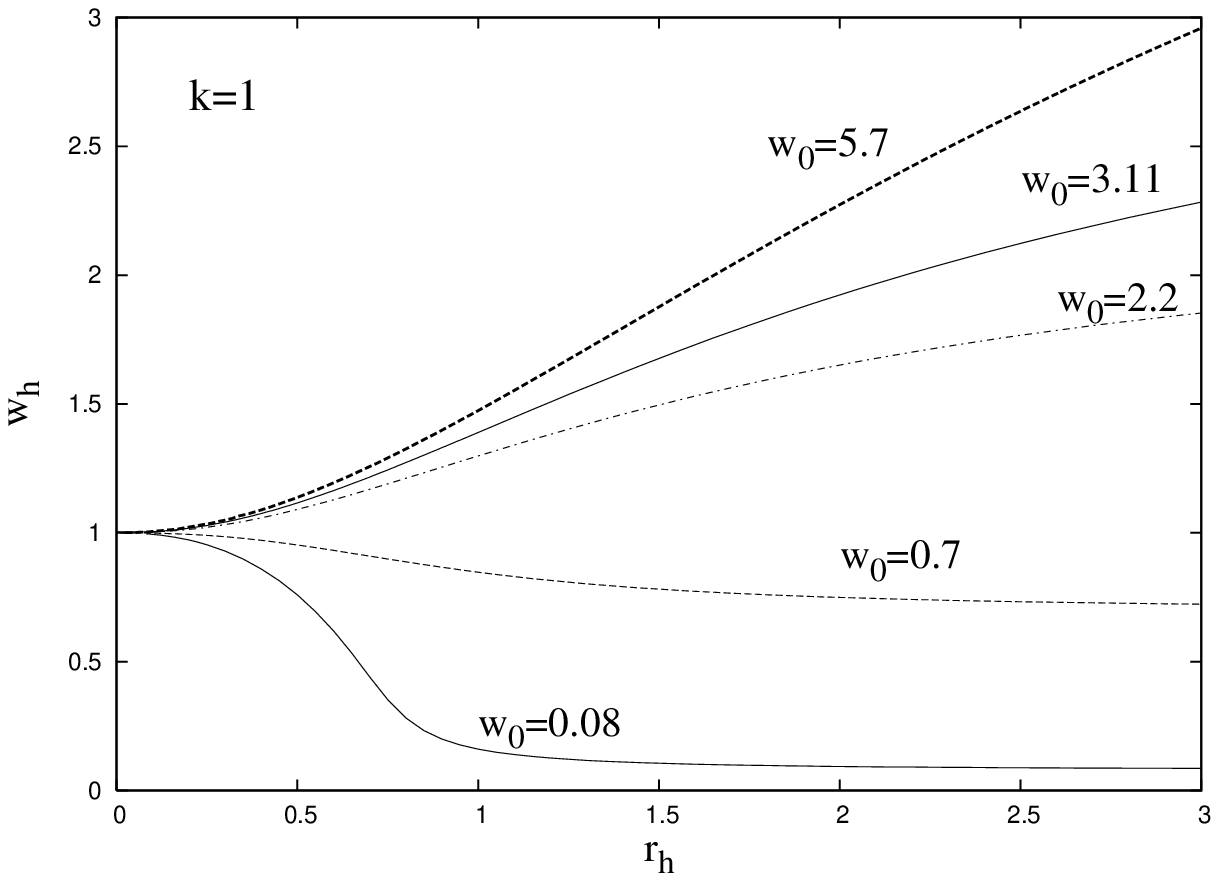,width=16cm}}
\end{picture}
\\
\\
\\
{\small {\bf Figure 7a.}}
\newpage
\setlength{\unitlength}{1cm}

\begin{picture}(16,16)
\centering
\put(-1,0){\epsfig{file=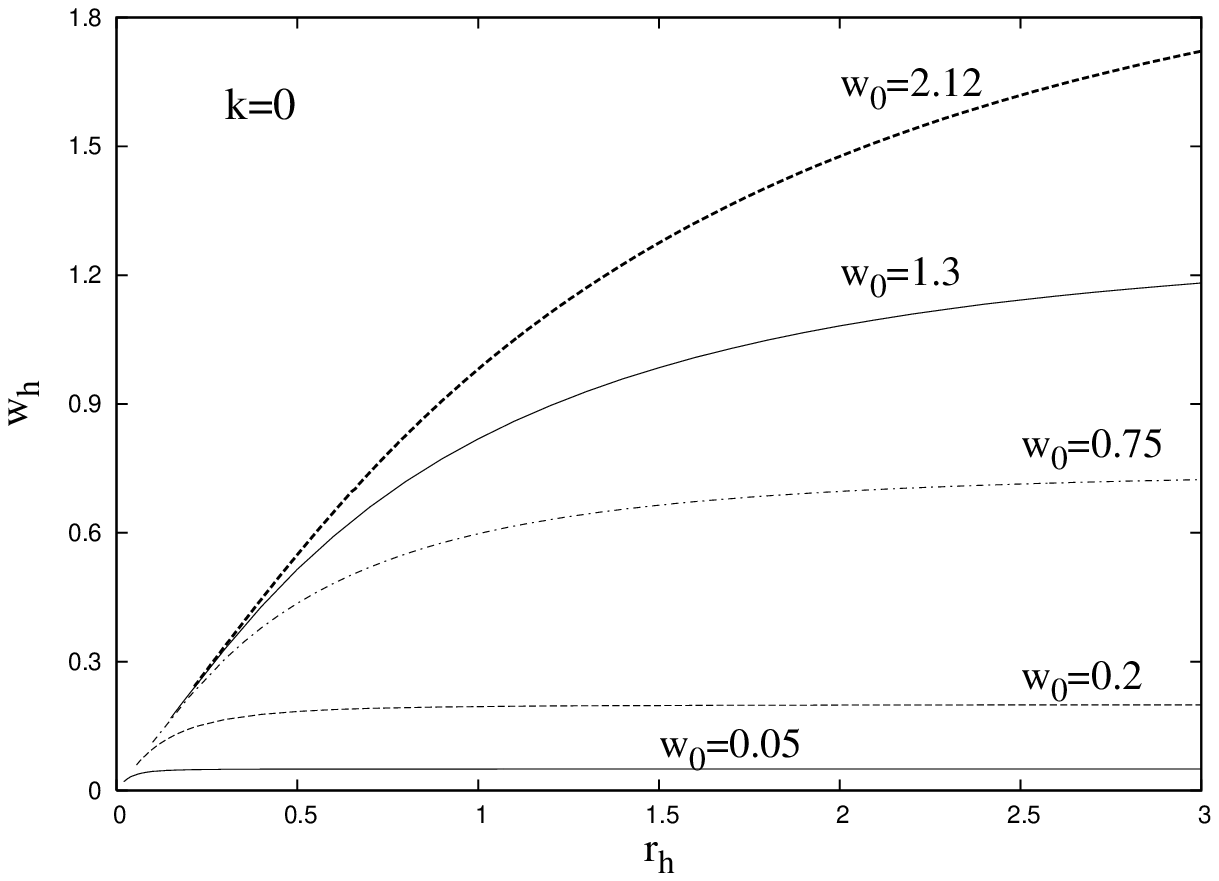,width=16cm}}
\end{picture}
\\
\\
\\
{\small {\bf Figure 7b.}}
\newpage
\setlength{\unitlength}{1cm}

\begin{picture}(16,16)
\centering
\put(-1,0){\epsfig{file=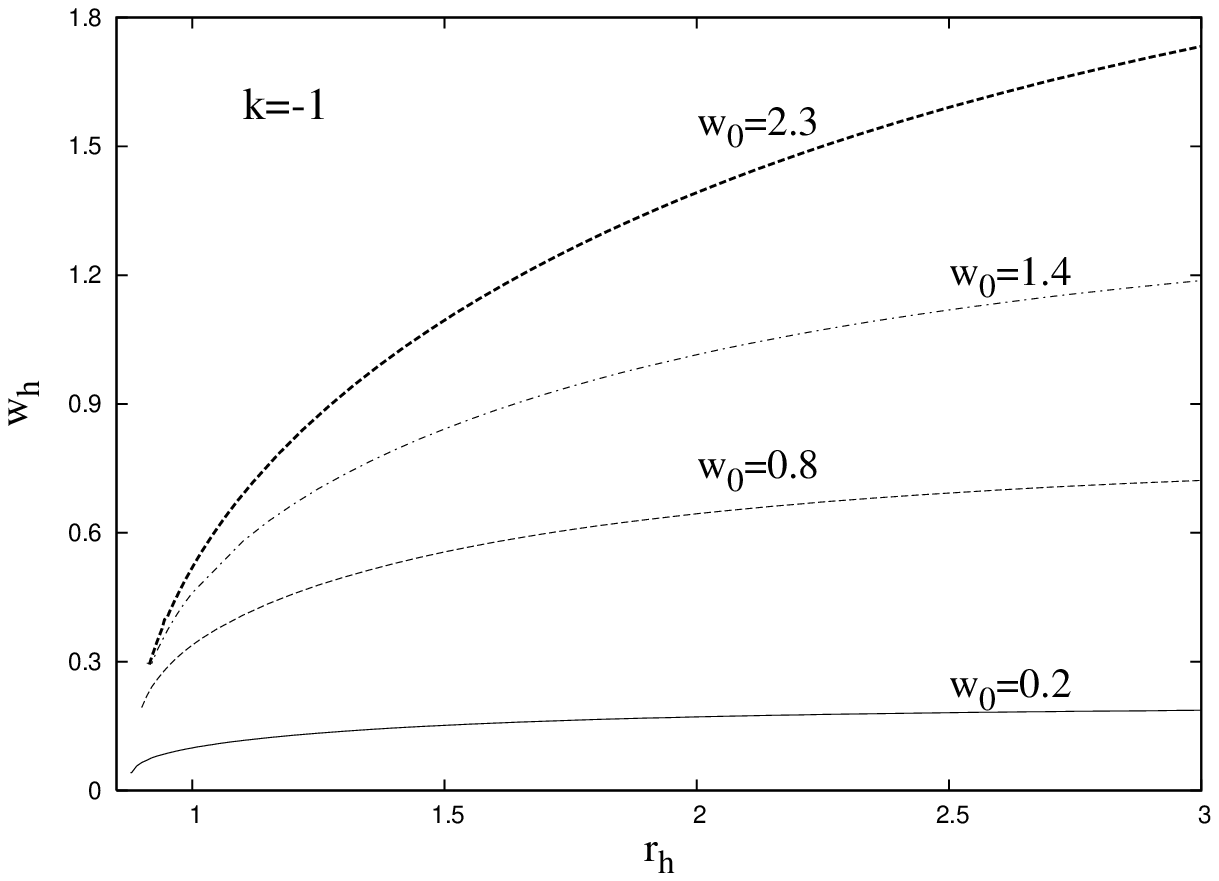,width=16cm}}
\end{picture}
\\
\\
\\
{\small {\bf Figure 7c.}
The value of the nonabelian gauge potential
at the event horizon $w_h$ is plotted as a function of the event horizon radius
$r_h$
for spherically symmetric and topological black hole monopole solutions and several values
of the magnetic potential at infinity.}

\newpage
\setlength{\unitlength}{1cm}

\begin{picture}(16,16)
\centering
\put(-1,0){\epsfig{file=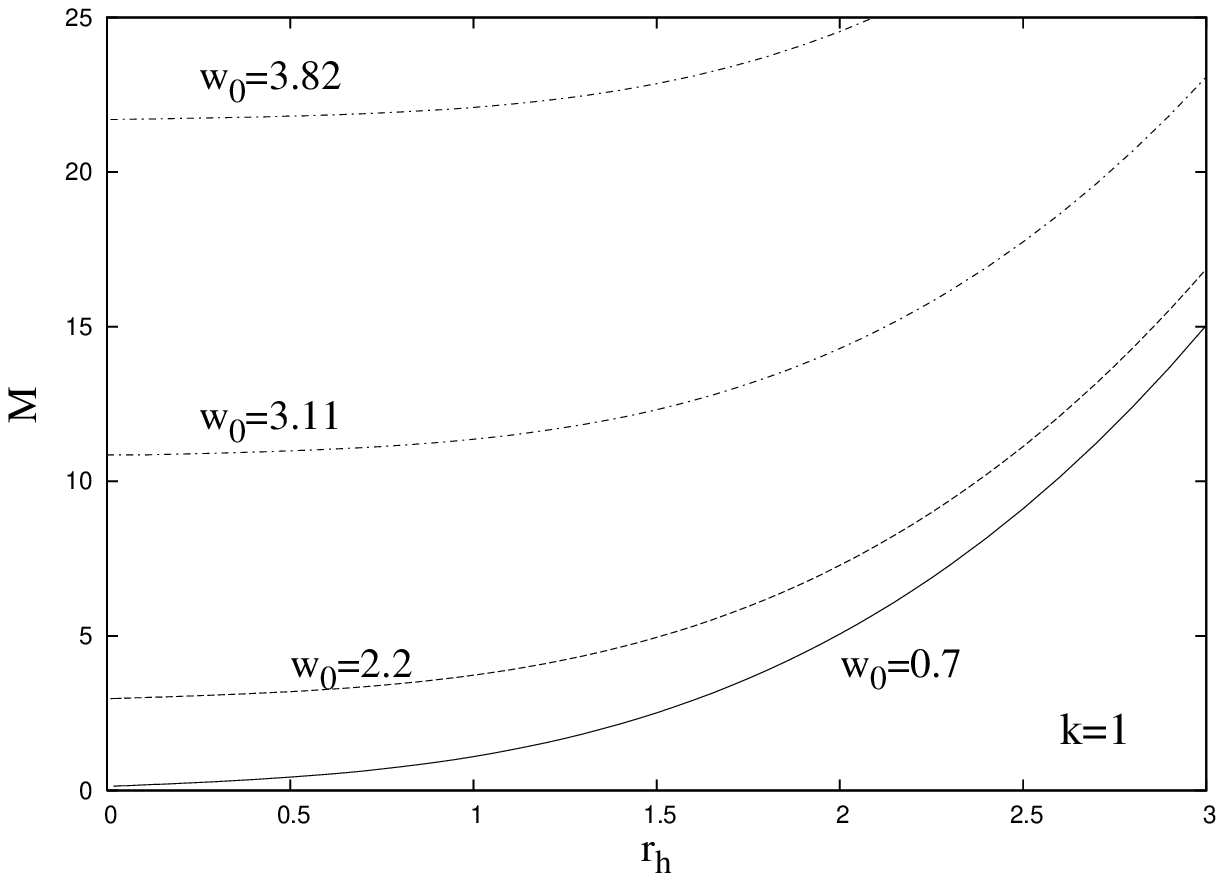,width=16cm}}
\end{picture}
\\
\\
\\
{\small {\bf Figure 8a.}}
\newpage
\setlength{\unitlength}{1cm}

\begin{picture}(16,16)
\centering
\put(-1,0){\epsfig{file=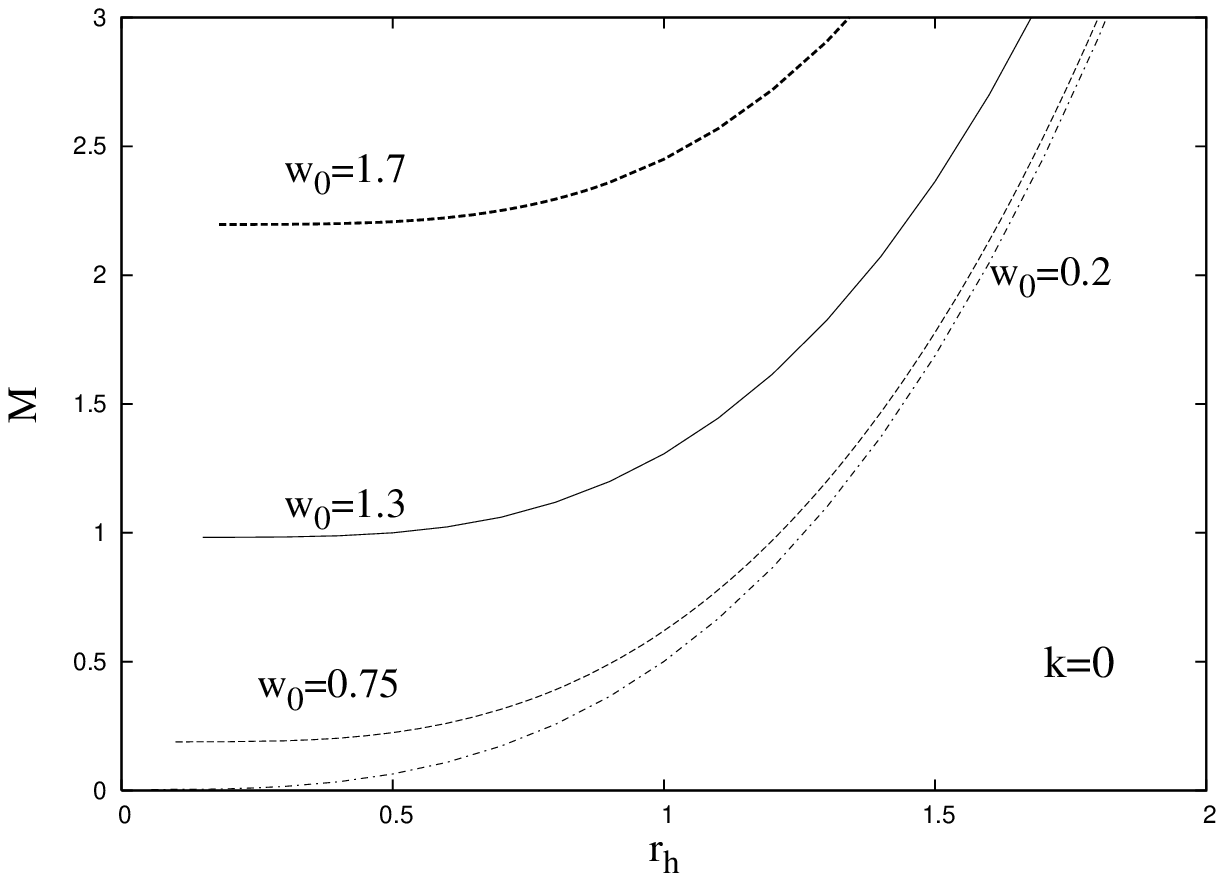,width=16cm}}
\end{picture}
\\
\\
\\
{\small {\bf Figure 8b.}}
\newpage
\setlength{\unitlength}{1cm}

\begin{picture}(16,16)
\centering
\put(-1,0){\epsfig{file=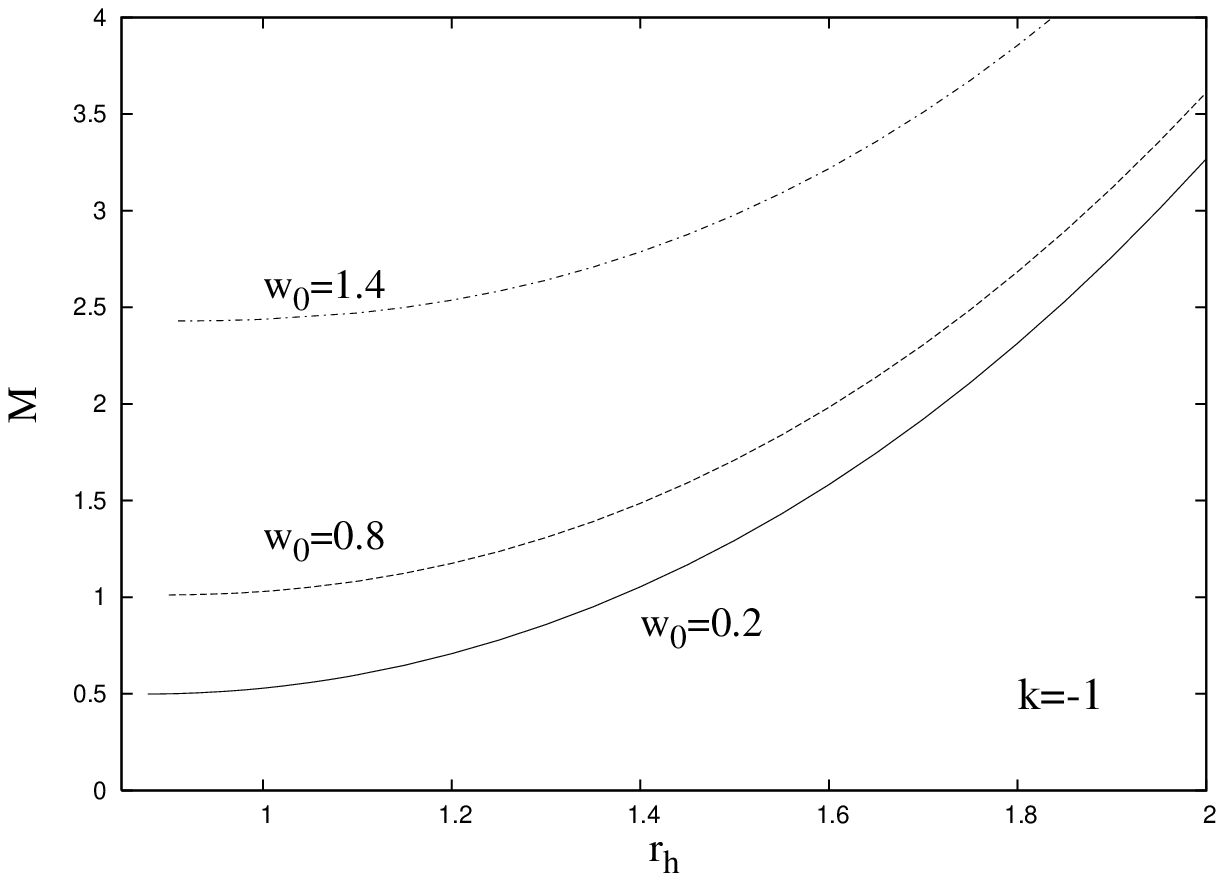,width=16cm}}
\end{picture}
\\
\\
\\
{\small {\bf Figure 8c.}
The black hole mass $M$ is plotted as a function of event horizon radius
$r_h$
for $k=1,0,-1$ black hole monopole solutions with $n=1$ and several values
of the magnetic potential at infinity.}

\newpage
\setlength{\unitlength}{1cm}

\begin{picture}(16,16)
\centering
\put(-1,0){\epsfig{file=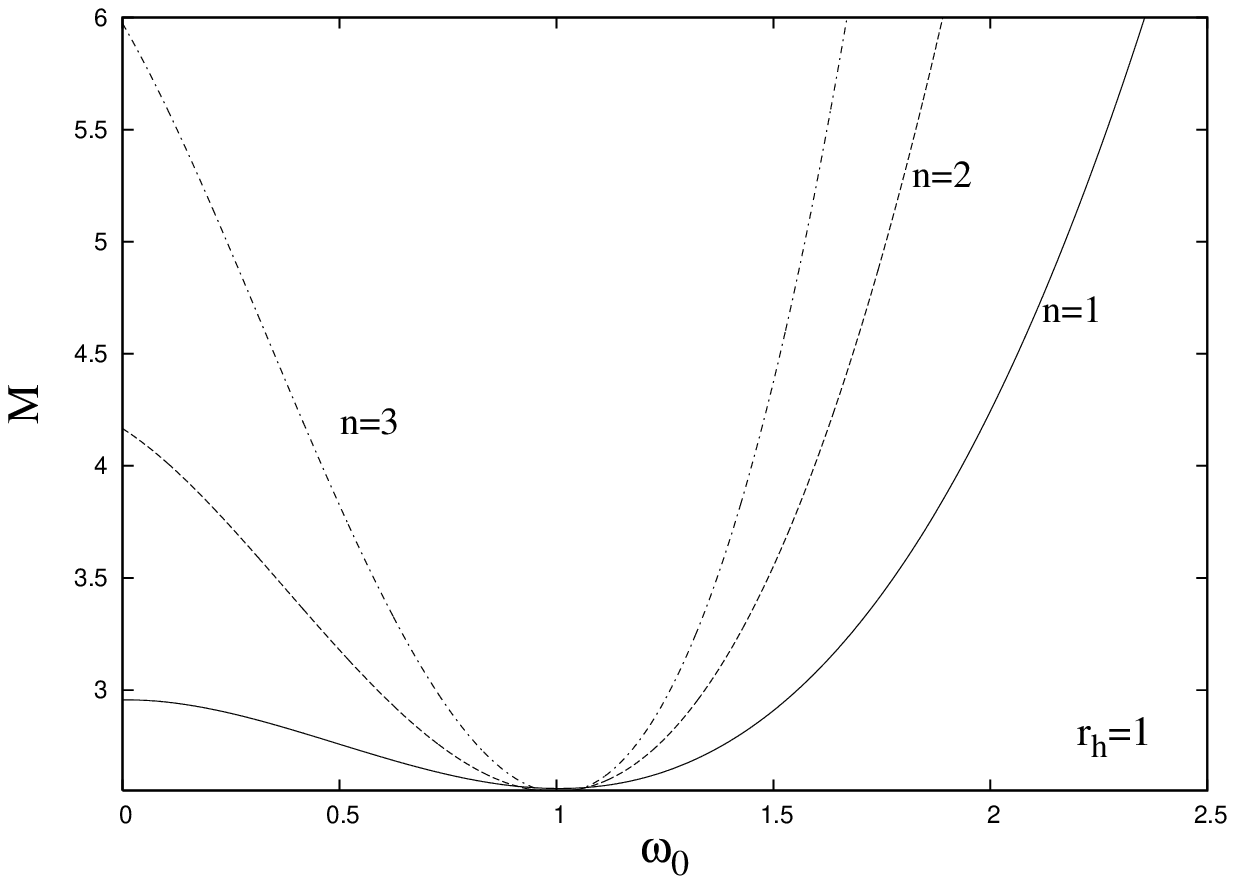,width=16cm}}
\end{picture}
\\
\\
\\
{\small {\bf Figure 9a.}}


\newpage
\setlength{\unitlength}{1cm}

\begin{picture}(16,16)
\centering
\put(-1,0){\epsfig{file=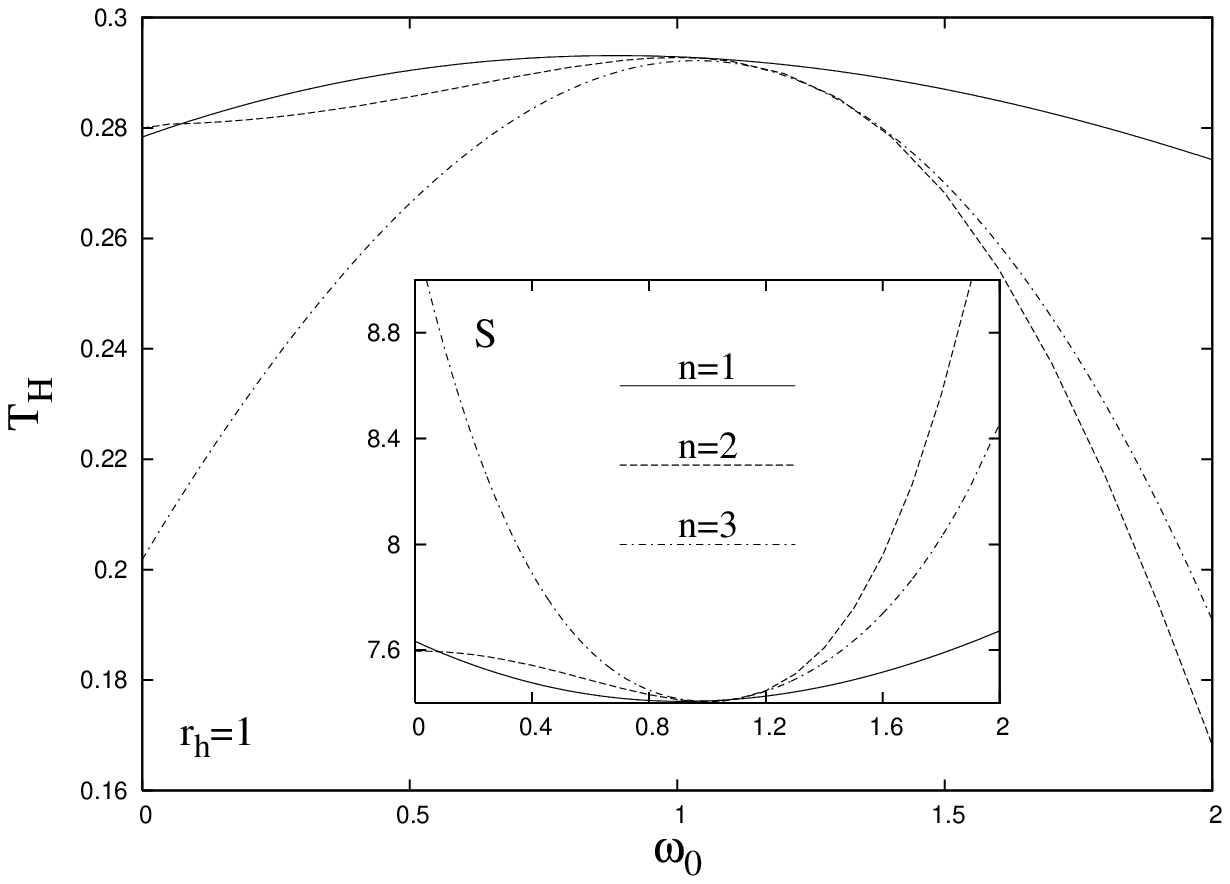,width=16cm}}
\end{picture}
\\
\\
\\
{\small {\bf Figure 9b.}  \\
The mass $M$ (Figure 9a) and the Hawking temperature and the entropy
(Figure 9b)  are plotted as a function of $\omega_0$
for static black hole   monopole solutions with $r_h=1$.
The winding number $n$ is also  marked.
The configurations with $n=2,3$ are axially symmetric.}


\newpage
\setlength{\unitlength}{1cm}

\begin{picture}(16,16)
\centering
\put(-1,0){\epsfig{file=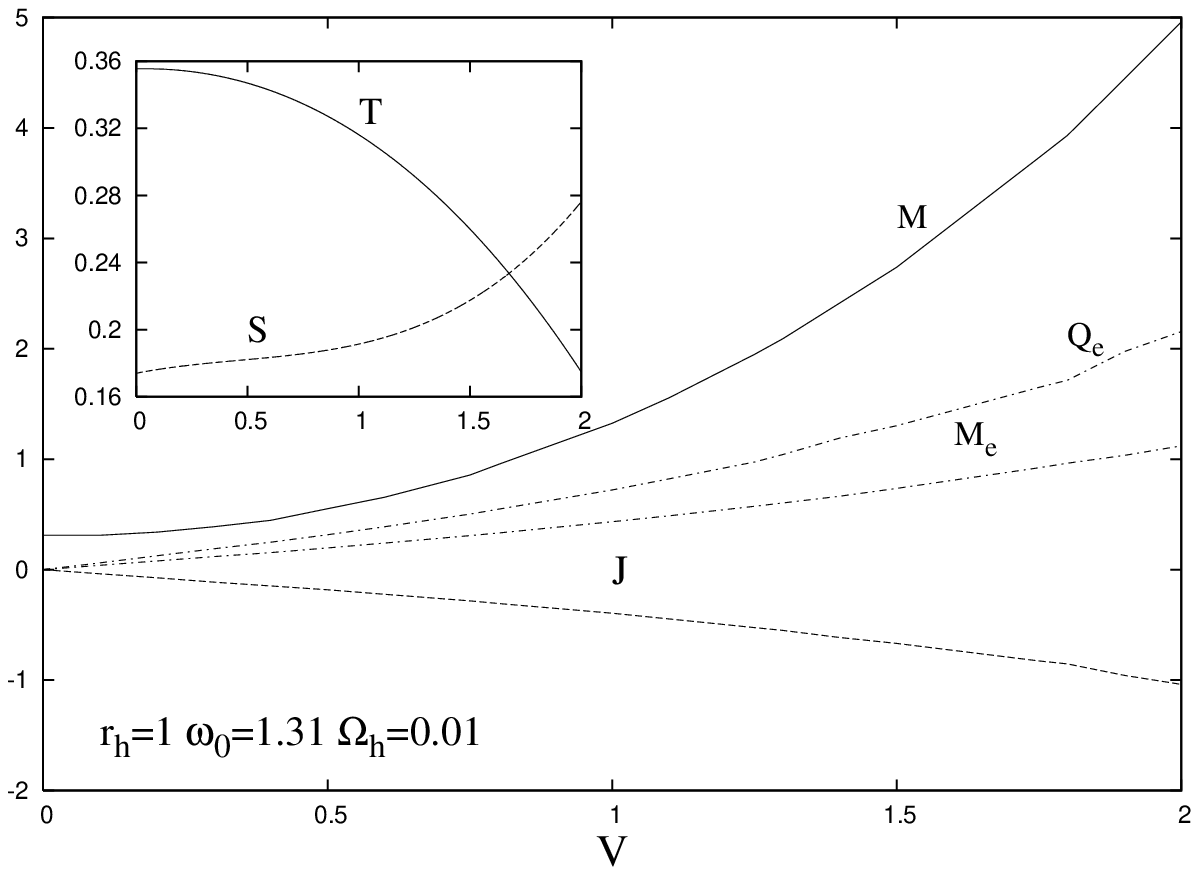,width=16cm}}
\end{picture}
\\
\\
\\
{\small {\bf Figure 10.}\\
The mass $M$, the angular momentum $J$, the electric charge $Q_e$ and the contribution $M_e$
of the electric field  to the total energy
of the system
of non-Abelian black hole rotating solutions
are shown as a function on the parameter $V$ for fixed values of $r_h,~\omega_0$ and $\Omega_h$.
Also shown are  the entropy and the Hawking temperature.
}

\newpage
\setlength{\unitlength}{1cm}

\begin{picture}(16,16)
\centering
\put(-1,0){\epsfig{file=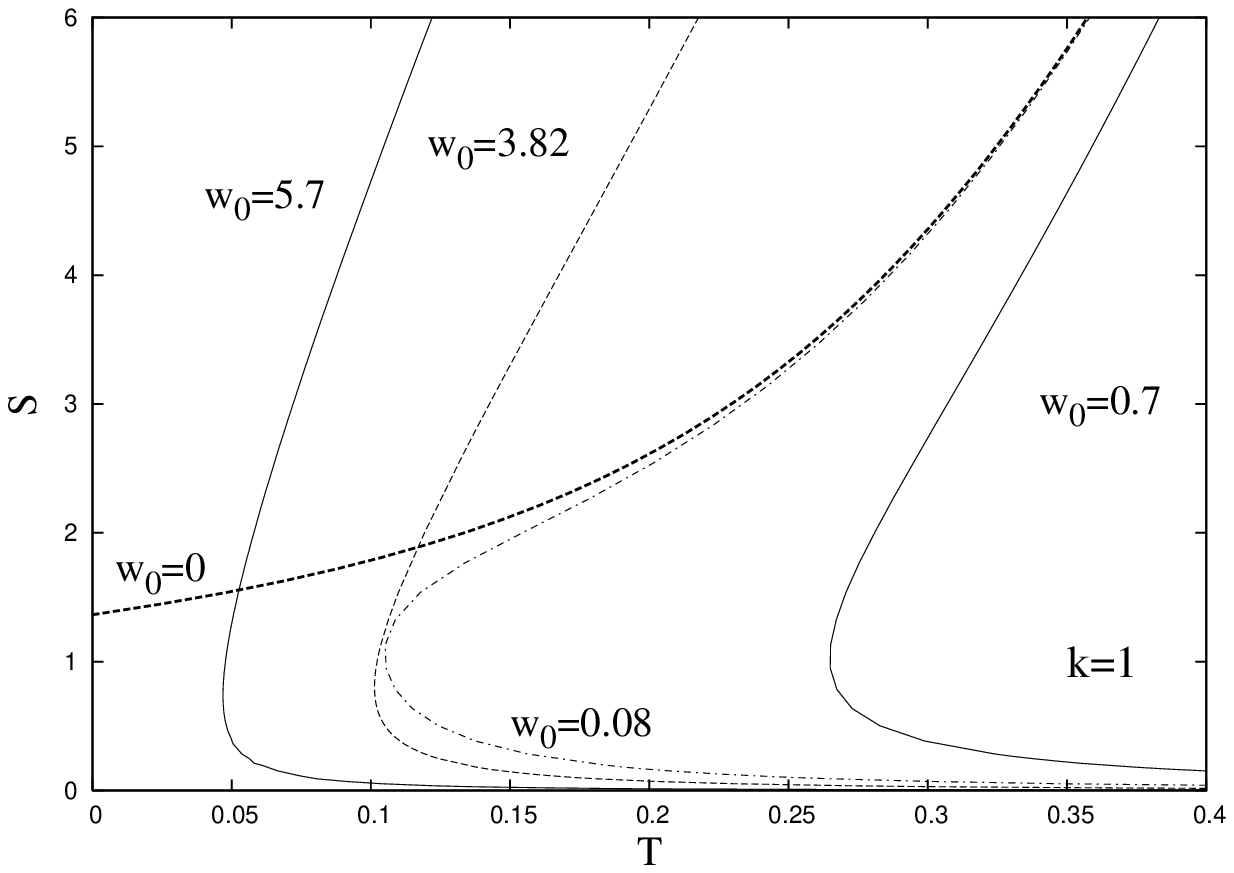,width=16cm}}
\end{picture}
\\
\\
\\
{\small {\bf Figure 11a.}}
\newpage
\setlength{\unitlength}{1cm}

\begin{picture}(16,16)
\centering
\put(-1,0){\epsfig{file=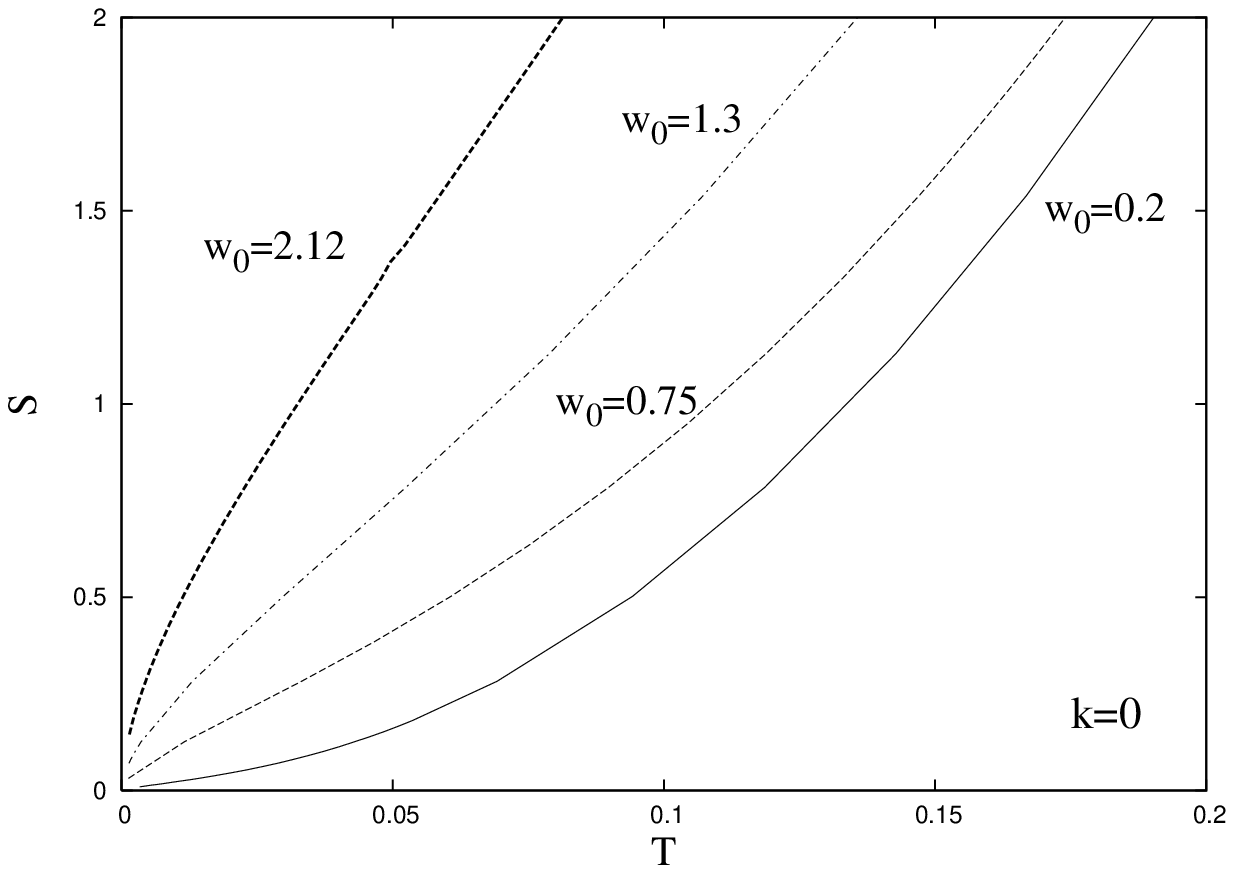,width=16cm}}
\end{picture}
\\
\\
\\
{\small {\bf Figure 11b.}}
\newpage
\setlength{\unitlength}{1cm}

\begin{picture}(16,16)
\centering
\put(-1,0){\epsfig{file=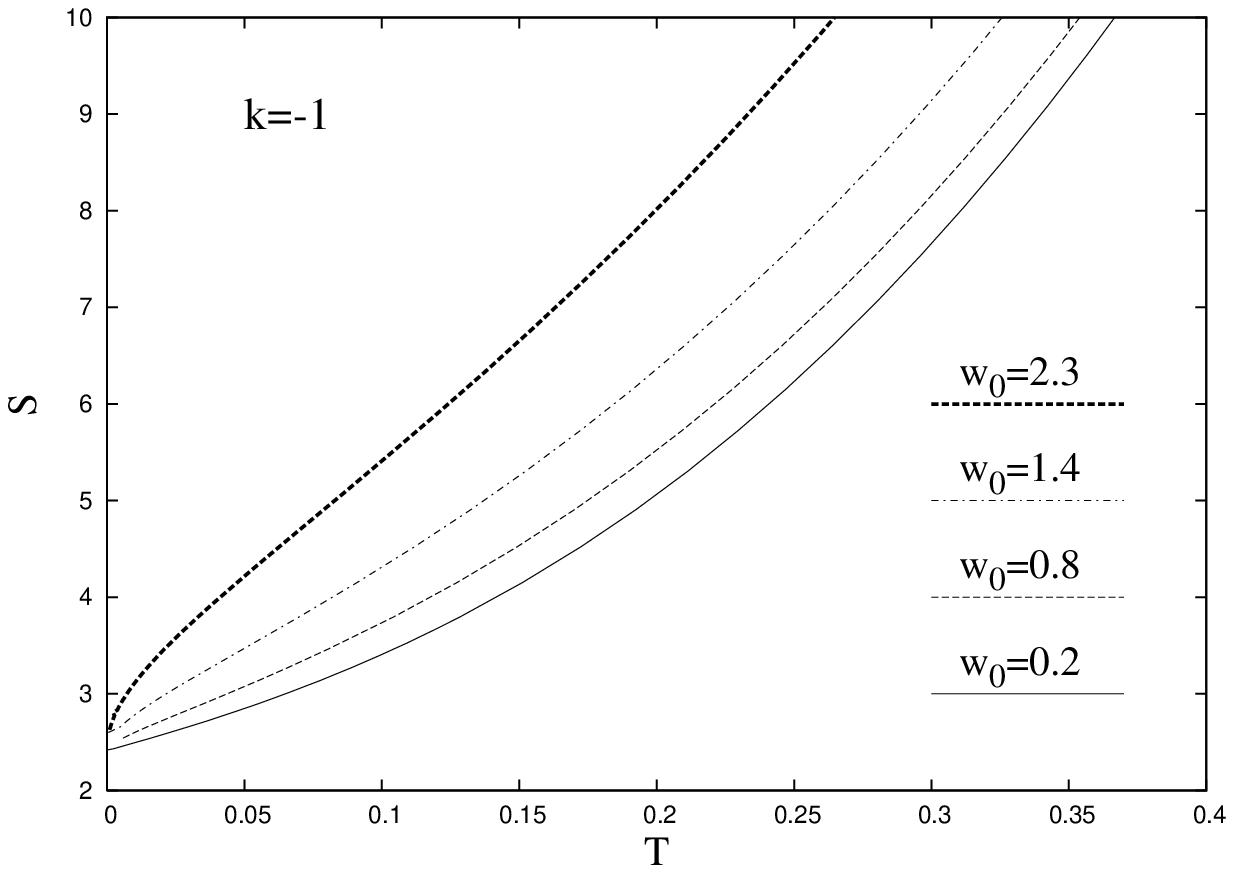,width=16cm}}
\end{picture}
\\
\\
\\
{\small {\bf Figure 11c.}
The entropy is plotted as a function of temperature
for $k=1,0,-1$ black hole monopole solutions and several values
of the magnetic potential at infinity.
Here and in Figure 12a, the $k=1$ curve with $w_0=0$ corresponds to
RNAdS abelian black holes
with unit magnetic charge.}

\newpage
\setlength{\unitlength}{1cm}

\begin{picture}(16,16)
\centering
\put(-1,0){\epsfig{file=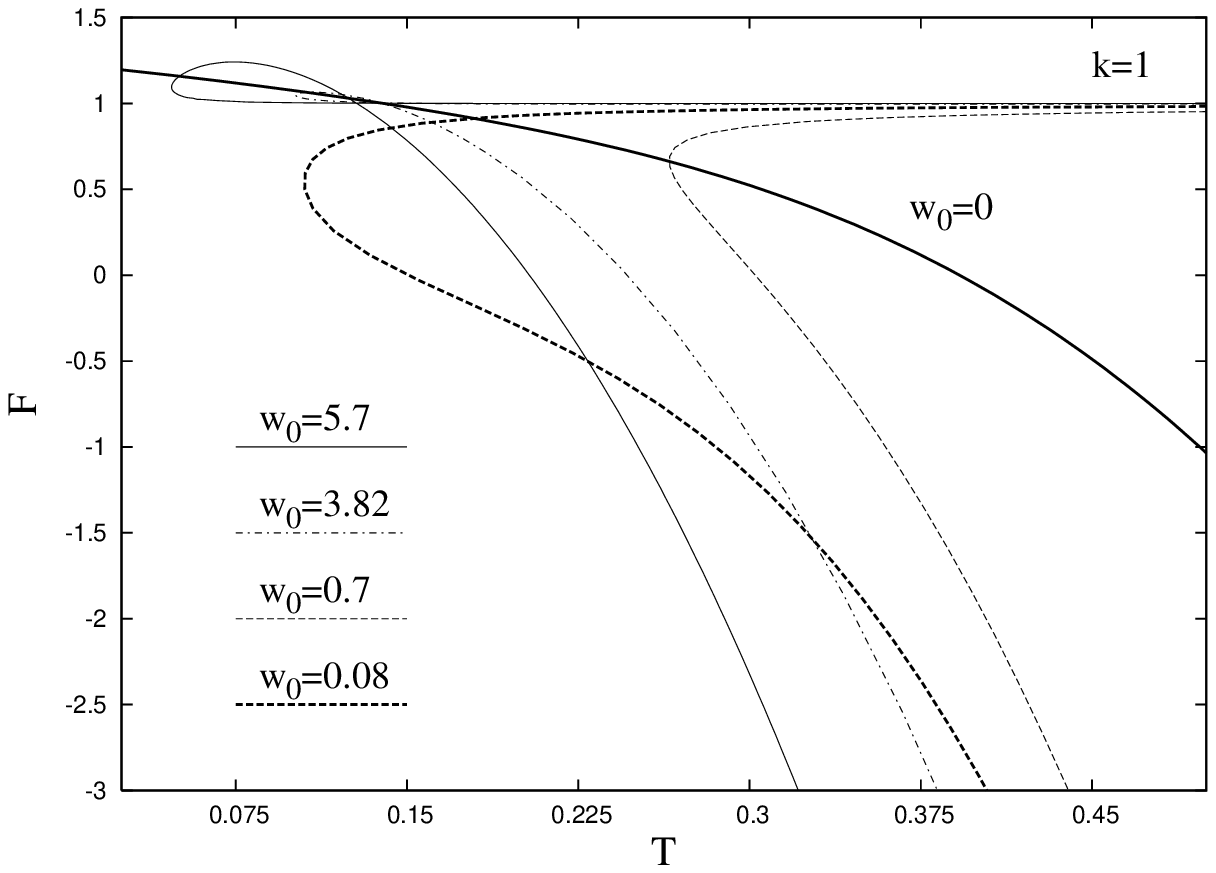,width=16cm}}
\end{picture}
\\
\\
\\
{\small {\bf Figure 12a.}}
\newpage
\setlength{\unitlength}{1cm}

\begin{picture}(16,16)
\centering
\put(-1,0){\epsfig{file=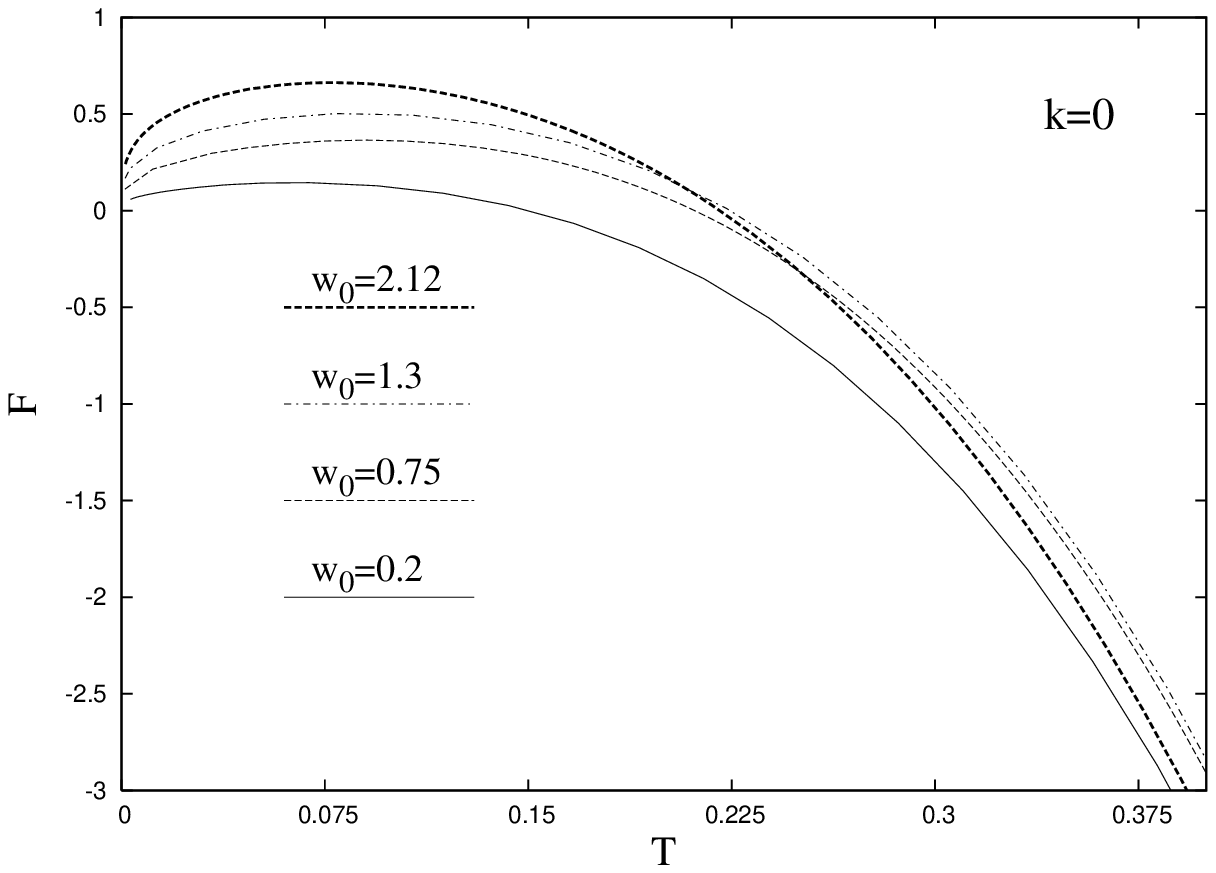,width=16cm}}
\end{picture}
\\
\\
\\
{\small {\bf Figure 12b.}}
\newpage
\setlength{\unitlength}{1cm}

\begin{picture}(16,16)
\centering
\put(-1,0){\epsfig{file=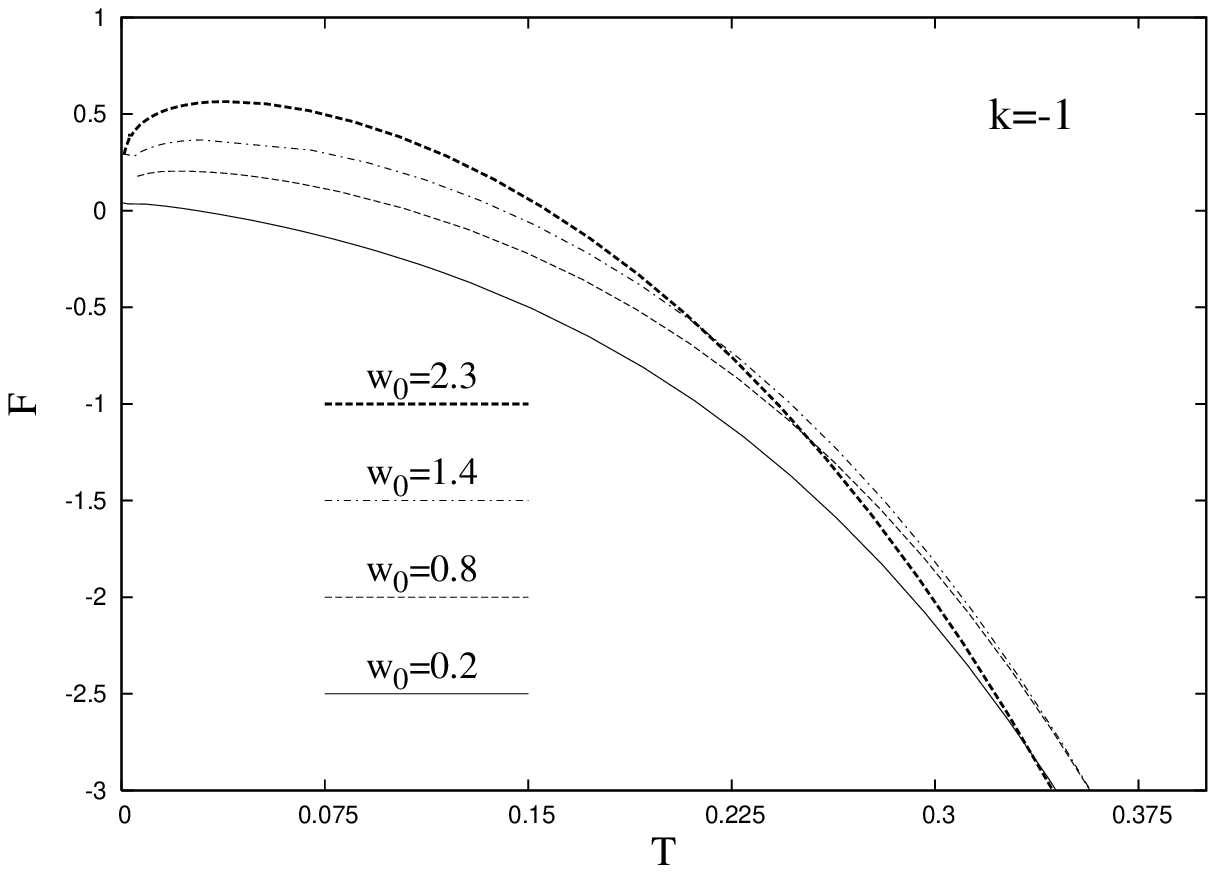,width=16cm}}
\end{picture}
\\
\\
\\
{\small {\bf Figure 12c.}
The free energy $F=M-TS$ is plotted as a function of temperature
for $k=1,0,-1$ black hole monopole solutions and several values
of the magnetic potential at infinity.}


\begin{thebibliography}{99}
\bibitem{Witten:1998qj}
E.~Witten,
Adv.\ Theor.\ Math.\ Phys.\  {\bf 2} (1998) 253
[arXiv:hep-th/9802150].
\bibitem{Maldacena:1997re}
J.~M.~Maldacena,
Adv.\ Theor.\ Math.\ Phys.\  {\bf 2} (1998) 231
[Int.\ J.\ Theor.\ Phys.\  {\bf 38} (1999) 1113]
[arXiv:hep-th/9711200].
\bibitem{Gubser:1998bc}
  S.~S.~Gubser, I.~R.~Klebanov and A.~M.~Polyakov,
  Phys.\ Lett.\ B {\bf 428} (1998) 105
  [arXiv:hep-th/9802109].
\bibitem{Bartnik:1988am}
  R.~Bartnik and J.~Mckinnon,
  Phys.\ Rev.\ Lett.\  {\bf 61} (1988) 141.
\bibitem{Volkov:sv}
M.~S.~Volkov and D.~V.~Galtsov,
Sov.\ J.\ Nucl.\ Phys.\  {\bf 51} (1990) 747
[Yad.\ Fiz.\  {\bf 51} (1990) 1171];
\\
P.~Bizon,
Phys.\ Rev.\ Lett.\  {\bf 64} (1990) 2844;
\\
H.~P.~Kuenzle and A.~K.~M.~Masood- ul- Alam,
J.\ Math.\ Phys.\  {\bf 31} (1990) 928.
\bibitem{Winstanley:1998sn}
E.~Winstanley,
Class.\ Quant.\ Grav.\  {\bf 16} (1999) 1963
[arXiv:gr-qc/9812064].
\bibitem{Bjoraker:2000qd}
J.~Bjoraker and Y.~Hosotani,
Phys.\ Rev.\ D {\bf 62} (2000) 043513
[arXiv:hep-th/0002098].

\bibitem{Sarbach:2001mc}
O.~Sarbach and E.~Winstanley,
Class.\ Quant.\ Grav.\  {\bf 18} (2001) 2125
[arXiv:gr-qc/0102033].

\bibitem{Breitenlohner:2003qj}
P.~Breitenlohner, D.~Maison and G.~Lavrelashvili,
Class.\ Quant.\ Grav.\  {\bf 21} (2004) 1667
[arXiv:gr-qc/0307029].
\bibitem{Hosotani:2001iz}
Y.~Hosotani,
J.\ Math.\ Phys.\  {\bf 43} (2002) 597
[arXiv:gr-qc/0103069].
\bibitem{Radu:2001ij}
E.~Radu,
Phys.\ Rev.\ D {\bf 65} (2002) 044005
[arXiv:gr-qc/0109015].
\bibitem{Radu:2004gu}
  E.~Radu and E.~Winstanley,
  Phys.\ Rev.\ D {\bf 70} (2004) 084023
  [arXiv:hep-th/0407248].
\bibitem{Radu:2002rv}
E.~Radu,
Phys.\ Lett.\ B {\bf 548} (2002) 224
[arXiv:gr-qc/0210074].
\bibitem{Radu:2002hf}
E.~Radu,
Phys.\ Rev.\ D {\bf 67} (2003) 084030
[arXiv:hep-th/0211120].
\bibitem{VanderBij:2001ia}
J.~J.~Van der Bij and E.~Radu,
Phys.\ Lett.\ B {\bf 536} (2002) 107
[arXiv:gr-qc/0107065].

\bibitem{Okuyama:2002mh}
N.~Okuyama and K.~i.~Maeda,
Phys.\ Rev.\ D {\bf 67} (2003) 104012
[arXiv:gr-qc/0212022].
\bibitem{Radu:2005mj}
  E.~Radu and D.~H.~Tchrakian,
  Phys.\ Rev.\ D {\bf 73} (2006) 024006
  [arXiv:gr-qc/0508033].
\bibitem{Cvetic:1999au}
  M.~Cvetic, H.~Lu and C.~N.~Pope,
  Nucl.\ Phys.\ B {\bf 574} (2000) 761
  [arXiv:hep-th/9910252].
\bibitem{Pope:1985bu}
C.~N.~Pope,
Class.\ Quant.\ Grav.\  {\bf 2} (1985) L77.
\bibitem{Chamseddine:1997nm}
  A.~H.~Chamseddine and M.~S.~Volkov,
  Phys.\ Rev.\ Lett.\  {\bf 79} (1997) 3343
  [arXiv:hep-th/9707176];
\\
  A.~H.~Chamseddine and M.~S.~Volkov,
  Phys.\ Rev.\ D {\bf 57} (1998) 6242
  [arXiv:hep-th/9711181].
\bibitem{Fuster:2005qt}
  A.~Fuster and J.~W.~van Holten,
  Phys.\ Rev.\ D {\bf 72} (2005) 024011
  [arXiv:gr-qc/0505159].
\bibitem{Maldacena:2004rf}
  J.~Maldacena and L.~Maoz,
  JHEP {\bf 0402} (2004) 053
  [arXiv:hep-th/0401024].
\bibitem{Gibbons:1976ue}
G.~W.~Gibbons and S.~W.~Hawking,
Phys.\ Rev.\ D {\bf 15} (1977) 2752.
\bibitem{Forgacs:1980zs}
P.~Forgacs and N.~S.~Manton,
Commun.\ Math.\ Phys.\ {\bf 72} (1980) 15;
\\
P.~G.~Bergmann and E.~J.~Flaherty,
J.\ Math.\ Phys.\ {\bf 19} (1978) 212.
\bibitem{Yasskin}
P.~Yasskin,
Phys. Rev. D {\bf 12} (1975)  2212.

\bibitem{Kleihaus:1997mn}
  B.~Kleihaus and J.~Kunz,
  Phys.\ Rev.\ D {\bf 57} (1998) 834
  [arXiv:gr-qc/9707045].
\bibitem{Brodbeck:1997ek}
  O.~Brodbeck, M.~Heusler, N.~Straumann and M.~S.~Volkov,
  Phys.\ Rev.\ Lett.\  {\bf 79} (1997) 4310
  [arXiv:gr-qc/9707057].
\bibitem{VanderBij:2001nm}
J.~J.~Van der Bij and E.~Radu,
Int.\ J.\ Mod.\ Phys.\ A {\bf 17} (2002) 1477
[arXiv:gr-qc/0111046];
\\
  J.~J.~van der Bij and E.~Radu,
  Int.\ J.\ Mod.\ Phys.\ A {\bf 18} (2003) 2379
  [arXiv:hep-th/0210185].
\bibitem{Paturyan:2004ps}
  V.~Paturyan, E.~Radu and D.~H.~Tchrakian,
  Phys.\ Lett.\ B {\bf 609} (2005) 360
  [arXiv:hep-th/0412011];
  \\
  B.~Kleihaus, J.~Kunz and U.~Neemann,
  Phys.\ Lett.\ B {\bf 623} (2005) 171
  [arXiv:gr-qc/0507047].
\bibitem{Galtsov:1989ip}
  D.~V.~Galtsov and A.~A.~Ershov,
  Phys.\ Lett.\ A {\bf 138} (1989) 160.
\bibitem{wald}
D. Sudarsky, R. M. Wald, Phys.Rev. {\bf D46} 1453 (1992).
\bibitem{Kleihaus:1997ws}
  B.~Kleihaus and J.~Kunz,
  Phys.\ Rev.\ D {\bf 57} (1998) 6138
  [arXiv:gr-qc/9712086].
\bibitem{Kleihaus:2002ee}
  B.~Kleihaus, J.~Kunz and F.~Navarro-Lerida,
  Phys.\ Rev.\ D {\bf 66} (2002) 104001
  [arXiv:gr-qc/0207042].


\bibitem{Balasubramanian:1999re}
V.~Balasubramanian and P.~Kraus,
Commun.\ Math.\ Phys.\  {\bf 208} (1999) 413
[arXiv:hep-th/9902121].
\bibitem{MannMisner}R.B. Mann, Phys.\ Rev.\ D {\bf 60} (1999) 104047
[arXiv:hep-th/9903229]
\bibitem{Radu:2004xp}
  E.~Radu and D.~H.~Tchrakian,
  Class.\ Quant.\ Grav.\  {\bf 22} (2005) 879
  [arXiv:hep-th/0410154].
\bibitem{Brihaye:2006nk}
  Y.~Brihaye and E.~Radu,
  Phys.\ Lett.\ B {\bf 636} (2006) 212
  [arXiv:gr-qc/0602069].
\bibitem{Mann:2002fg}
R.~B.~Mann,
Found.\ Phys.\  {\bf 33} (2003) 65
[arXiv:gr-qc/0211047].
\bibitem{Myers:1999qn}
R.~C.~Myers,
Phys.\ Rev.\ D {\bf 60} (1999) 046002.
\bibitem{Klebanov:2002ja}
  I.~R.~Klebanov and A.~M.~Polyakov,
  Phys.\ Lett.\ B {\bf 550} (2002) 213
  [arXiv:hep-th/0210114].
\bibitem{Hawking:1995ap}
  S.~W.~Hawking and S.~F.~Ross,
  Phys.\ Rev.\ D {\bf 52} (1995) 5865
  [arXiv:hep-th/9504019].
\bibitem{Leigh:2003gk}
  R.~G.~Leigh and A.~C.~Petkou,
  JHEP {\bf 0306} (2003) 011
  [arXiv:hep-th/0304217];
\\
  R.~G.~Leigh and A.~C.~Petkou,
  JHEP {\bf 0312} (2003) 020
  [arXiv:hep-th/0309177].

\bibitem{Ibadov:2004rt}
  R.~Ibadov, B.~Kleihaus, J.~Kunz and Y.~Shnir,
  Phys.\ Lett.\ B {\bf 609} (2005) 150
  [arXiv:gr-qc/0410091];
\\
  R.~Ibadov, B.~Kleihaus, J.~Kunz and M.~Wirschins,
  Phys.\ Lett.\ B {\bf 627} (2005) 180
  [arXiv:gr-qc/0507110].

\bibitem{Kleihaus:2003tn}
B.~Kleihaus, J.~Kunz and K.~Myklevoll,
Phys.\ Lett.\ B {\bf 582}, 187 (2004)
[arXiv:hep-th/0310300];
\\
  B.~Kleihaus, J.~Kunz and K.~Myklevoll,
  Phys.\ Lett.\ B {\bf 638} (2006) 367
  [arXiv:hep-th/0601124].
\bibitem{Manton:1977ht}
N.~S.~Manton,
Nucl.\ Phys.\ B {\bf 135} (1978) 319.
\bibitem{Rebbi:1980yi}
C.~Rebbi and P.~Rossi,
Phys.\ Rev.\ D {\bf 22} (1980) 2010.
\bibitem{ymh}
B.~Kleihaus and J.~Kunz,
Phys.\ Rev.\ D {\bf 61} 025003 (2000)
[hep-th/9909037];
\\
Y.~Brihaye and J.~Kunz,
Phys.\ Rev.\ D {\bf 50}, 4175 (1994)
[hep-ph/9403392];
\\
B.~Kleihaus and J.~Kunz,
Phys.\ Rev.\ D {\bf 50} 5343 (1994)
[hep-ph/9405387].


\bibitem{Chrusciel:1987jr}
P.~T.~Chrusciel and W.~Kondracki,
Phys.\ Rev.\ D {\bf 36} (1987) 1874;
\newline
D.~Sudarsky and R.~M.~Wald,
Phys.\ Rev.\ D {\bf 46} (1992) 1453.
\bibitem{Abbott:1982jh}
L.~F.~Abbott and S.~Deser,
Phys.\ Lett.\ B {\bf 116} (1982) 259;
\\
C.~H.~Oh, C.~P.~Soo and C.~H.~Lai,
Phys.\ Rev.\ D {\bf 36} (1987) 2532;
\\
J.~Tafel and A.~Trautman,
J.\ Math.\ Phys.\  {\bf 24} (1983) 1087.
\bibitem{Corichi:1999nw}
A.~Corichi and D.~Sudarsky,
Phys.\ Rev.\ D {\bf 61} (2000) 101501;
\\
A.~Corichi, U.~Nucamendi and D.~Sudarsky,
Phys.\ Rev.\ D {\bf 62} (2000) 044046.
\bibitem{Creighton:1995au}
J.~D.~Creighton and R.~B.~Mann,
Phys.\ Rev.\ D {\bf 52} (1995) 4569.
\bibitem{FIDISOL}
W. Sch\"onauer and R. Wei\ss, J. Comput. Appl. Math. {\bf 27}, 279 (1989);
\newline
M. Schauder, R. Wei\ss\ and W. Sch\"onauer,
The CADSOL Program Package,
 Universit\"at Karlsruhe, Interner Bericht Nr. 46/92 (1992);
\newline
W. Sch\"onauer and E. Schnepf,  ACM Trans. on Math. Soft. {\bf13}, 333 (1987).

\bibitem{hawking1} S.~W.~Hawking and D.~N.~Page, 
Commun.\ Math.\ Phys.\ \textbf{87} (1983)
577.

\bibitem{eli2} E. Elizalde, \textit{Ten physical applications of spectral
zeta functions} (Springer, Berlin, 1995); \newline
E. Elizalde, S.D. Odintsov, A. Romeo, A.A. Bytsenko and S. Zerbini, \textit{%
Zeta regularization techniques with applications} (World Sci., Singapore,
1994).
\bibitem{Elizalde:1988fg}
  E.~Elizalde,
  J.\ Phys.\ A {\bf 22} (1989) 931.
\bibitem{Cooper:1994eh}
  F.~Cooper, A.~Khare and U.~Sukhatme,
  Phys.\ Rept.\  {\bf 251} (1995) 267
  [arXiv:hep-th/9405029].
\end{thebibliography}
\end{document}